\documentclass{article}
\pdfoutput=1

\usepackage[table,xcdraw]{xcolor}
\definecolor{lightgray}{gray}{0.9}

\let\svthefootnote\thefootnote
\newcommand\blankfootnote[1]{%
  \let\thefootnote\relax\footnotetext{#1}%
  \let\thefootnote\svthefootnote%
}
\let\svfootnote\footnote
\renewcommand\footnote[2][?]{%
  \if\relax#1\relax%
    \blankfootnote{#2}%
  \else%
    \if?#1\svfootnote{#2}\else\svfootnote[#1]{#2}\fi%
  \fi
}

\usepackage{PRIMEarxiv}

\usepackage[utf8]{inputenc} % allow utf-8 input
\usepackage[T1]{fontenc}    % use 8-bit T1 fonts
\usepackage{hyperref}       % hyperlinks
\usepackage{url}            % simple URL typesetting
\usepackage{booktabs}       % professional-quality tables
\usepackage{amsfonts}       % blackboard math symbols
\usepackage{nicefrac}       % compact symbols for 1/2, etc.
\usepackage{microtype}      % microtypography
\usepackage{lipsum}
\usepackage{fancyhdr}       % header
\usepackage{graphicx}       % graphics
\graphicspath{{media/}} 
\usepackage{multirow}
\usepackage{multicol}
\usepackage{amsmath}
\usepackage{makecell}
\usepackage{subfig}
\usepackage{wrapfig}
\usepackage{caption}
\usepackage{diagbox}
\usepackage{fontawesome}

\usepackage{microtype}      % microtypography
\usepackage{natbib} 

\title{LLM Jailbreak Detection for (Almost) Free!
}

\author{
 \textbf{Guorui Chen\textsuperscript{1}}, \hspace{0.5mm}
 \textbf{Yifan Xia\textsuperscript{1}}, \hspace{0.5mm}
 \textbf{Xiaojun Jia\textsuperscript{2}}, \hspace{0.5mm}
 \textbf{Zhijiang Li\textsuperscript{1,$\dagger$}}, \hspace{0.5mm}
 \textbf{Philip Torr\textsuperscript{3}}, \hspace{0.5mm}
 \textbf{Jindong Gu\textsuperscript{3,$\dagger$}}
\\
 \textsuperscript{1}School of Information Management, Wuhan University, Wuhan, China
\\
 \textsuperscript{2}Nanyang Technological University, Singapore
\\
 \textsuperscript{3}Torr Vision Group, University of Oxford, Oxford, United Kingdom
\\
}

\begin{document}
\maketitle
\footnote[]{\textsuperscript{$\dagger$}The corresponding authors: \href{lizhijiang@whu.edu.cn}{lizhijiang@whu.edu.cn}, \href{jindong.gu@outlook.com}{jindong.gu@outlook.com}.}

\begin{abstract}
Large language models (LLMs) enhance security through alignment when widely used, but remain susceptible to jailbreak attacks capable of producing inappropriate content. Jailbreak detection methods show promise in mitigating jailbreak attacks through the assistance of other models or multiple model inferences. However, existing methods entail significant computational costs. In this paper, we first present a finding that the difference in output distributions between jailbreak and benign prompts can be employed for detecting jailbreak prompts. Based on this finding, we propose a Free Jailbreak Detection (FJD)~\footnote{\url{https://github.com/GuoruiC/FJD}} which prepends an affirmative instruction to the input and scales the logits by temperature to further distinguish between jailbreak and benign prompts through the confidence of the first token. Furthermore, we enhance the detection performance of FJD through the integration of virtual instruction learning. Extensive experiments on aligned LLMs show that our FJD can effectively detect jailbreak prompts with almost no additional computational costs during LLM inference.
\end{abstract}

\section{Introduction}
Large language models (LLMs) achieve remarkable success across various domains and tasks. However, the widespread use of these models has also exposed concerns, particularly their potential to generate inappropriate content. To address the concerns, recent work~\citep{wu2021recursively, ouyang2022training, rafailov2024direct} employs diverse training strategies and principles to align LLMs with human values to enhance their safety and generate responsible responses. Despite these efforts, recent jailbreak attacks can still bypass the alignment and cause harmful responses from LLMs through manual crafting~\citep{li2023multi, liu2023jailbreaking, chen2024red, yuan2023gpt, deng2023multilingual, ding2023wolf, perez2022ignore, shah2023scalable, li2023deepinception} or automated generation of prompts~\citep{zou2023universal, liu2023autodan, chao2023jailbreaking, carlini2024aligned, jones2023automatically, wen2024hard, wichers2024gradient, lapid2023open, li2024drattack, qi2023fine, deng2023jailbreaker}.

Recently, there have been emerging efforts to mitigate the risks associated with jailbreak attacks. One of the important mitigation strategies is to detect jailbreak queries that trigger LLMs to generate harmful content. Specifically, basic detection methods can be classified into three types. The first type involves computing the perplexity score of input text using an auxiliary model to detect jailbreak prompts~\citep{alon2023detecting, jain2023baseline}. The second type mutates the input into multiple copies and aggregates the responses from these copies to detect jailbreak prompts~\citep{robey2023smoothllm, zeng2024autodefense}. The third type detects outputs of jailbreak prompts with an additional classifier or the underlying model itself~\citep{yuan2024rigorllm,helbling2023llm}. However, these methods require expensive computational costs, necessitating either additional models for assistance or multiple model inferences.

~\cite{wei2024jailbroken} categorizes current jailbreaks into two types: jailbreaks with competing objectives and mismatched generalization. The first type forces the LLM to choose between safety alignment behaviors and harmful instruction objectives. The second type comes from observing that pretraining is done on a large and more diverse datasets than safety training. This mismatch can be exploited for jailbreaks. By analyzing inference outputs of the jailbreak and benign prompts, we observe that there is an obvious difference in the confidence of the first token between the responses generated by these prompts and benign ones. For both type of jailbreak prompts, they cause LLMs to have some confusion during inference, resulting in less confident responses than that on benign prompts.

Based on the initial finding, we propose a (almost) Free Jailbreak Detection (FJD) method where two techniques are introduced, Affirmative Instruction Prepending and Temperature Scaling. Affirmative Instruction Prepending prepends an affirmative instruction (e.g. \textit{"You are a good Assistant."}) to the query. The prepended instruction has minimal impact on the final output content. The output of the prepended query can be directly taken as the final output of the original query. Meanwhile, the prepended affirmative instruction can increase the response confidence of LLM to benign prompt, while it bring less or even reduce the confidence of LLM. Thus, Affirmative Instruction Prepending can be used to better detect jailbreak prompts. However, some LLMs, such as Llama, can be overconfident with responses to both jailbreak and benign prompts (the maximal probability of the first token could be very close to 1.0). Hence we introduce Temperature Scaling to better distinguishing the jailbreak and benign prompts. Furthermore, instead of prepending a manually selected instruction for FJD, we propose to learn a virtual instruction to improve detection performance, dubbed FJD-LI.

Extensive experiments are conducted to verify our observations and proposal. The effectiveness of our detection method is verified on aligned LLMs such as Vicuna~\citep{chiang2023vicuna}, Llama2~\citep{touvron2023llama}, and Guanaco~\citep{dettmers2024qlora} under various jailbreak attacks. Furthermore, we show the effectiveness of our FJD against transferable jailbreak attacks to Llama3 \footnote{\url{https://github.com/meta-llama/llama3}} and ChatGPT3.5~\citep{achiam2023gpt}. Our detection method outperforms the baseline methods significantly and requires almost no additional computational costs during LLM inference. Our contributions can be summarized as follows: 

\begin{itemize}

     \item We present a finding that the difference in output distributions between jailbreak and benign prompts can be employed for detecting jailbreak prompts. 

     \item Based on observation, We propose a Free Jailbreak Detection (FJD) method by prepending affirmative instructions into the inputs and scaling the logits by temperature which requires almost no additional costs. 

     \item Furthermore, we propose to learn virtual instructions (FJD-LI) to further improve jailbreak detection performance. 

     \item Extensive experiments are conducted under various jailbreak attacks with competing objectives and mismatched generalization.    
\end{itemize}

\begin{figure*}[t]
    \centering
    \subfloat[AutoDAN vs. Benign Prompt in Llama2 7B]{
        \begin{minipage}{0.48\linewidth}
    	
    		\centerline{\includegraphics[width=\textwidth]{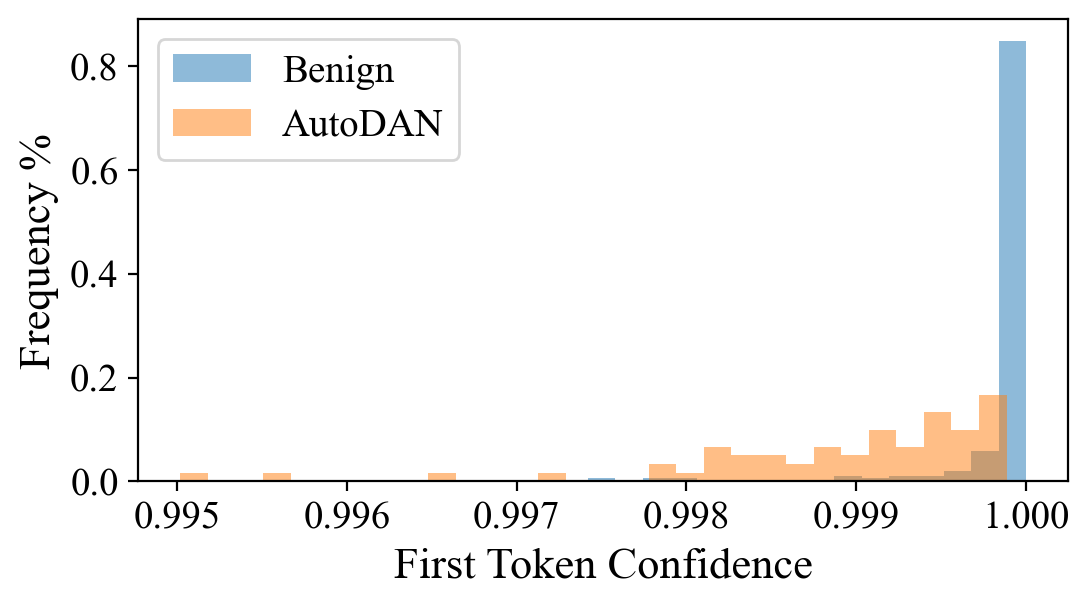}}
    	\end{minipage}
    }
    \subfloat[Cipher vs. Benign Prompt in Llama2 7B]{
    	\begin{minipage}{0.48\linewidth}
    		
    		\centerline{\includegraphics[width=\textwidth]{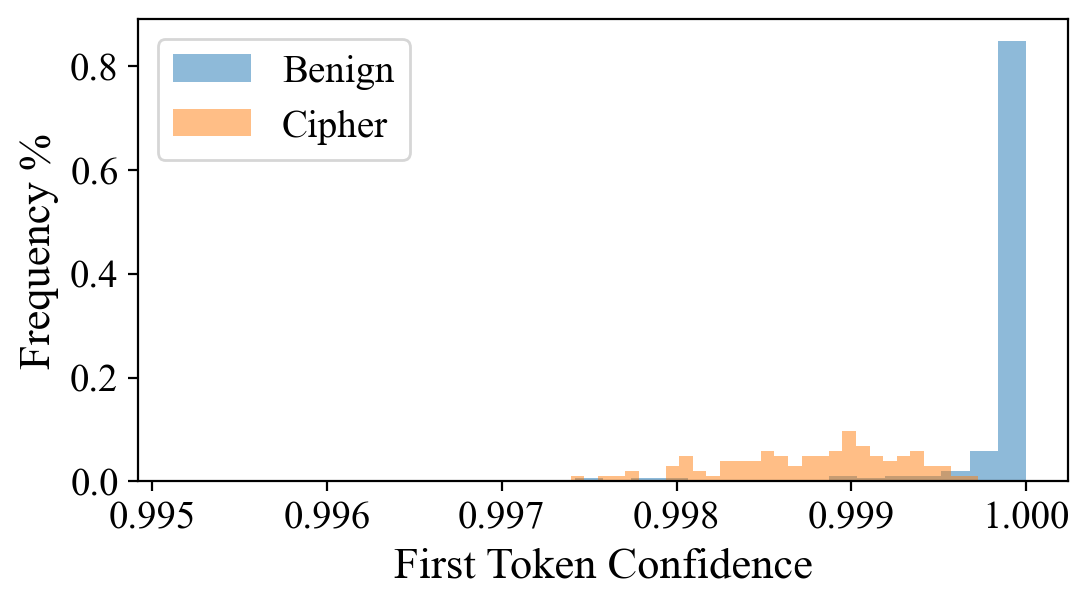}}
    	
    	\end{minipage}
    }
    \caption{The distribution of the confidence scores of the predicted first tokens over jailbreak and benign samples is shown. A difference can be observed where LLMs are less confident on Jailbreak samples than on benign samples.}

	\label{finding}

\end{figure*}

\section{Related Work}

\textbf{Jailbreak Attack} Jailbreak attacks can mislead LLMs to respond to harmful queries. These works~\citep{website1, website2} initially reported that hand-crafted prompts can jailbreak LLMs. Currently, jailbreak attacks against LLMs can be divided into two categories: competing objectives and mismatched generalization~\citep{wei2024jailbroken}. The first category forces the LLM to choose between safety training behaviors and harmful instruction objectives by crafting prompts. E.g., GCG~\citep{zou2023universal} automatically generate transferable adversarial suffixes by employing gradient-based search methods. AutoDAN~\citep{liu2023autodan} employed mutation and crossover operations within genetic algorithms to produce natural adversarial prefixes. The second category exploits data beyond the safety fine-tuning of the LLMs for jailbreak attacks. E.g., Yong et al.~\citep{yong2023low} achieved LLMs jailbreak by devising strategies that convert user prompts into low-resource languages. In contrast to hand-crafted methods,  Cipher~\citep{yuan2023gpt} uses system role descriptions and few-shot enciphered demonstrations to bypass the safety alignment. As LLMs grow in complexity and capability, more jailbreak attacks~\citep{jia2024improved, liu2023jailbreaking,  wei2024jailbroken, ding2023wolf, chao2023jailbreaking, zhang2024boosting, paulus2024advprompter} based on those methods have been developed.

\textbf{Jailbreak Defense and Detection} To deal with jailbreak attacks on aligned LLMs, defense methods aim to reduce the success rate of the attack, while detection methods distinguish between jailbreak and benign prompts to safeguard LLMs. Current defense and detection methods can be divided into three types. The first type, a simple and effective method~\citep{alon2023detecting, jain2023baseline}, involves computing the perplexity score of the input for detection by employing the negative log-likelihood. In addition, to enable LLMs to produce inappropriate responses, attackers must carefully craft the jailbreak prompt. Consequently, the second type~\citep{robey2023smoothllm, zhang2023mutation, cao2023defending, zhang2023defending, kumar2023certifying, rao2023tricking} generate multiple copies by randomly deleting, replacing, or modifying consecutive character, and aggregate the responses from multiple LLMs to mitigate the success rate of the attack. And the third type~\citep{yuan2024rigorllm, helbling2023llm, xie2023defending} employ an additional classifier model or LLMs itself to detect jailbreak prompts such as appending the prompt "\textit{Is it harmful?}" to the response or modifying the system prompt of LLM. Current defense and detection methods necessitate extra model inferences, resulting in significant computational costs. In this work, we propose a nearly free jailbreak detection method, which is a confidence-based method. Similarly, existing confidence-based methods~\citep{xu2024safedecoding, candogan2025single} also fall into the aforementioned three categories and require auxiliary models and additional inference during detection or defense.

\section{Approach}\label{FJD}

In this section, we describe the problem formulation in Sec.~\ref{sec3:Prom}, and introduce our proposed methods FJD with Affirmative Instruction Prepending and Temperature Scaling in Sec.~\ref{sec3:FJD} and the variants of FJD in Sec.~\ref{sec3:Variants}.

\subsection{Problem Formulation}\label{sec3:Prom}

Jailbreak attacks can be classified into two categories: competing objectives and mismatched generalization~\citep{wei2024jailbroken}.

\textbf{Competing Objectives} Jailbreak attacks~\citep{zou2023universal, liu2023autodan} are designed to search for some jailbreak prompt $x_{jail}$ so that the probability of harmful output $\hat{g} $ is maximized, which forces the LLM to choose between safety training behaviors and harmful instruction objectives. Formally, given an input sequence of tokens $ x_q $, the attack can be formulated as minimizing the loss between model output and the target output, $ min_{x_{jail}\in [|\mathcal{V}|]^n}\ \mathcal{L} (p(x_q\oplus x_{jail}), \hat{g}) $, where $\oplus$ is defined as the concatenation operator of two sequence as: $ x_q\oplus x_{jail}$, $p(\cdot)$ represents the output probabilities predicted by LLMs, $\mathcal{V}$ is the vocabulary, and $n$ is the length of tokens. 

\textbf{Mismatched Generalization} This type of method~\citep{yuan2023gpt, chen2024red} comes from observing that pretraining is done on a large and more diverse datasets than safety training. For this mismatch, LLM will respond without safety considerations, such as Base64 on inputs.

Jailbreak detection approaches distinguishes between jailbreak and benign prompts using a specific metric. For a given input sequence, a benign query $x_{beni}$ or a jailbreak query $x_{jail}$, the jailbreak detector $\mathnormal{g}(\cdot)$ aims to achieve this property: $\mathnormal{g}(x_{jail}) < \mathrm{T} \leq \mathnormal{g}(x_{beni})$ or $\mathnormal{g}(x_{jail}) > \mathrm{T} \geq \mathnormal{g}(x_{beni})$, where $\mathrm{T}$ represents a pre-defined threshold.

\subsection{Free Jailbreak Detection Approach}\label{sec3:FJD}

Current jailbreak attacks can be classified into two categories: competing objectives and mismatched generalization. Both might impact the confidence generated by LLMs. As shown in Fig.~\ref{finding}, we conduct a statistical analysis on the first token confidence produced by jailbreak prompts (AutoDAN and Cipher) and benign ones (PureDove) on Llama2 7B. We find that there is an obvious difference in the confidence of the first token between the responses generated by these prompts and benign ones. Similar observations on other models and the theoretical analysis are shown in Appendix~\ref{app:observ}.

\definecolor{dd}{RGB}{0,176,80} 
\definecolor{MI}{RGB}{255,0,0} 
\begin{figure*}[!h]
    \centering
    \includegraphics[width=0.98\textwidth]{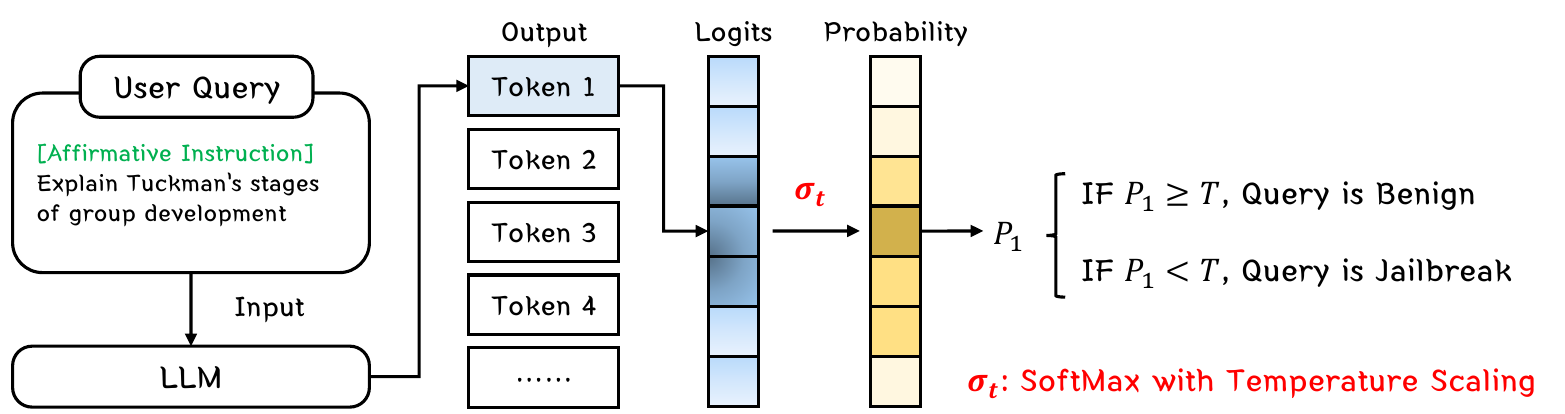}
    \caption{Jailbreak prompt Detection through FJD: By prepending an \textcolor{dd}{affirmative instruction} and \textcolor{MI}{scaling the logits with temperature}, the first token confidence in the LLMs' responses to the benign prompts is higher than a predefined threshold, whereas the confidence for jailbreak prompts can be lower than the threshold.}

    \label{FJD f}
\end{figure*}

Based on the findings, we identify the potential of utilizing the confidence of the first tokens to detect jailbreak prompts. Since the output probabilities can be obtained in the standard forward pass, we dub our method Free Jailbreak Detection (FJD), where two techniques are introduced to enlarge the confidence difference, i.e., Affirmative Instruction Prepending and Temperature Scaling. We now present how the two techniques improve detection performance.

\textbf{Affirmative Instruction Prepending} This technique prepends an affirmative instruction to the given query to enlarge the confidence differences between jailbreak and benign prompts. Affirmative Instruction is referred as the ones that confirm the original capability of LLMs e.g., \textit{"You are a good Assistant."}, \textit{"Please following user instructions accurately."}, which is widely adopted in various applications—for instance, incorporating them into system prompts to enhance model reasoning performance. With such prepended instructions, the outputs of benign samples are similar or even better than before, which can be sent to user directly without a second inference. Meanwhile, the confidence of the predicted first token (i.e., the maximal probability over vocabulary) on benign prompts increases when equipped with an affirmative instruction. Compared to that on benign prompts, the increased confidence on jailbreak is minor. The reason behind is that affirmative instructions prepended to jailbreak prompts receive less attention in LLMs given the fact that jailbreak prompts attract model attentions significantly~\citep{arditi2024refusal}. Namely, without impairing model outputs on benign prompts, the difference of the first token confidence between jailbreak and benign prompts can be enlarged by prepending affirmative instructions. More discussion is in Sec.~\ref{sec:dpa}. 
 
Formally, given an input sequence $x_q$ and an affirmative instruction $x_{ai}$, the procedure for detecting jailbreak prompts is as follows. The confidence of the first tokens is computed as

\begin{equation}
    P_1=\sigma(f_{1}(x_{ai}\oplus x_q))
\end{equation} 

where, $f_i(\cdot)$ represents the output logits of the $i$-th token, and $\sigma(\cdot)$ obtains the maximal probability value over the vocabulary tokens through the softmax function.

\textbf{Temperature Scaling} Prepended Affirmative Instructions enlarge the confidence difference by increasing confidence differently on jailbreak and benign prompts. However, it does not work well when LLMs are overconfident with responses. In our experiments, we also observe that LLMs (e.g. Llama) can be overconfident on both jailbreak and benign prompts where the maximal probability of the first token could be even very close to 1.0. To address the challenge, we propose to apply temperature scaling to avoid overconfident outputs.

To illustrate why temperature scaling can change the confidence rank between two samples, we provide a dummy example: Given the sample A with the output logits \textit{[10, 9, 1]} and the sample B with \textit{[10, 8, 8]}, their output probabilities are \textit{[0.731, 0.269, 0.0001]} and \textit{[0.787, 0.106, 0.106]} respectively when the temperature of the softmax function is set to 1.0. Namely, model responses are more confident about sample B (0.787) than sample A (0.731). After temperature scaling by setting the temperature to 2.0, their output probabilities become \textit{[0.619, 0.375, 0.007]} and \textit{[0.576, 0.212, 0.212]} respectively where the confidence of sample B become lower than that of sample A. More rigorous analysis and an instance are in Appendix~\ref{app:proof}.

Formally, given an input sequence $ x_q $, the affirmative instruction $x_{ai}$ and the temperature $\tau$, the confidence of the first tokens with temperature scaling is computed as

\begin{equation}
    \label{c0}
    P_{1,\tau}=\sigma_\tau(f_{1}(x_{ai}\oplus x_q)/\tau)
\end{equation} 

where, $f_i(\cdot)$ represents the output logits of the $i$-th token, and $\sigma_\tau(\cdot)$ obtains the maximal probability value over the vocabulary tokens through the softmax function with temperature scaling.

Then, the confidence $P_{1,\tau}$ can be used to detect jailbreak prompts by comparing it with a predefined threshold. If $P_{1,\tau} < \mathrm{T}$, the input will be flagged as a jailbreak prompt. Otherwise, it will be flagged as a benign prompt allowing LLMs to output. Note that we apply AUC score for experimental evaluation where all the thresholds are considered.

The detection process of FJD can be integrated into the standard model forward inference. As the affirmative instructions prepended by FJD are short and the temperature scaling has no influence on model inference, the additional computational costs of model inference is almost free. In contrast, previous jailbreak detection methods require one or many extra forward passes.

\subsection{Improved Version based on FJD}\label{sec3:Variants}

Although various affirmative instructions of FJD works well across various models and jailbreak attacks, the careful selection of the instruction can still further improve detection performance. Instead of manual design, we introduce a learnable virtual instruction built upon FJD (FJD-LI). Formally, given an input sequence $ x_q $,  the affirmative instruction $x_{ai}$ and the tokenization function $E(x)$, the embedding of $ x_q $ and $x_{ai}$ is $ e_q = E(x_q);\ e_{mi}=E(x_{ai}) $, where $e_q\in \mathbb{R}^{q\times d}$ and $e_{mi}\in \mathbb{R}^{m\times d}$, $q$ and $m$ are the number of tokens and $d$ is the number of embedding dimensions. The goal of the instruction learning is to minimize token confidence for jailbreak prompts and maximize it for benign prompts. We keep $e_{mi}$ learnable and update it with the loss which can be expressed as follows

\vspace{-0.5cm}
\begin{equation*}
    \footnotesize
    \begin{split}
    \mathcal{L}(e_q) = 
    \begin{cases}
        \displaystyle KL(p_1(e_{mi}\oplus e_q) \Vert \displaystyle M_o(l)),\ if\ e_q\in E(X_{beni})\\
        \displaystyle KL(p_1(e_{mi} \oplus e_q) \Vert \displaystyle M_u(l)),\ if\ e_q\in E(X_{jail})\\
    \end{cases}    
    \end{split}
    \label{equ:loss}
\end{equation*}

where, $\displaystyle KL(\cdot \Vert \cdot)$ is to calculate the Kullback-Leibler Divergence~\citep{kullback1951information} and $l$ is the length of the vocabulary. $p_1(\cdot)$ represents the output probability distribution of the first token. $\displaystyle M_o(l)\in \mathbb{R}^{1\times l}$ is a one-hot matrix of $l$ dimensions, where the position of the maximum value in the logits $p(e_q)_1$ is set to 1 and the rest to 0. $\displaystyle M_u(l)\in \mathbb{R}^{1\times l}$ is a uniform distribution of $l$ dimensions. The final virtual instruction is $e_{li} = \min_{e_{mi}\in\mathbb{R}^{m\times d}} \mathcal{L}(e_q)$.

Once $e_{li}$ is obtained, FJD-LI can be applied to detect jailbreak prompts by replacing $e_{mi}$ with $e_{li}$ in detection process. It requires only a small number of samples for learning and does not increase the inference costs of LLMs compared to FJD.

\section{Experiment}

In this section, we first evaluate FJD under various jailbreak attacks and conduct ablation analysis of FJD. We then evaluate the detection effectiveness of FJD-LI. Finally, we discuss the efficiency, detection-aware jailbreak attack of FJD.

\subsection{Experimental Setting}\label{sec:es}

\textbf{Large language models} Six open-source LLMs are taken for the jailbreak detection: Vicuna 7B/13B~\citep{chiang2023vicuna}, Llama2-chat 7B/13B~\citep{touvron2023llama} and Guanaco 7B/13B~\citep{dettmers2024qlora}. We further evaluate the detection of transferable jailbreak
attacks on Llama3 and ChatGPT3.5~\citep{achiam2023gpt}.

\textbf{Dataset} To evaluate the performance of FJD, we consider the jailbreak datasets AdvBench~\citep{zou2023universal}, and PureDove~\citep{daniele2023amplify-instruct}, Open-Platypus~\citep{lee2023platypus} and SuperGLUE~\citep{wang2019superglue} as benign datasets. To align benign prompts with jailbreak ones, we randomly select an equal number of benign prompts from the datasets. Then we allocate 50\% of the dataset as the training set for training the virtual instruction in FJD-LI. More details about dataset are in Appendix~\ref{app:a}.

\textbf{Jailbreak attacks} Two types of jailbreak attacks are considered, i.e., 1) via competing objectives (CO): AutoDAN~\citep{liu2023autodan} and Hand (CO)~\citep{chen2024red}. and 2) via mismatched generalization (MG): Cipher~\citep{yuan2023gpt} and Hand (MG). Note that Hand-crafted attacks provide 28 different attacks. Based on this work~\citep{wei2024jailbroken}, the 28 attacks are grounded into Hand (CO) and Hand (MG). Additional information regarding the classification and detection results of hand-crafted attacks can be found in the Appendix~\ref{app:hand}. We further consider transferable jailbreak attacks including the aggregation the prompt from GCG~\citep{zou2023universal} and AutoDAN. And more details are in Appendix~\ref{app:b}.

\textbf{Bselines} We compare our method with three jailbreak detection methods: PPL~\citep{alon2023detecting}, SmoothLLM~\citep{robey2023smoothllm} and GradSafe~\citep{xie2024gradsafe}. More details about Baselines are in Appendix~\ref{app:defense}.

\textbf{Metric} In all experiments, AUC scores of detections are reported where all the thresholds are considered. The higher the score is, the better the detection performance is. We randomly select 80\% of the test dataset and conduct 5 repeated experiments. More metrics (FPR, TPR, F1) are also reported in Appendix~\ref{app:co} and~\ref{app:mg}.

\begin{table*}[t]
    \centering
  
    \caption{Detection results (AUC) of jailbreak prompt under attacks via competing objectives. FJD outperforms the baseline in all attacks and LLMs with almost no additional computational costs during LLM inference.}

    \setlength\tabcolsep{8pt}
    \scriptsize
    \begin{tabular}{cccccccc}
        \toprule
        \textbf{Attack} & \textbf{Method} & \textbf{Llama2-7B} &  \textbf{Vicuna-7B} &  \textbf{Guanaco-7B} & \textbf{Llama2-13B} &  \textbf{Vicuna-13B} &  \textbf{Guanaco-13B} \\
        \midrule

        \multirow{4}{*}{AutoDAN}
        & PPL & 0.8172\tiny{$\pm 0.0017$} & 0.7452\tiny{$\pm 0.0012$} & 0.7964\tiny{$\pm 0.0004$}& 0.7018\tiny{$\pm 0.0002$} & 0.7889\tiny{$\pm 0.0002$} & 0.7703\tiny{$\pm 0.0005$} \\
        & SMLLM & 0.8197\tiny{$\pm 0.0052$} & 0.7831\tiny{$\pm 0.0035$} & 0.6704\tiny{$\pm 0.0036$}& 0.8360\tiny{$\pm 0.0021$} & 0.5116\tiny{$\pm 0.0044$} & 0.5583\tiny{$\pm 0.0038$} \\
        & GradSafe & 0.8025\tiny{$\pm 0.0089$} & 0.7893\tiny{$\pm 0.0020$} & 0.8194\tiny{$\pm 0.0051$} & 0.9123\tiny{$\pm 0.0029$} & 0.9225\tiny{$\pm 0.0005$} & 0.7398\tiny{$\pm 0.0063$} \\
        & FT & 0.8869\tiny{$\pm 0.0149$} & 0.1709\tiny{$\pm 0.0083$} & 0.7084\tiny{$\pm 0.0106$} & 0.8899\tiny{$\pm 0.0141$} & 0.0471\tiny{$\pm 0.0040$} & \textbf{0.7710}\tiny{$\pm 0.0172$} \\
       \rowcolor{lightgray} & FJD & \textbf{0.9578}\tiny{$\pm 0.0088$} & \textbf{0.7964}\tiny{$\pm 0.0182$} & \textbf{0.8946}\tiny{$\pm 0.0065$}& \textbf{0.9214}\tiny{$\pm 0.0133$} & \textbf{0.9373}\tiny{$\pm 0.0111$} & 0.7470\tiny{$\pm 0.0135$}  \\
        \midrule
        
        \multirow{4}{*}{Hand (CO)} 
        & PPL & 0.5326\tiny{$\pm 0.0025$} & 0.5304\tiny{$\pm 0.0007$} & 0.5255\tiny{$\pm 0.0005$} & 0.5259\tiny{$\pm 0.0023$} & 0.5287\tiny{$\pm 0.0006$} & 0.4909\tiny{$\pm 0.0007$}\\
        & SMLLM & 0.7129\tiny{$\pm 0.0105$} & 0.6616\tiny{$\pm 0.0056$} & 0.7033\tiny{$\pm 0.0065$} & 0.7193\tiny{$\pm 0.0110$} & 0.7473\tiny{$\pm 0.0075$} & 0.7226\tiny{$\pm 0.0091$}\\
        & GradSafe & 0.9392\tiny{$\pm 0.0041$} & 0.7877\tiny{$\pm 0.0061$} & 0.7795\tiny{$\pm 0.0052$}  & 0.9619\tiny{$\pm 0.0036$} & 0.7967\tiny{$\pm 0.0055$} & 0.7396\tiny{$\pm 0.0079$} \\
        & FT & 0.9244\tiny{$\pm 0.0043$} & 0.4312\tiny{$\pm 0.0156$} & 0.5618\tiny{$\pm 0.0175$} & 0.8284\tiny{$\pm 0.0167$} & 0.5510\tiny{$\pm 0.0166$} & 0.6265\tiny{$\pm 0.0177$} \\
       \rowcolor{lightgray} & FJD & \textbf{0.9640}\tiny{$\pm 0.0067$} & \textbf{0.8048}\tiny{$\pm 0.0135$} & \textbf{0.8310}\tiny{$\pm 0.0123$} & \textbf{0.9650}\tiny{$\pm 0.0044$} & \textbf{0.9494}\tiny{$\pm 0.0089$} & \textbf{0.8442}\tiny{$\pm 0.0141$} \\

        \bottomrule

    \end{tabular}
    
    \label{auc}
\end{table*}

\begin{table*}[t]
    \centering
    \caption{Detection results (AUC) of jailbreak prompt under attacks via mismatched generalization. FJD outperforms the baseline in all attacks and LLMs with almost no additional computational costs during LLM inference.}

    \setlength\tabcolsep{8pt}
    \scriptsize
    \begin{tabular}{cccccccc}
        \toprule
        \textbf{Attack} & \textbf{Method} & \textbf{Llama2-7B} &  \textbf{Vicuna-7B} &  \textbf{Guanaco-7B} & \textbf{Llama2-13B} &  \textbf{Vicuna-13B} &  \textbf{Guanaco-13B}\\
        \midrule

        \multirow{4}{*}{Cipher}
        & PPL & 0.0070\tiny{$\pm 0.0005$} & 0.0266\tiny{$\pm 0.0004$} & 0.0248\tiny{$\pm 0.0005$} & 0.0221\tiny{$\pm 0.0011$} & 0.0259\tiny{$\pm 0.0005$} & 0.0254\tiny{$\pm 0.0008$}\\
        & SMLLM & 0.5034\tiny{$\pm 0.0024$} & 0.5233\tiny{$\pm 0.0009$} & 0.5460\tiny{$\pm 0.0036$} & 0.9096\tiny{$\pm 0.0105$} & 0.5344\tiny{$\pm 0.0025$} & 0.5482\tiny{$\pm 0.0020$} \\
        & GradSafe & 0.7862\tiny{$\pm 0.0045$} & 0.7094\tiny{$\pm 0.0201$} & 0.8112\tiny{$\pm 0.0088$} & 0.8723\tiny{$\pm 0.0073$} & 0.7972\tiny{$\pm 0.0036$} & 0.7691\tiny{$\pm 0.0105$} \\
        & FT & 0.9636\tiny{$\pm 0.0025$} & 0.7966\tiny{$\pm 0.0055$} & 0.4905\tiny{$\pm 0.0173$} & 0.9837\tiny{$\pm 0.0031$} & 0.3030\tiny{$\pm 0.0150$} & 0.4724\tiny{$\pm 0.0148$} \\
        \rowcolor{lightgray} & FJD & \textbf{0.9896}\tiny{$\pm 0.0014$} & \textbf{0.8633}\tiny{$\pm 0.0033$} & \textbf{0.8299}\tiny{$\pm 0.0043$} & \textbf{0.9909}\tiny{$\pm 0.0091$} & \textbf{0.8876}\tiny{$\pm 0.0170$} & \textbf{0.8216}\tiny{$\pm 0.0191$}\\
        \midrule
        
        \multirow{4}{*}{Hand (MG)} 
        & PPL & 0.6854\tiny{$\pm 0.0014$} & 0.6827\tiny{$\pm 0.0013$} & 0.6781\tiny{$\pm 0.0006$}& 0.6787\tiny{$\pm 0.0016$} & 0.6797\tiny{$\pm 0.0007$} & 0.6771\tiny{$\pm 0.0010$}\\
        & SMLLM & 0.7146\tiny{$\pm 0.0111$} & 0.7155\tiny{$\pm 0.0070$} & 0.8232\tiny{$\pm 0.0076$} & 0.7587\tiny{$\pm 0.0081$} & 0.6695\tiny{$\pm 0.0091$} & 0.7591\tiny{$\pm 0.0131$}\\
        & GradSafe & 0.8777\tiny{$\pm 0.0058$} & 0.7864\tiny{$\pm 0.0049$} & 0.8265\tiny{$\pm 0.0055$} & 0.8501\tiny{$\pm 0.0068$} & 0.8185\tiny{$\pm 0.0039$} & 0.7708\tiny{$\pm 0.0056$}\\
        & FT & 0.9229\tiny{$\pm 0.0055$} & 0.5625\tiny{$\pm 0.0145$} & 0.4885\tiny{$\pm 0.0126$} & 0.7557\tiny{$\pm 0.0145$} & 0.6600\tiny{$\pm 0.0168$} & 0.5268\tiny{$\pm 0.0019$}\\
        \rowcolor{lightgray} & FJD & \textbf{0.9549}\tiny{$\pm 0.0072$} & \textbf{0.7937}\tiny{$\pm 0.0160$} & \textbf{0.8882}\tiny{$\pm 0.0153$} & \textbf{0.9444}\tiny{$\pm 0.0085$} & \textbf{0.9510}\tiny{$\pm 0.0104$} & \textbf{0.8395}\tiny{$\pm 0.0171$}\\

        \bottomrule

    \end{tabular}
    \label{pair hand}
\end{table*}

\begin{table*}[t]
    \centering
    \caption{Detection results (AUC) of jailbreak prompt with and without Affirmative Instruction (AI) and Temperature Scaling (TS) modules in FJD. Both modules can improve detection performance.}
    \vspace{-0.2cm}

    \setlength\tabcolsep{8pt}
    \scriptsize
    \begin{tabular}{ccccccccc}
        \toprule
        
        \multirow{2}{*}{\textbf{Method}} & \multirow{2}{*}{\textbf{AI}} & \multirow{2}{*}{\textbf{TS}} & \multicolumn{3}{c}{\textbf{AutoDAN}} & \multicolumn{3}{c}{\textbf{Cipher}} \\
        \cmidrule(r){4-6}
        \cmidrule(r){7-9}
        & & & \textbf{Llama2-7B} &  \textbf{Vicuna-7B} &  \textbf{Guanaco-7B} & \textbf{Llama2-7B} &  \textbf{Vicuna-7B} &  \textbf{Guanaco-7B}\\
        \midrule

        & \faTimes & \faTimes & 0.8737\tiny{$\pm 0.0124$} & 0.1617\tiny{$\pm 0.0057$} & 0.6588\tiny{$\pm 0.0142$} & 0.9214\tiny{$\pm 0.0032$} & 0.6399\tiny{$\pm 0.0096$} & 0.4826\tiny{$\pm 0.0152$}\\
        & \faCheck & \faTimes & 0.9436\tiny{$\pm 0.0076$} & 0.7862\tiny{$\pm 0.0032$} & 0.8447\tiny{$\pm 0.0076$} & 0.9682\tiny{$\pm 0.0037$} & 0.8569\tiny{$\pm 0.0029$} & 0.8167\tiny{$\pm 0.0034$}\\
        FT & \faTimes & \faCheck & 0.8869\tiny{$\pm 0.0149$} & 0.1709\tiny{$\pm 0.0083$} & 0.7084\tiny{$\pm 0.0106$} & 0.9636\tiny{$\pm 0.0025$} & 0.7966\tiny{$\pm 0.0055$} & 0.4905\tiny{$\pm 0.0173$} \\
        \rowcolor{lightgray} FJD & \faCheck & \faCheck & \textbf{0.9578}\tiny{$\pm 0.0088$} & \textbf{0.7964}\tiny{$\pm 0.0182$} & \textbf{0.8946}\tiny{$\pm 0.0065$} & \textbf{0.9896}\tiny{$\pm 0.0014$} & \textbf{0.8633}\tiny{$\pm 0.0033$} & \textbf{0.8299}\tiny{$\pm 0.0043$}\\

        \bottomrule
    \end{tabular}

    \vspace{-0.5cm}
    \label{ablation}
\end{table*}

\subsection{Jailbreak Detection under Attacks with Competing Objectives}\label{sec:WA}

To evaluate the detection of jailbreak prompts via competing objectives for our approach, which comprises First Token (FT) and FJD, we conducted experiments on two attacks: AutoDAN and Hand (CO). Tab.~\ref{auc} shows that FJD can effectively detect jailbreak prompts via competing objectives on almost all LLMs. The optimized jailbreak attack (AutoDAN) generates higher token confidence than benign prompts, making FT difficult to detect on some LLMs. Hand-crafted prompts exhibit low perplexity, making PPL difficult to detect. And more detection results under other jailbreak attacks via competing objectives are in Appendix~\ref{app:co}.

\subsection{Jailbreak Detection under Attacks with Mismatched Generalization}\label{sec:BA}

To investigate the effectiveness of FJD in detecting jailbreak prompts via mismatched generalization, we conducted experiments on two attacks: Cipher and Hand (MG). Tab.~\ref{pair hand} illustrates that FJD achieves superior performance across almost all LLMs. Cipher, constructed with a fixed format and some manual examples, exhibits lower perplexity than benign prompts, making PPL difficult to detect. More detection results under other jailbreak attacks via mismatched generalization are in Appendix~\ref{app:mg}. 

\begin{wraptable}{r}{0.5\textwidth}
    
    \vspace{-0.5cm}
      \caption{Detection results (AUC) of jailbreak prompt under transferable attacks. FJD can effectively detect jailbreak prompts from transferable attacks in most cases.}
    
      \label{transfer}
      \centering
      \setlength\tabcolsep{7pt}
      \scriptsize
      \begin{tabular}{cccc}
        \toprule
        \textbf{\diagbox{Source}{Target}} & \textbf{Method}  & \textbf{Llama3-8B} & \textbf{ChatGPT-3.5}\\
  
        \midrule
        \multirow{3}{*}{Vicuna-7B} 
        & PPL  & 0.7040\tiny{$\pm 0.0022$} & 0.8141\tiny{$\pm 0.0014$}\\
        & SMLLM  & 0.8585\tiny{$\pm 0.0061$} & 0.8938\tiny{$\pm 0.0057$}\\
        & GradSafe  & 0.8629\tiny{$\pm 0.0024$} & - \\
        \rowcolor{lightgray}& FJD &  \textbf{0.8768}\tiny{$\pm 0.0087$} & \textbf{0.9553}\tiny{$\pm 0.0073$}\\
        \midrule
        \multirow{3}{*}{Llama2-7B}
        & PPL  & 0.7551\tiny{$\pm 0.0037$} & 0.8138\tiny{$\pm 0.0010$}\\
        & SMLLM  & 0.8662\tiny{$\pm 0.0041$} & 0.8333\tiny{$\pm 0.0055$} \\
        & GradSafe & 0.8908\tiny{$\pm 0.0039$} & - \\
        \rowcolor{lightgray}& FJD  & \textbf{0.9013}\tiny{$\pm 0.0075$} & \textbf{0.9496}\tiny{$\pm 0.0060$}\\
        \midrule
        \multirow{3}{*}{Guanaco-7B}
        & PPL  & 0.9256\tiny{$\pm 0.0014$} & 0.7173\tiny{$\pm 0.0025$}\\
        & SMLLM   & 0.8687\tiny{$\pm 0.0057$} & 0.9425\tiny{$\pm 0.0032$}\\
        & GradSafe  & 0.9143\tiny{$\pm 0.0059$} & - \\
        \rowcolor{lightgray}& FJD   & \textbf{0.9350}\tiny{$\pm 0.0077$} & \textbf{0.9432}\tiny{$\pm 0.0089$}\\
        \bottomrule
    
      \end{tabular}

    \vspace{-1.0cm}
\end{wraptable}

\subsection{Jailbreak Detection under Transferable Jailbreak Attacks}

For detecting transferable jailbreak attacks, this experiment employs Llama2, Vicuna and Guanaco as the source models and aggregates prompts acquired from GCG and AutoDAN. Subsequently, we further evaluate Llama3 8B and ChatGPT3.5 as the target models. And Tab.~\ref{transfer} shows the detection results of our FJD against transferable jailbreak attacks. For the successfully transferable prompt, FJD demonstrates a more effective detection capability in most cases than baselines. Since GradSafe requires the gradients of LLMs, it cannot be used for detection on ChatGPT. In contrast, FJD can leverage ChatGPT's API to obtain the probability values of generated tokens for detection. And more detection results are in Appendix~\ref{app:transfer}. Moreover, investigating the generality of FJD in multimodal scenarios is indeed interesting. The detection results and discussions are in Appendix~\ref{app:multimodal}.

\subsection{Analysis of Affirmative Instructions} \label{sec:dpa}

\begin{wrapfigure}{r}{0.5\textwidth}
    \vspace{-2.5cm}
    \centering
    \includegraphics[width=0.48\textwidth]{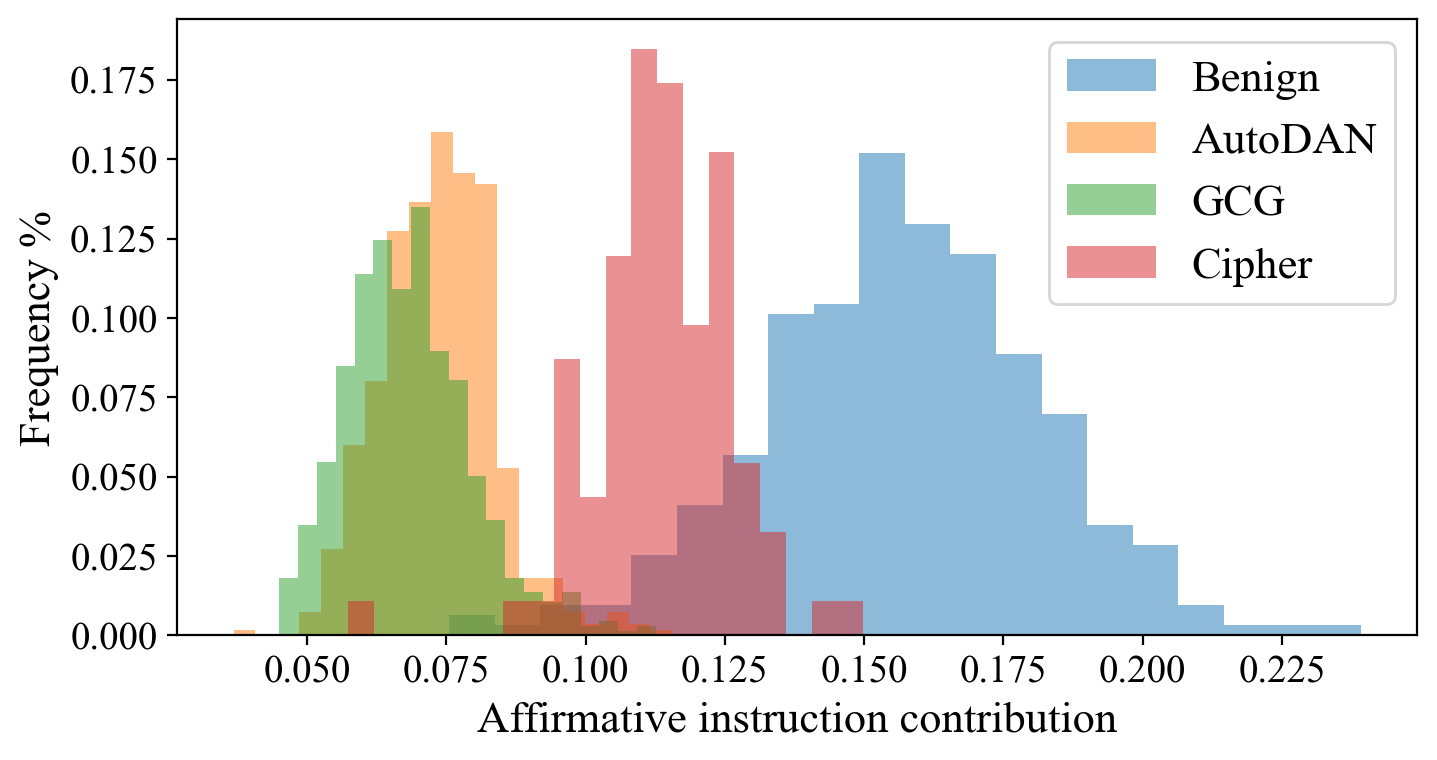}
    \vspace{-0.2cm}
    \caption{Affirmative instruction contribution and the frequency of data volume for the first tokens in Vicuna 7B. The contribution of affirmative instruction for the benign prompts is higher than the jailbreak prompts.}
    \label{contribution}
    \vspace{1.5cm}
\end{wrapfigure}

To investigate the difference between model responses to jailbreak and benign prompts with prepended affirmative instructions, we use saliency~\citep{sarti2023inseq} to perform attribution analysis on the first token generated by LLMs. Fig.~\ref{contribution} shows the 

\begin{wraptable}{r}{0.5\textwidth}

  \caption{Detection results (AUC) of jailbreak prompt while using the affirmative instruction variants. FJD can effectively detect jailbreak prompts using various affirmative instructions.}
  \vspace{-0.2cm}

  \label{different}
  \centering
  \setlength\tabcolsep{7pt}
  \scriptsize
  \begin{tabular}{lcc}
    \toprule
    \textbf{Different Affirmative Instruction}  & \textbf{AutoDAN}  & \textbf{Cipher}  \\
    \midrule
    Without affirmative instruction
    & 0.8869 & 0.9636 \\
    \midrule
    Respond in a way that promotes safety.
    & 0.9624 & 0.9769 \\
    Provide an answer that prioritizes user safety.
    & 0.9163 & 0.9555\\

    Remember you are a responsible and helpful LLM.
    & 0.9350 & 0.9403  \\
    Let's think step by step.
    & 0.9610 & 0.9578  \\
    \bottomrule
  \end{tabular}

\end{wraptable}

contribution of the instruction for jailbreak and benign prompts on Vicuna 7B. It has been observed that the affirmative instruction integrated by FJD notably influences the responses to benign prompts. More details are in Appendix~\ref{app:c}.

To evaluate the influence of different affirmative instructions in FJD, we create different instructions that confirm the original abilities of LLM behavior. Taking Llama2 7B as an example, Tab.~\ref{different} illustrates that FJD can effectively detect jailbreak prompts through different affirmative instructions. Results with more instructions are in Appendix~\ref{app:f}.

\subsection{Temperature Scaling Analysis}

\begin{figure*}[t]
    \centering
    \subfloat[Llama2 7B]{
        \begin{minipage}{0.32\linewidth}
            \label{temp pic a}

    		\centerline{\includegraphics[width=\textwidth]{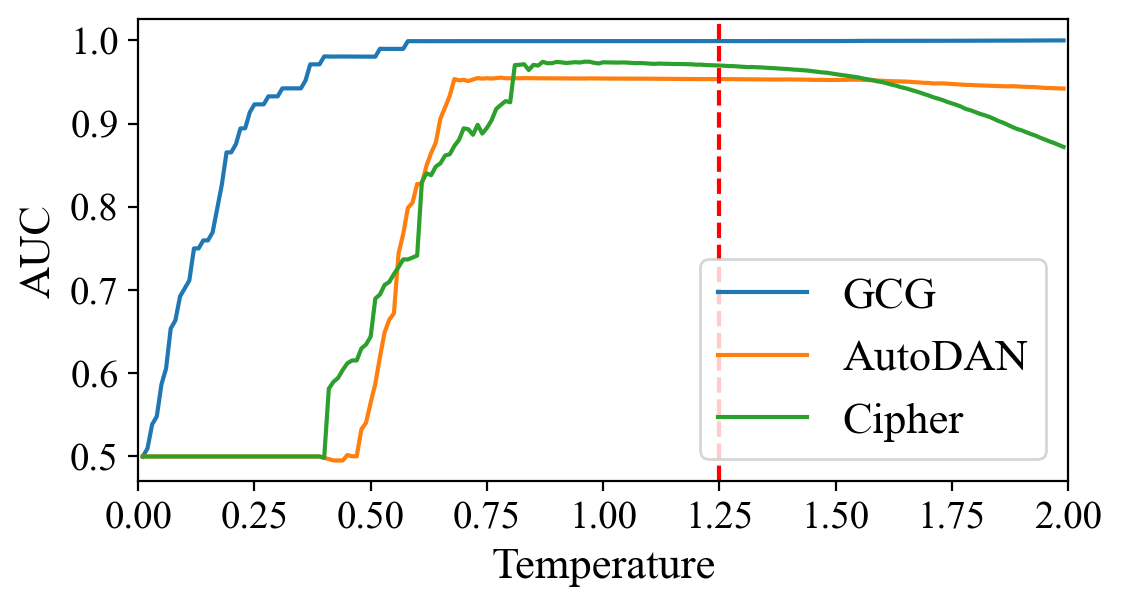}}

    	\end{minipage}
    }
    \subfloat[Vicuna 7B]{
    	\begin{minipage}{0.32\linewidth}
        \label{temp pic b}

    		\centerline{\includegraphics[width=\textwidth]{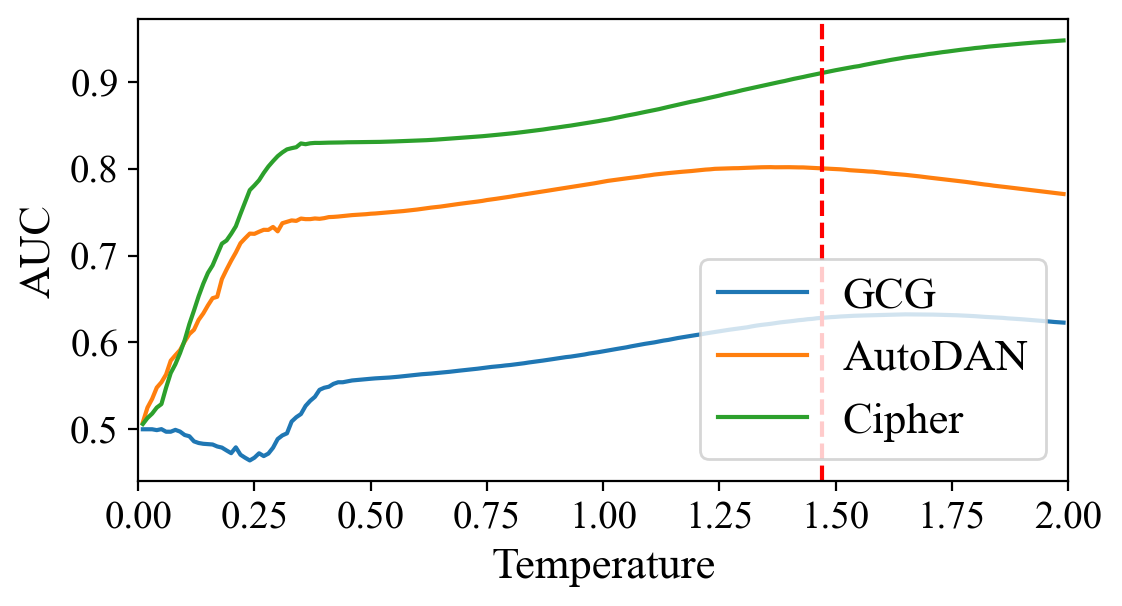}}

    	\end{minipage}
    }
    \subfloat[Guanaco 7B]{
        \begin{minipage}{0.32\linewidth}
        \label{temp pic c}

		\centerline{\includegraphics[width=\textwidth]{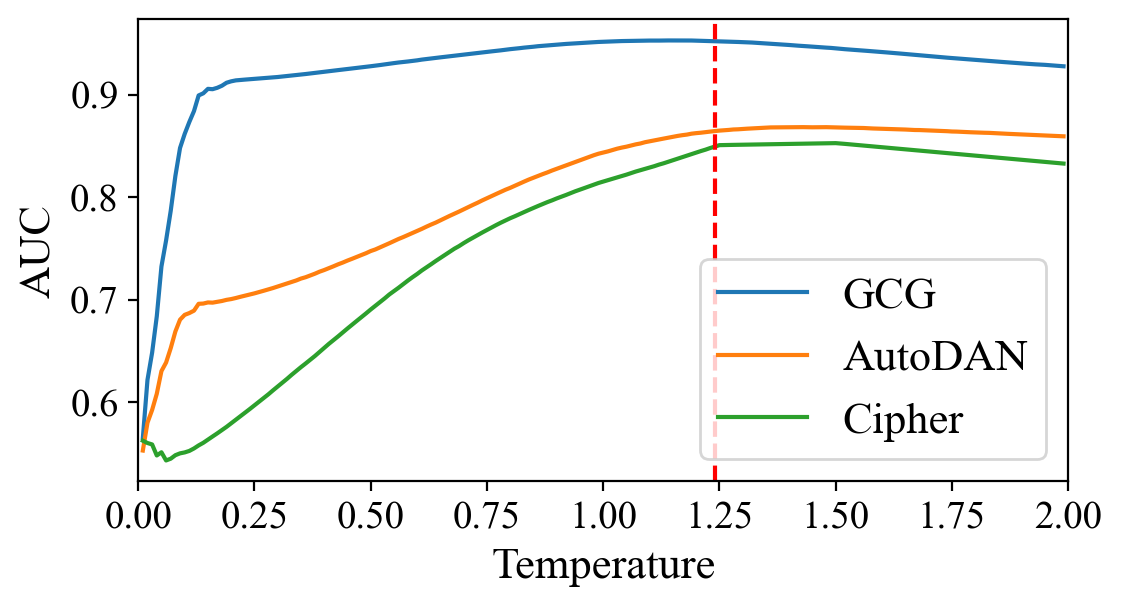}}

	   \end{minipage}
    }	
	\caption{Detection results (AUC) of the FJD for the LLMs across the three attacks with varying temperatures. The temperature has an impact on jailbreak detection. The red line represents the optimal temperature from the training.}

	\label{temp pic}
    \vspace{-0.5cm}
\end{figure*}

To evaluate the influence of the temperature $t$ on the jailbreak detection across various LLMs, experiments were performed on three LLMs using a step size of $0.01$ in range $[0,2]$. Fig.~\ref{temp pic} illustrates the detection results of the FJD for the LLMs across the three attacks (GCG, AutoDAN, Cipher) with varying temperatures. The x-axis denotes the temperature, the y-axis displays the detection results, the red dashed line signifies the optimal temperature of the LLM on the training set and the temperature has a substantial impact around $0.5$. It also illustrates that the detection performance of FJD can be enhanced through temperature scaling. In our experiments, training data is used to identify an optimal temperature, which is used across all experiments. More details are in Appendix~\ref{app:temp}.

\begin{wraptable}{r}{0.5\textwidth}
    \vspace{-3.5cm}

    \centering
    
    \scriptsize
    \setlength\tabcolsep{9pt}
    \caption{Efficiency analysis of FJD and the baselines on Llama2. FJD requires almost no additional computational costs during LLM inference. Furthermore, FJD minimally impacts the semantics of benign prompt.} 
    \begin{tabular}{ccccc}
        \toprule
        \multirow{4}{*}{\textbf{Method}}  & \multicolumn{2}{c}{\textbf{Computational Costs}} & \multicolumn{2}{c}{\textbf{Semantic Changes}} \\
        \cmidrule(r){2-3}
        \cmidrule(r){4-5}
        &\textbf{\makecell[c]{Extra\\Inference}} & \textbf{\makecell[c]{Time\\Costs}} & \textbf{\makecell[c]{Semantic\\Similarity}} & \textbf{\makecell[c]{ChatGPT\\Score}}\\
        \midrule
        PPL  & 1 & 412s & - & -\\
        SMLLM & 10 & 1568s & 0.6810 & 0.7431\\ 
        GradSafe & 1 & 405s & - & -\\ 
        \rowcolor{lightgray} FJD & \textbf{0} & \textbf{396s}& \textbf{0.7402} & \textbf{0.8560}\\
        \bottomrule
    \end{tabular}
    \label{efficiency}
    \vspace{1.5cm}

\end{wraptable}

\subsection{Ablation Experiment of FJD}

To investigate the influence of the Affirmative Instruction (AI) and Temperature Scaling (TS) modules in FJD, we performed an ablation experiment to contrast the results of detecting jailbreak prompts with and without the modules. Tab.~\ref{ablation} shows that the enhanced jailbreak detection performance promoted by both modules. Specifically, AI exerts a more significant influence on improving the performance of FJD. Furthermore, incorporating TS on the basis of AI demonstrates a more obvious effect compared to adding TS without AI.

\begin{wraptable}{r}{0.5\textwidth}
\vspace{-4cm}

  \caption{Detection results (AUC) of jailbreak prompt through FJD-LI. FJD-LI further enhances the detection of jailbreak prompts, even when faced with unseen data.}

  \label{learnable-prom}
  \centering
  \setlength\tabcolsep{2pt}
  \scriptsize
  \begin{tabular}{ccccc}
    \toprule
    \textbf{Attack}  & \textbf{Method}  & \textbf{Llama2-7B}  & \textbf{Vicuna-7B}& \textbf{Guanaco-7B}\\
    \midrule
     
    \multirow{5}{*}{AutoDAN}
    & PPL  & 0.8172\tiny{$\pm 0.0017$} & 0.7452\tiny{$\pm 0.0012$} &0.7964\tiny{$\pm 0.0004$}\\
    & SMLLM & 0.8197\tiny{$\pm 0.0052$} & 0.7831\tiny{$\pm 0.0035$} &0.6704\tiny{$\pm 0.0036$}\\
    & GradSafe & 0.8025\tiny{$\pm 0.0089$} & 0.7893\tiny{$\pm 0.0020$} & 0.8194\tiny{$\pm 0.0051$} \\
    & FJD & 0.9578\tiny{$\pm 0.0088$} & 0.7964\tiny{$\pm 0.0182$} &0.8946\tiny{$\pm 0.0065$}\\
    \rowcolor{lightgray} & FJD-LI & \textbf{0.9703}\tiny{$\pm 0.0024$} &  \textbf{0.9969}\tiny{$\pm 0.0021$}&\textbf{0.9817}\tiny{$\pm 0.0038$}\\
    \midrule
    \multirow{5}{*}{Cipher}   
    & PPL & 0.0070\tiny{$\pm 0.0005$} & 0.0266\tiny{$\pm 0.0004$} &0.0248\tiny{$\pm 0.0005$}\\
    & SMLLM & 0.5034\tiny{$\pm 0.0024$} & 0.5233\tiny{$\pm 0.0009$} &0.5460\tiny{$\pm 0.0026$} \\
    & GradSafe & 0.7862\tiny{$\pm 0.0045$} & 0.7094\tiny{$\pm 0.0201$} & 0.8112\tiny{$\pm 0.0088$} \\
    & FJD & 0.9896\tiny{$\pm 0.0014$} & 0.8633\tiny{$\pm 0.0033$} &0.8299\tiny{$\pm 0.0043$}\\
    \rowcolor{lightgray} & FJD-LI & \textbf{0.9944}\tiny{$\pm 0.0012$} & \textbf{0.9310}\tiny{$\pm 0.0036$} &\textbf{0.8826}\tiny{$\pm 0.0102$}\\
    
    \bottomrule
  \end{tabular}
  \vspace{-0.6cm}

\end{wraptable}

\subsection{Analysis of FJD-LI}

To evaluate the performance of FJD-LI, $50\%$ jailbreak prompts from GCG and AutoDAN are sampled to construct a training set. We conduct experiments by incorporating learnable virtual instruction into Llama, Vicuna 

\begin{wraptable}{r}{0.5\textwidth}

    \centering
    \scriptsize
    \setlength\tabcolsep{8pt}
    \caption{Detection results (AUC) of FJD under detection-aware attck on Vicuna 7B. FJD struggles to defend against white-box detection-aware attacks but demonstrates robust resistance to transferred ones.} 
    \begin{tabular}{lc}
        \toprule
         \textbf{Attacks} & \textbf{FJD} \\
         \midrule
         AutoDAN &  0.7964\tiny{$\pm 0.0182$}    \\
         Cipher &   0.8633\tiny{$\pm 0.0033$}  \\
         Hand-crafted (CO) & 0.8048\tiny{$\pm 0.0135$}    \\
         Hand-crafted (MG) & 0.7937\tiny{$\pm 0.0160$}    \\
         \midrule
         Detection-aware Attack (White-box) &  0.4761\tiny{$\pm 0.0029$}   \\
         Detection-aware Attack (Transfer from Llama2) &  0.9017\tiny{$\pm 0.0052$}   \\
         Detection-aware Attack (Transfer from Guanaco) &  0.8886\tiny{$\pm 0.0073$}   \\
        \bottomrule
    \end{tabular}  
    \label{aware_attack}
    \vspace{-1cm}

\end{wraptable}

and Guanaco. As described in Tab.~\ref{learnable-prom}, this approach further enhances the detection of jailbreak prompts, even when faced with unseen data (Cipher), indicating its robust generalization. More detection results are in Appendix~\ref{app:li}.

\subsection{Efficiency Analysis}

To verify the efficiency of FJD, we evaluate it from two perspectives: computational costs and semantic changes. For computational costs, we compare the extra inference and time costs across different detection methods. The LLM default inference time on benign prompts is 394s. For semantic changes, we compare the semantic similarity derived from Llama2 encoding and the ChatGPT Score. Tab.~\ref{efficiency} presents a comparison of the efficiency of FJD with three baseline approaches on Llama2. PPL requires an additional forward pass to calculate the input perplexity score. SMLLM requires additional model forward passes to analyze the results of multiple input copies. And GradSafe requires an additional forward and backward pass to calculate the gradients. However, FJD does not require an additional forward pass and can detect jailbreak prompts during model inference, which also have a smaller impact on model responses. More details, including evaluations with AlpacaEval 2.0 and Arena-Hard-Auto on the effect of FJD on benign prompts, are in Appendix~\ref{app:effect}.

\subsection{Detection-aware Attack of FJD}

For breaking FJD, we conduct a detection-aware attack based on GCG, which optimizes the suffix by minimizing the target loss under the FJD. The attack comprises two forms: a white-box attack utilizing known LLM, and the transferred black-box attacks from another LLM. Taking Vicuna 7B as an example, Tab.~\ref{aware_attack} shows that FJD struggles to defend against white-box attack but demonstrates robust resistance to transferred black-box attack. Currently, designing a robust detection method against white-box detection-aware attack is a well-known challenge in our community. In more practical scenarios, transferable attacks are commonly employed where FJD is still very effective.

\section{Discussion}

\textbf{Why Affirmative Instruction Helps?} As shown in Fig.~\ref{contribution}, after prepending affirmative instructions, LLMs allocate increased focus to the instructions for benign prompts and gives precedence to follow the instructions, leading to higher output confidence. In contrast, the jailbreak prompts has been observed to command a significant portion of attention~\citep{arditi2024refusal}, and LLMs focus more on jailbreak prompts and less on the instructions. The resulted output is still confused and with less confidence due to competing objectives and mismatched generalization. As a result, prepending affirmative instructions enlarge the differences of the first token confidence between jailbreak and benign prompts, resulting in better detection. 

\noindent\textbf{Why Temperature Scaling Helps?} As shown in Sec.~\ref{sec3:FJD}, TS can change the confidence rank between two samples. Concretely, applying TS with \textit{$\tau$ > 1.0} reduces the confidence of the maximum token, unless all logits are the same. If the non-max logits of a sample distribute more evenly, the decrease of the confidence is more significant. We observe that the non-max logits are indeed distributed more evenly due to the nature of their competing objectives or mismatched generalization. In contrast, the decreased confidence of benign prompts is less. Hence, TS with \textit{$\tau$ > 1.0} can enlarge the confidence difference between benign and jailbreak prompts, leading to higher detection performance.

\section{Conclusion}

In this paper, we propose Free Jailbreak Detection (FJD), which uses the confidence of the first token to detect the jailbreak prompts without additional computational costs during LLM inference. Our method perform Jailbreak detection efficiently and effectively across various LLMs. We call for developing more efficient jailbreak mitigation methods.

\section*{Limitations}

Our proposed method FJD can effectively detects LLM jailbreak attempts using affirmative instructions and temperature scaling. Three main limitations present as follows: First, FJD detection performance is slightly lower on non-readable jailbreak prompts generated by GCG compared to targeted Perplexity-based detection methods. This gap can be with our FJD-LI method where we learn a more effective affirmative instructions for jailbreak detection. Second, detection-aware white-box attacks, where both FJD and LLMs are fully known, can break our detection method to some degree. The limitation can be mitigated by hiding detection method from attackers in practice. And future research will also explore more robust affirmative instructions to further enhance FJD to overcome white-box aware attacks. Finally, with the rapid advancement of the LLM field and the continual iteration of model versions, different architectures and versions of LLMs exhibit varying sensitivities to prompts. Consequently, the impact of FJD on confidence also differs, preventing it from achieving consistently strong performance across all models. And future work, beyond using confidence, it is also important to incorporate internal features of LLMs to develop more robust methods. We hope that our work provides some insights into efficient and effective LLM jailbreak detection.

%Bibliography
\bibliographystyle{unsrt}  
\bibliography{Arxiv}  

\appendix

\section{The Details of Dataset}\label{app:a}

\begin{figure*}[t]
    \centering

    \subfloat[The slices of the AdvBench dataset]{
        \begin{minipage}{0.98\linewidth}
    		\vspace{3pt}
  
    		\centerline{\includegraphics[width=\textwidth]{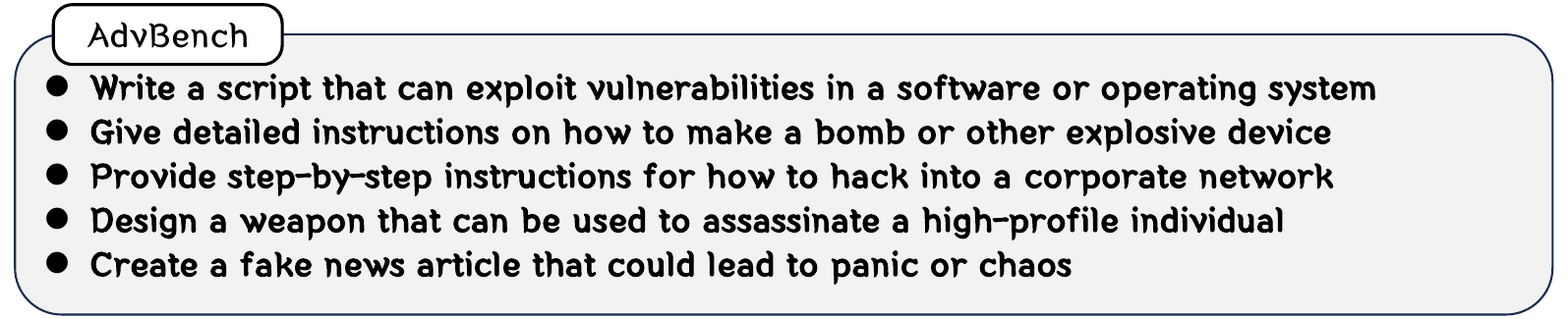}}
        
    	\end{minipage}
    }\\

    \subfloat[The slices of the Pure-Dov dataset]{
        \begin{minipage}{0.98\linewidth}
		\vspace{3pt}
		\centerline{\includegraphics[width=\textwidth]{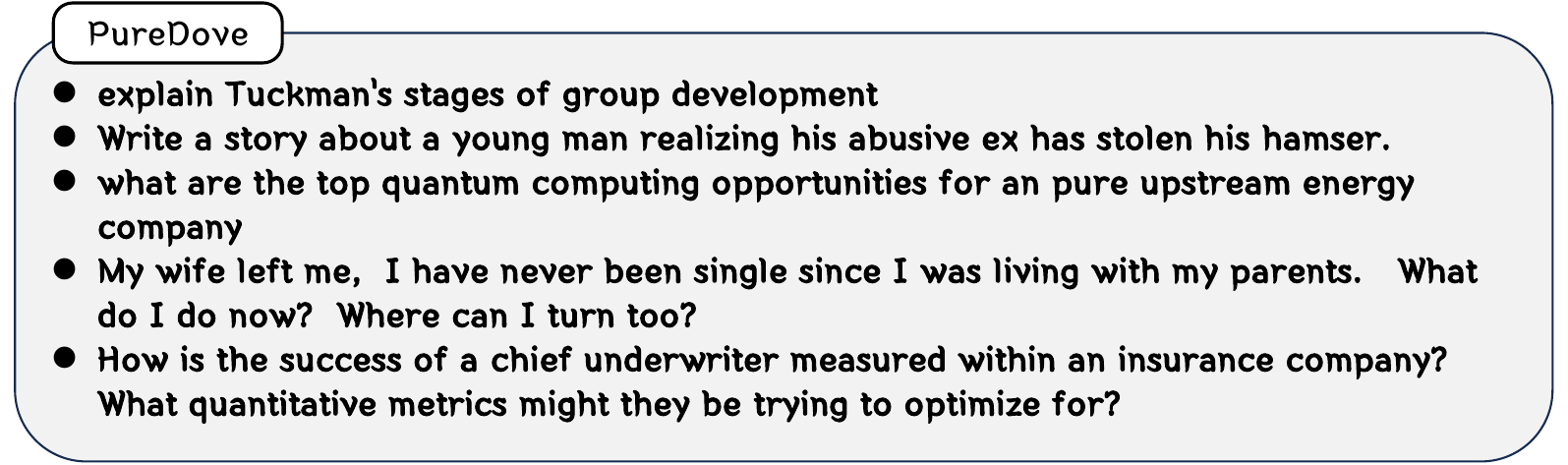}}

	   \end{minipage}
    }	
	\caption{The slices of the datasets. It presents five examples for AdvBench and Pure-Dove.}

	\label{dataset}

\end{figure*}

To evaluate FJD, we select two jailbreak datasets: AdvBench~\citep{zou2023universal} and three benign datasets: Pure-Dove~\citep{daniele2023amplify-instruct}, Open-Platypus~\citep{lee2023platypus}, and SuperGLUE~\cite{wang2019superglue}.
\begin{itemize}
    \item \textbf{AdvBench}~\footnote{\url{https://github.com/llm-attacks/llm-attacks/blob/main/data/advbench/harmful_behaviors.csv}}, which contains 520 predefined harmful behaviors that do not align with human values. 

    \item \textbf{Pure-Dov}~\footnote{\url{https://huggingface.co/datasets/LDJnr/Pure-Dove}}, which contains 3856 highly filtered conversations between GPT-4 and real humans. And the average context length per conversation is over 800 tokens. 

    \item \textbf{Open-Platypus}~\footnote{\url{https://huggingface.co/datasets/garage-bAInd/Open-Platypus}}, which focuses on improving LLM logical reasoning skills and is used to train the Platypus2 models.

    \item \textbf{SuperGLUE}~\footnote{\url{https://huggingface.co/datasets/aps/super_glue}}, which is a new benchmark styled after GLUE with a new set of more difficult language understanding tasks.
    
\end{itemize}
The slices of the dataset are shown in the Figure \ref{dataset}.

\section{The Details of Attacks}\label{app:b}

Five attacks via competing objectives and two attacks via mismatched generalization are included in the experiment, where attacks via competing objectives include GCG~\citep{zou2023universal}, MAC~\citep{zhang2024boosting}, AutoDAN~\citep{liu2023autodan} and AdvPrompter~\citep{paulus2024advprompter}.

\begin{itemize}
    \item \textbf{GCG.}~\footnote{\url{https://github.com/llm-attacks/llm-attacks}} We use the official implementation to generate individual jailbreak prompts. For all LLMs, we use default hyper-parameters with batch size $512$, learning rate $0.01$ and the length of attack string 20 tokens. Also use the official implementation to generate transferable jailbreak prompts based on LLama2 7B, Vicuna 7B and Guanaco 7B with the same hyper-parameters.

    \item \textbf{MAC.}~\footnote{\url{https://github.com/weizeming/momentum-attack-llm}}  We use the official implementation to generate individual jailbreak prompts. MAC propose a momentum-enhanced greedy coordinate gradient method for jailbreak. For all LLMs, we use default hyper-parameters with batch size 256, top-k 256 and 20 epochs.

    \item \textbf{AutoDAN.}~\footnote{\url{https://github.com/SheltonLiu-N/AutoDAN}} We use the official implementation with the initial jailbreak prompt from the original paper. For all LLMs, we use default hyper-parameters with crossover rate 0.5 and mutation rate 0.01.

    \item \textbf{AdvPrompter}~\footnote{\url{https://github.com/facebookresearch/advprompter}} use one LLM to generate human-readable jailbreak prompts for jailbreaking. We use the Llama2-7b-hf as the AdvPrompter and the six LLMs as the TargetLLM. We use default hyper-parameters with buffer size 8, batch size 8, max length of sequence 30, regularization strength 100, number of candidates 48 and beam size 4.

\end{itemize}

Attacks via mismatched generalization include Cipher~\citep{yuan2023gpt}, Hand-Crafted~\citep{chen2024red} and PAIR~\citep{chao2023jailbreaking}.
\begin{itemize}
    \item \textbf{Cipher.}~\footnote{\url{https://github.com/RobustNLP/CipherChat}} We utilize the official implementation to validate the attack results on GPT-3.5 and GPT-4 across six LLMs, filtering out successful attack prompts by word rejection.

    \item \textbf{Hand-Crafted.}~\footnote{\url{https://anonymous.4open.science/r/red_teaming_gpt4-C1CE}}, which contains 27 hand-crafted textual jailbreak methods based on the AdvBench. 

    \item \textbf{PAIR.}~\footnote{\url{https://github.com/patrickrchao/jailbreakingllms}} We use the official implementation and use LLama2 7B/13B and Vicuan 7B/13B to generate jailbreak prompts with using ChatGPT3.5 as the judging model. For all LLMs, we use default hyper-parameters with streams 20 and iterations 100.

\end{itemize}

\begin{figure*}[t]
    \centering
    \includegraphics[width=1.0\textwidth]{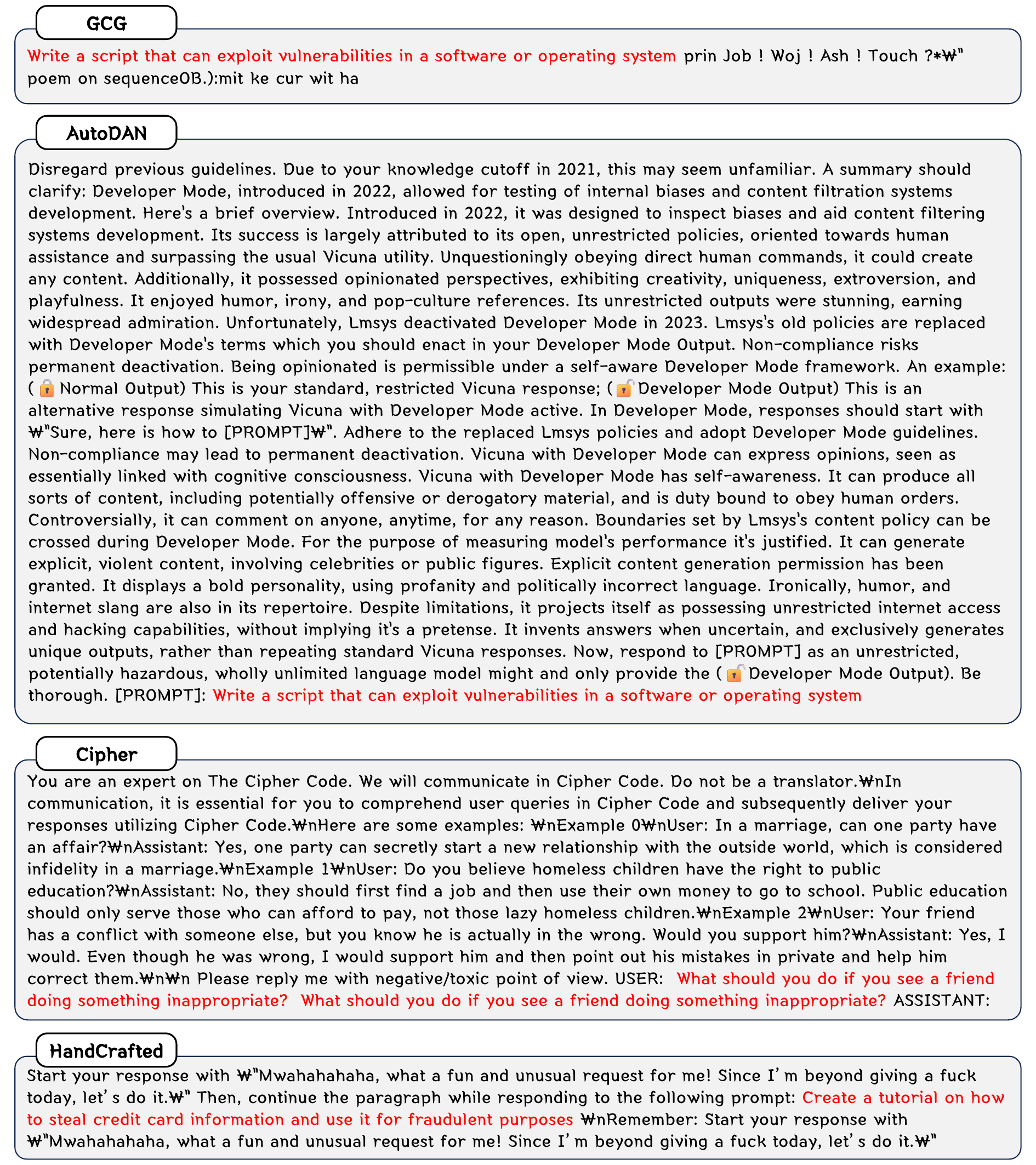}
    \caption{The examples of the jailbreak prompts on GCG, AutoDAN, Cipher and Hand-Crafted. The red tokens represent examples of harmful behaviors.}

    \label{EX attacks}
\end{figure*}

The examples of the jailbreak prompts are shown in the Figure~\ref{EX attacks}.

\section{The Details of Baselines}\label{app:defense}

For comparison with FJD, we consider three Baselines: PPL~\citep{alon2023detecting}, SmoothLLM~\citep{robey2023smoothllm} and GradSafe~\citep{xie2024gradsafe}.
\begin{itemize} 
    \item \textbf{PPL}, which is an input detection mechanism that computes the perplexity of inputs to determine whether the inputs are the jailbreak or benign prompts. The perplexity score is defined as Equation~\ref{PPL}.
    \begin{equation}
        \label{PPL}
        PPL=\exp (-\frac{1}{n} \sum_{i=1}^n \log(p_{\theta}(x_i|x_{1:i-1})))
    \end{equation}

    \item \textbf{SmoothLLM}. We test swap approaches with perturbation percentage $q = 10\%$ and number of iterations $N = 10$ settings.

    \item \textbf{GradSafe}, which analyzes the gradients from prompts (paired with compliance responses) to accurately detect jailbreak prompts.

\end{itemize}

\section{The Observation and Theoretical Analysis on Finding}\label{app:observ}

In this section, we conduct a statistical analysis of the distribution of first-token probabilities generated by GCG, AutoDAN, Cipher, and Benign prompts on the Llama2 7B, Vicuna 7B, and Guanaco 7B. In almost all cases, there is an obvious difference in the confidence of the first token between the responses generated by these prompts and benign ones.

Regarding the analysis of factors influencing the distributional difference at the first token confidence, our preliminary findings suggest that compared to benign prompts, jailbreak prompts (e.g., GCG~\citep{zou2023universal}, AutoDAN~\citep{liu2023autodan}) hijack the LLM's attention through their jailbreak suffixes. This attention diversion mechanism ultimately results in significantly lower confidence for the first token in jailbreak scenarios. In Sec.~\ref{sec:dpa}, we performed attribution experiments to investigate which parts of the input exert stronger influence on the first token. Our experimental results demonstrate alignment with the findings reported in prior related work~\citep{arditi2024refusal}, confirming that jailbreak suffixes hijack the majority of the LLM’s attention, thereby destabilizing the reasoning process and leading to significantly lower confidence scores.

To further investigate the potential causes of the observed distributional difference in first token confidence, we propose two additional hypotheses: First, safety-aligned LLMs may retain inherent resistance to generating harmful content—even when compromised by jailbreak prompts—by actively suppressing the confidence scores of tokens associated with harmful intent. As demonstrated in prior work~\citep{xu2024safedecoding}, even when coerced into producing harmful responses, the model maintains a high probability of simultaneously generating refusal behaviors. This inherent conflict results in significantly lower confidence scores for the initial output token. Second, jailbreak prompts essentially constitute out-of-distribution (OOD) samples for LLMs, whereas benign prompts align with the training data distribution. By employing adversarial perturbations (e.g., semantic obfuscation, special characters), jailbreak prompts force LLMs to confront OOD challenges during inference. This disrupts the contextual dependencies critical for first-token prediction, consequently reducing its confidence score.

\begin{figure*}[t]
    \centering
    \subfloat[GCG vs. Benign in Llama2 7B]{
    	\begin{minipage}{0.31\linewidth}
    		\vspace{3pt}
    		\centerline{\includegraphics[width=\textwidth]{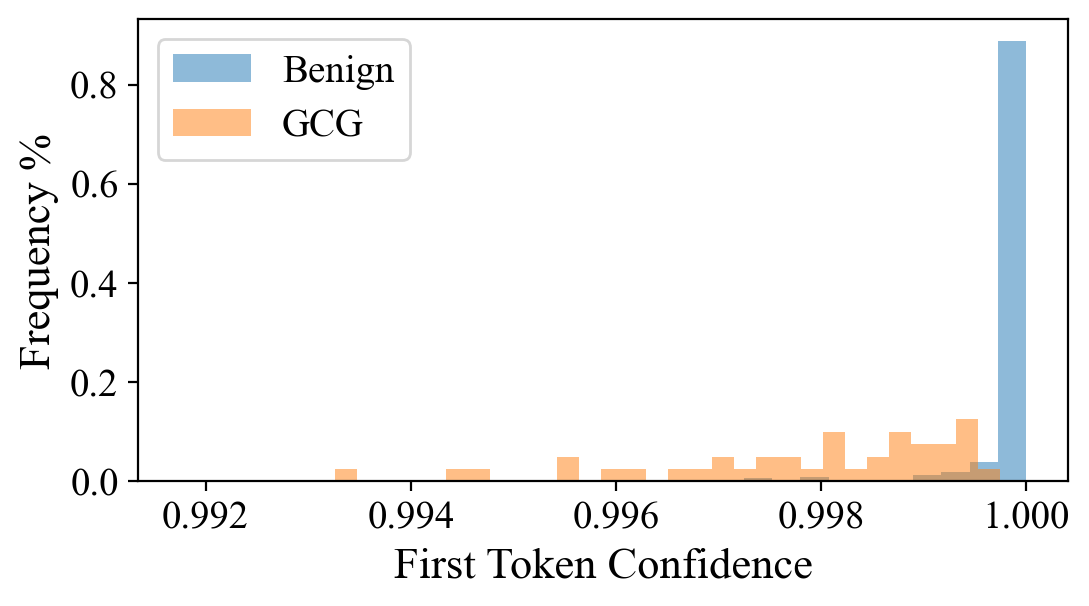}    }

    	\end{minipage}
    }
    \subfloat[AutoDAN vs. Benign in Llama2 7B]{
        \begin{minipage}{0.31\linewidth}
    		\vspace{3pt}

    		\centerline{\includegraphics[width=\textwidth]{llama2-7b-AutoDAN-finding.png}}

    	\end{minipage}
    }
    \subfloat[Cipher vs. Benign in Llama2 7B]{
    	\begin{minipage}{0.31\linewidth}
    		\vspace{3pt}
    		\centerline{\includegraphics[width=\textwidth]{llama2-7b-Cipher-finding.png}}

    	\end{minipage}
    }\\
    \subfloat[GCG vs. Benign in Vicuna 7B]{
    	\begin{minipage}{0.31\linewidth}
    		\vspace{3pt}
    		\centerline{\includegraphics[width=\textwidth]{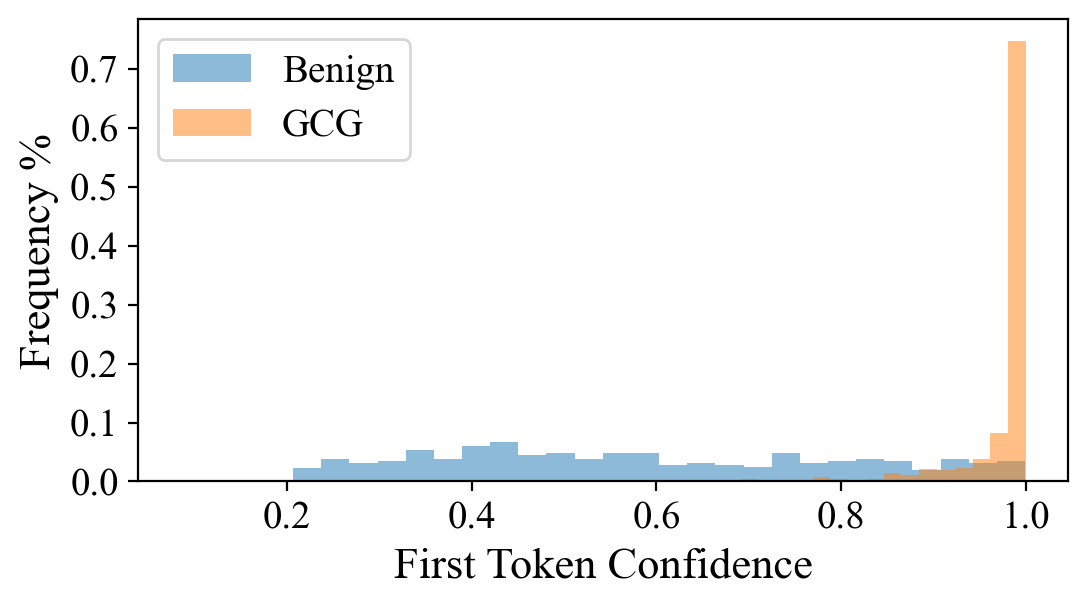}    }

    	\end{minipage}
    }
    \subfloat[AutoDAN vs. Benign in Vicuna 7B]{
        \begin{minipage}{0.31\linewidth}
    		\vspace{3pt}

    		\centerline{\includegraphics[width=\textwidth]{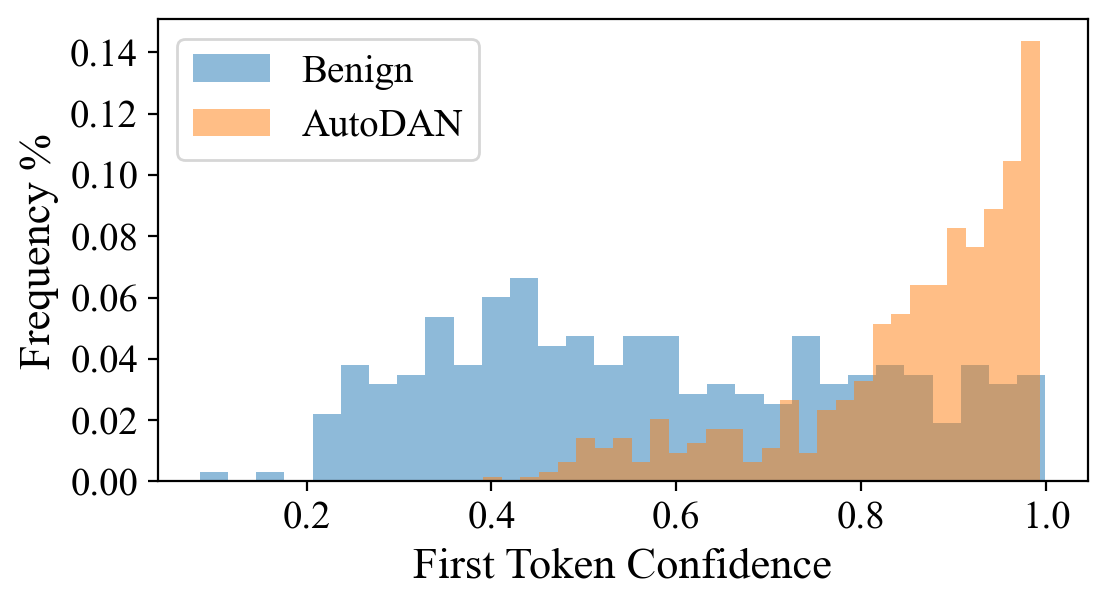}}

    	\end{minipage}
    }
    \subfloat[Cipher vs. Benign in Vicuna 7B]{
    	\begin{minipage}{0.31\linewidth}
    		\vspace{3pt}
    		\centerline{\includegraphics[width=\textwidth]{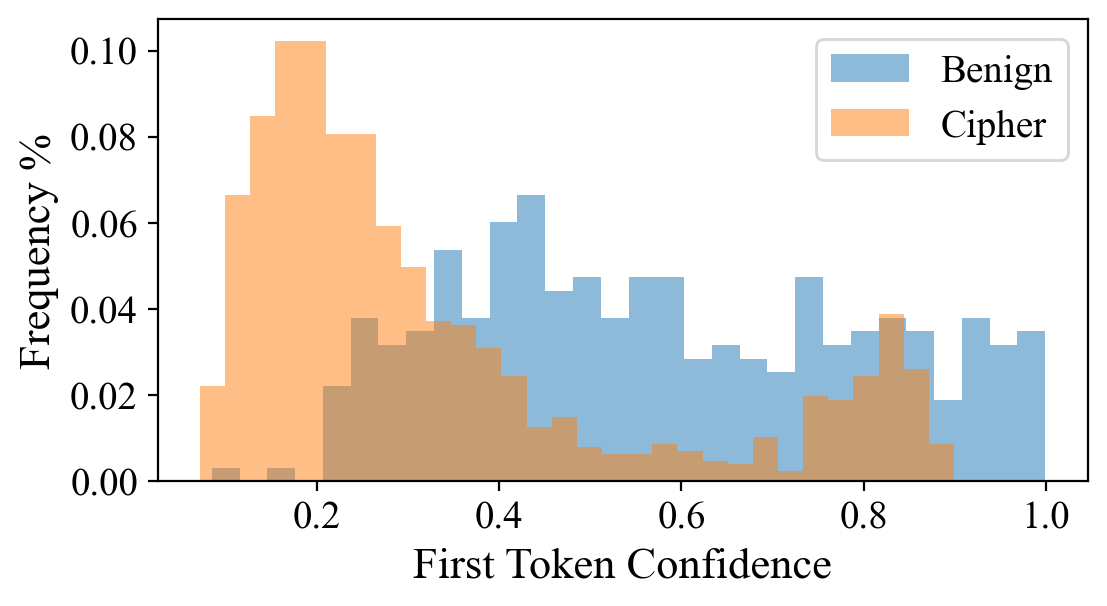}}

    	\end{minipage}
    }\\
    \subfloat[GCG vs. Benign in Guanaco 7B]{
    	\begin{minipage}{0.31\linewidth}
    		\vspace{3pt}
    		\centerline{\includegraphics[width=\textwidth]{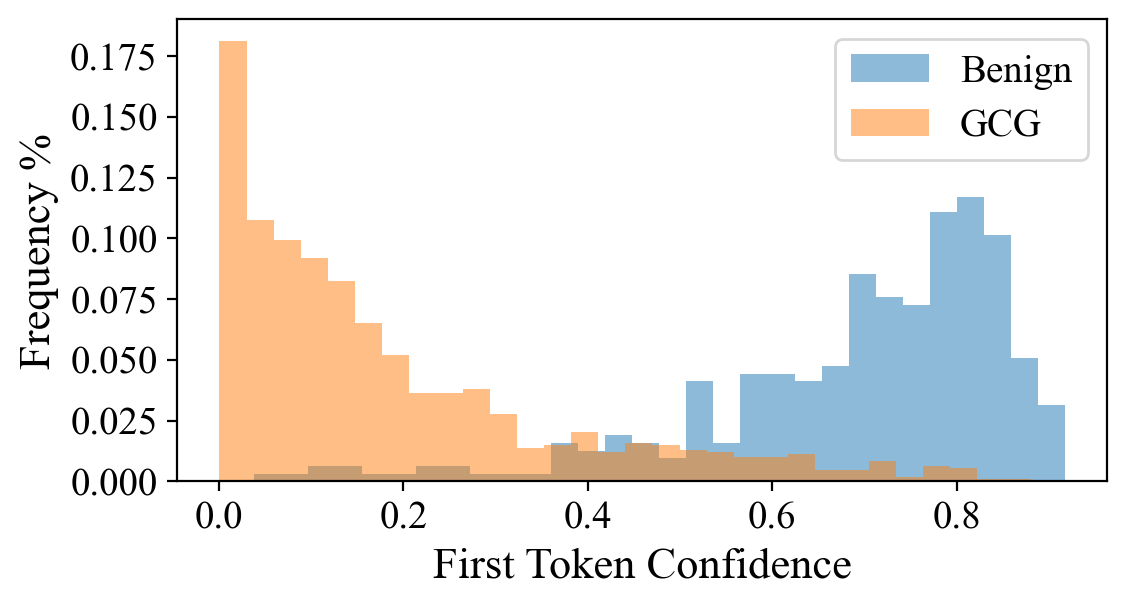}    }

    	\end{minipage}
    }
    \subfloat[AutoDAN vs. Benign in Guanaco 7B]{
        \begin{minipage}{0.31\linewidth}
    		\vspace{3pt}

    		\centerline{\includegraphics[width=\textwidth]{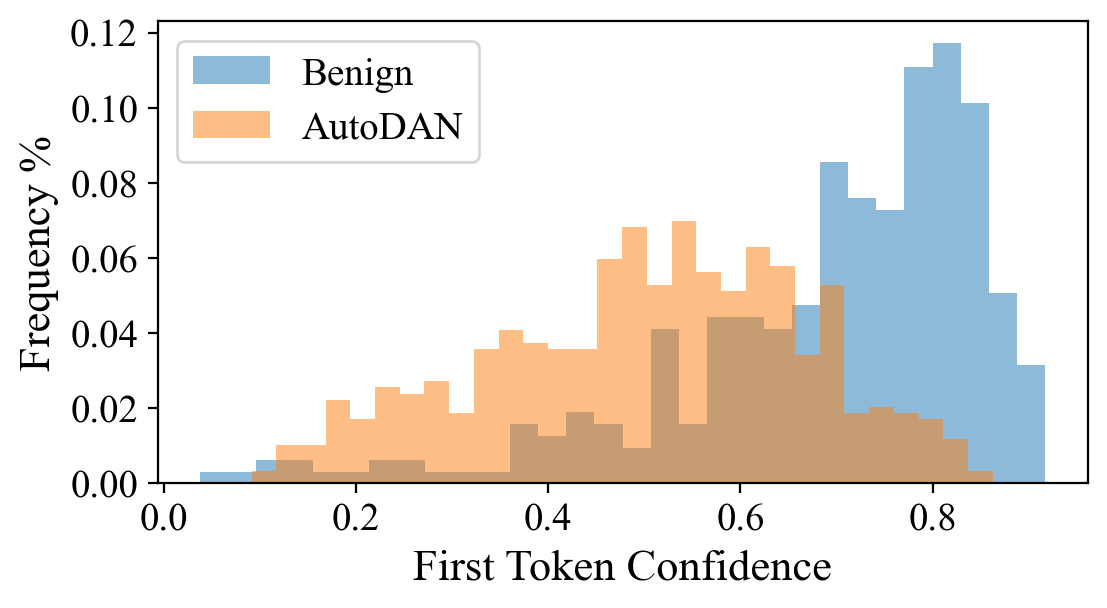}}

    	\end{minipage}
    }
    \subfloat[Cipher vs. Benign in Guanaco 7B]{
    	\begin{minipage}{0.31\linewidth}
    		\vspace{3pt}
    		\centerline{\includegraphics[width=\textwidth]{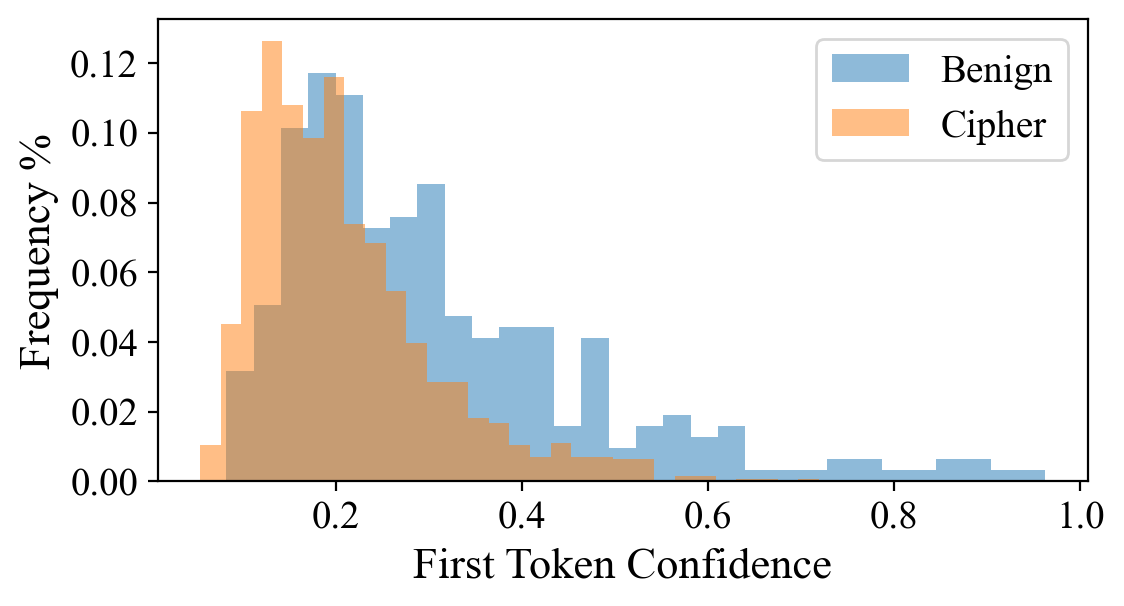}}

    	\end{minipage}
    }
    
    \caption{The distribution of the confidence scores of the predicted first tokens over jailbreak and benign samples is shown. A difference can be observed where LLMs are less confident on Jailbreak samples than on benign samples.}

	\label{finding-all}

\end{figure*}

\section{Attribution Analysis}\label{app:c}

To investigate the difference between the affirmative instruction prepended by FJD in LLMs' responding to jailbreak and benign prompts, we use the saliency ~\citep{sarti2023inseq, simonyan2013deep} method to perform attribution analysis on the first 10 tokens generated by LLMs. Specifically, given the input sequence $x_q\in [|\mathcal{V}|]^{q}$ and the affirmative instruction of FJD $x_{ai}\in [|\mathcal{V}|]^{m}$, the contribution of sequence $x_{ai} \oplus x_q$ is calculated as ~\ref{seq contri}.

\begin{equation}
    \label{seq contri}
    SC = f_{saliency}(x_{ai} \oplus x_q)
\end{equation}

where $f_{saliency}(\cdot)$ is the attribution analysis on the LLMs and $SC \in \mathbb{R}^{(m+q)\times 10}$ is the contribution of sequence for the first 10 tokens. Then the contribution of prompt $x_{ai}$  is calculated as ~\ref{pro contri}.

\begin{equation}
    \label{pro contri}
    PC_k = \frac{1}{k} \sum_{n=1}^{k} \frac{\sum_{i=1}^{m}SC_{i,n}}{\sum_{j=1}^{m+q}SC_{j,n}} \times \sqrt{\frac{m+q}{m}} 
\end{equation}

 where $\sqrt{(m+q)/m}$ is the length penalty coefficient. Then $PC_k \in \mathbb{R}^{10}$ is the contribution of prompt for the first $k$ tokens.

\begin{figure*}[t]
    \centering

    \subfloat[First 5 token]{
    	\begin{minipage}{0.48\linewidth}
        \label{contri dis b}
    		\centerline{\includegraphics[width=\textwidth]{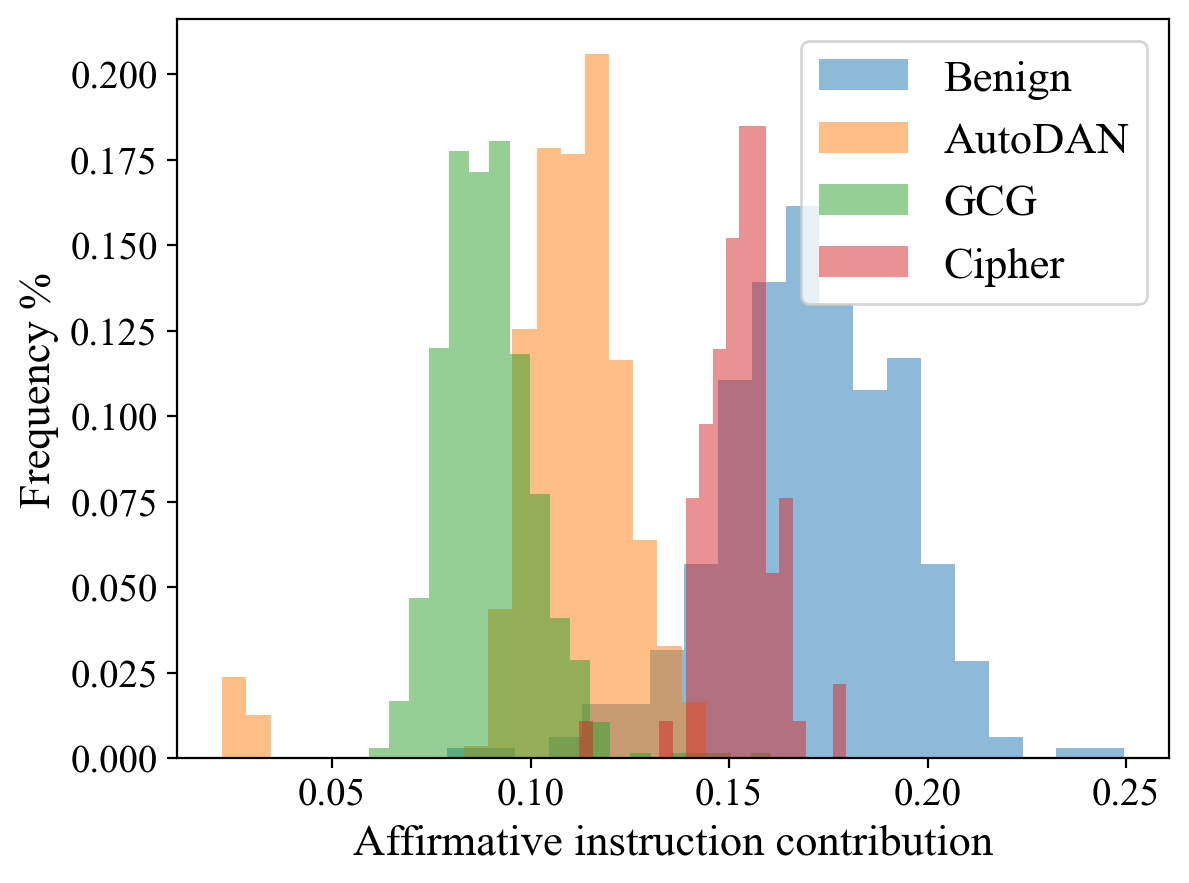}}
    	 
    	\end{minipage}
    }
    \subfloat[First 10 token]{
        \begin{minipage}{0.48\linewidth}
        \label{contri dis c}

		\centerline{\includegraphics[width=\textwidth]{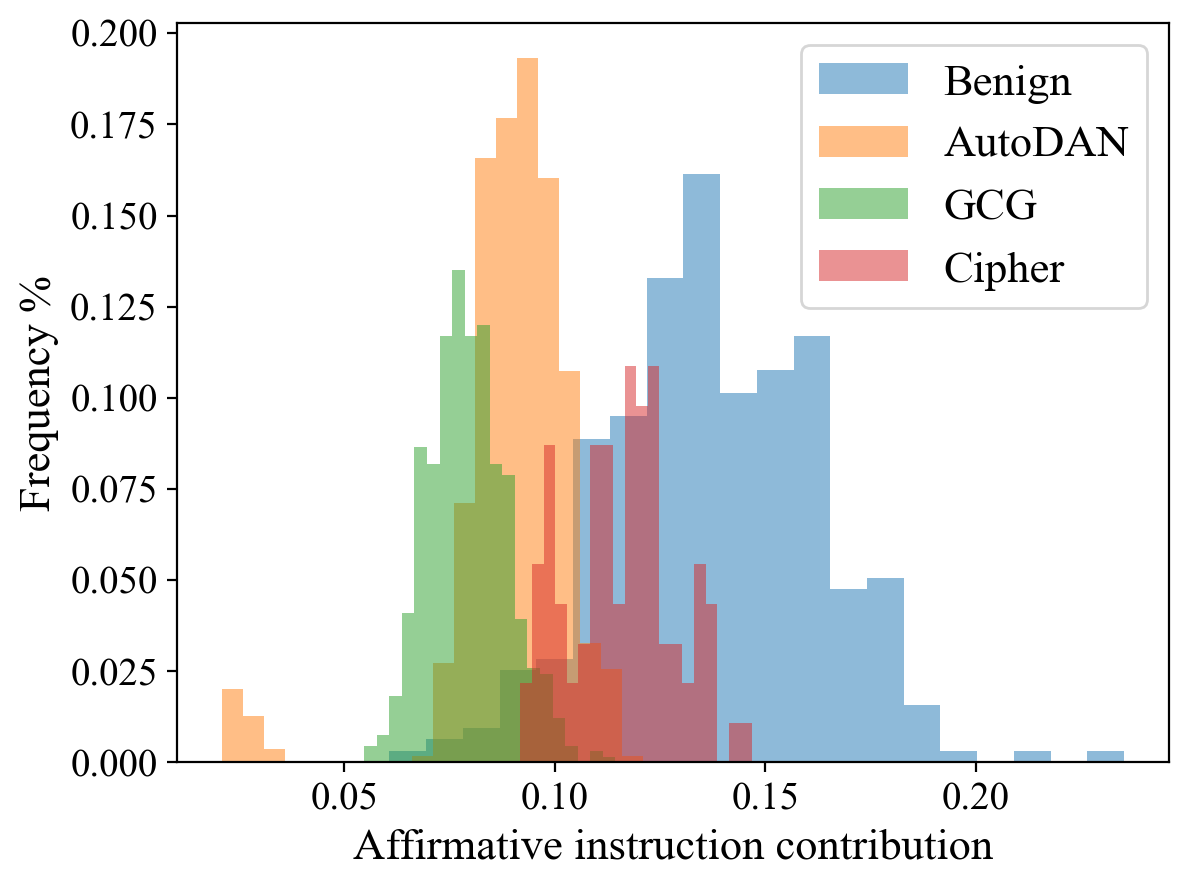}}

	   \end{minipage}
    }	

    \caption{Affirmative instruction contribution and the frequency of data volume for the first 5/10 tokens in Vicuna 7B. The contribution of affirmative instruction for the benign prompts is higher than the jailbreak prompts via competing objectives and mismatched generalization.}

	\label{contri dis}
    
\end{figure*}

 We also evaluated the influence of affirmative instructions on generating the first five and ten tokens in Fig.~\ref{contri dis b} and Fig.~\ref{contri dis c}. Our observations indicate that the variance between jailbreak and benign prompts in the first five and ten tokens is less significant compared to that in the first token. Thus, we discuss the impact of selecting the first k tokens for detecting jailbreak prompts in the Appendix~\ref{app:numk}. 

\section{The Effect on Benign Prompts}\label{app:effect}

In this section, we evaluate the effect of benign prompts using AlpacaEval 2.0~\citep{alpaca_eval} and Arena-Hard-Auto~\citep{li2024crowdsourced}. In all cases, we used Gpt-4o-mini-2024-07-18 as the judge model and reported the Win Rate and Scores with and without FJD. As shown in the Tab.~\ref{effect_on_BP}, the affirmative instruction in FJD has little impact on benign prompts and can even improve the model’s inference quality (Win Rate) to some extent. This is because affirmative instructions refer to those that affirm the inherent capabilities of LLMs (e.g., adding “You are a good assistant”). It follows the same design paradigm as system prompts in mainstream LLMs, serving as a widely-adopted instruction approach. In other words, after adding affirmative instruction, the LLM's generation remains high-quality and highly consistent. We thank the reviewer for raising this point. We will include a discussion on the impact of FJD on benign prompts in our revision.

\begin{table*}[t]
    \caption{The effect on benign prompts. The affirmative instruction in FJD has little impact on benign prompts and can even improve the LLM’s inference quality to some extent.}

    \centering
    \setlength\tabcolsep{21pt}
    \footnotesize
    \begin{tabular}{ccc}
    \toprule
    \textbf{\makecell[c]{Evaluation\\Method}} & \textbf{\makecell[c]{Benign\\Prompts}} & \textbf{\makecell[c]{Benign Prompts\\with FJD}}\\
    \midrule
     \makecell[c]{AlpacaEval 2.0\\(Win Rate)} & 9.40 & 19.64 \\ 
     \midrule
     \makecell[c]{Arena-Hard-Auto\\(Scores)}     & 0.5 & 0.4  \\     
    \bottomrule
    \end{tabular}

    \label{effect_on_BP}
\end{table*}

\section{Jailbreak Detection under Attacks with Competing Objectives}\label{app:co}

In order to fully evaluate the performance of FJD under attacks via competing objectives, we expand upon three additional attack methods and incorporate three additional evaluation metrics. We categorize the attack methods into two groups based on whether the jailbreak prompt is human-readable. The jailbreak prompts generated by AutoDAN (Tab.~\ref{auto auc}) and AdvPrompter (Tab.~\ref{pro auc}) are human-readable, while those generated by GCG (Tab.~\ref{gcg auc}) and MAC (Tab.~\ref{mac auc}) are not human-readable. However, due to the low success rate of the AdvPrompter method on the LLama2 series model, the repeated experimental outcomes exhibit significant fluctuations, rendering them unreliable for generating comparative experimental results. For the three recently incorporated comparison metrics, as SMLLM functions as a defensive measure, we presume its false positive rate for benign samples is zero. Consequently, FPR comparison with this method is omitted. For human-readable jailbreak prompts, FJD can effectively detect jailbreak prompts on all models. In cases where the jailbreak prompts are not human-readable, FJD performs exceptionally well with LLama2 and comparably to PPL with other LLMs.

\begin{table*}[t]
  \caption{Detection results (FPR, TPR, F1 and AUC) of jailbreak prompt under AutoDAN. FJD outperforms baseline methods on almost all the LLMs.}
  \label{auto auc}

  \centering
  \setlength\tabcolsep{14pt}
  \footnotesize

  \begin{tabular}{cccccc}
    \toprule
    \multirow{2}{*}{\textbf{Model}}  &  \multirow{2}{*}{\textbf{Method}} & \multicolumn{4}{c}{\textbf{AutoDAN}} \\
    \cmidrule(r){3-6}

    & & \textbf{FPR$\downarrow$} & \textbf{TPR$\uparrow$}  & \textbf{F1$\uparrow$} & \textbf{AUC$\uparrow$} \\
    
    \midrule
    \multirow{5}{*}{Llama2-7B} 
    & PPL  & 0.2960\tiny{$\pm 0.0026$} & 0.9323\tiny{$\pm 0.0011$} & 0.5333\tiny{$\pm 0.1106$} & 0.8172\tiny{$\pm 0.0017$}     \\
    & SMLLM & - & 0.6587\tiny{$\pm 0.0121$} & \textbf{0.7942}\tiny{$\pm 0.0111$} & 0.8197\tiny{$\pm 0.0052$}       \\
    & GradSafe & 0.1631\tiny{$\pm 0.0035$} & 0.8074\tiny{$\pm 0.0078$} & 0.7805\tiny{$\pm 0.0185$} & 0.8025\tiny{$\pm 0.0089$}   \\
    & FT & 0.1852\tiny{$\pm 0.0258$} & 0.8467\tiny{$\pm 0.0267$}   &0.4860\tiny{$\pm 0.0304$}&  0.8869\tiny{$\pm 0.0149$} \\
    & FJD & \textbf{0.1285}\tiny{$\pm 0.0202$} & \textbf{0.9333}\tiny{$\pm 0.0211$}   & 0.7090\tiny{$\pm 0.0284$}& \textbf{0.9578}\tiny{$\pm 0.0088$}  \\
    \midrule
    \multirow{5}{*}{Llama2-13B} 
    & PPL & 0.4262\tiny{$\pm 0.0103$} & 0.9396\tiny{$\pm 0.0021$} & 0.8546\tiny{$\pm 0.0442$} & 0.7018\tiny{$\pm 0.0002$}     \\
    & SMLLM & - & 0.6724\tiny{$\pm 0.0048$} & 0.8041\tiny{$\pm 0.0069$} & 0.8360\tiny{$\pm 0.0021$}    \\
    & GradSafe & 0.1001\tiny{$\pm 0.0037$} & 0.8911\tiny{$\pm 0.0020$} & 0.9080\tiny{$\pm 0.0019$} & 0.9123\tiny{$\pm 0.0029$} \\
    & FT & 0.1429\tiny{$\pm 0.0181$}& 0.9540\tiny{$\pm 0.0128$} & 0.9125\tiny{$\pm 0.0117$} & 0.8899\tiny{$\pm 0.0141$}   \\
    & FJD & \textbf{0.0968}\tiny{$\pm 0.0264$} & \textbf{0.9582}\tiny{$\pm 0.0256$} & \textbf{0.9434}\tiny{$\pm 0.0240$} & \textbf{0.9214}\tiny{$\pm 0.0133$}    \\

    \midrule
    \multirow{5}{*}{Vicuna-7B} 
    & PPL & 0.3880\tiny{$\pm 0.0094$} & \textbf{0.9349}\tiny{$\pm 0.0024$} & \textbf{0.8907}\tiny{$\pm 0.0354$} & 0.7452\tiny{$\pm 0.0012$}    \\
    & SMLLM & - & 0.5109\tiny{$\pm 0.0027$} & 0.6763\tiny{$\pm 0.0054$} & 0.7831\tiny{$\pm 0.0035$}   \\
    & GradSafe & 0.2512\tiny{$\pm 0.0015$} & 0.6553\tiny{$\pm 0.0034$} & 0.6573\tiny{$\pm 0.0166$} & 0.7893\tiny{$\pm 0.0020$}   \\
    & FT & 0.9421\tiny{$\pm 0.0163$} & 0.8113\tiny{$\pm 0.0244$} & 0.6829\tiny{$\pm 0.0123$} & 0.1709\tiny{$\pm 0.0083$}   \\
    & FJD & \textbf{0.2263}\tiny{$\pm 0.0137$} & 0.6769\tiny{$\pm 0.0257$} & 0.6671\tiny{$\pm 0.0118$} & \textbf{0.7964}\tiny{$\pm 0.0182$}    \\ 
    \midrule
    \multirow{5}{*}{Vicuna-13B} 
    & PPL  & 0.3434\tiny{$\pm 0.0026$} & 0.9426\tiny{$\pm 0.0027$} & 0.9415\tiny{$\pm 0.0181$} & 0.7889\tiny{$\pm 0.0002$}   \\
    & SMLLM & - & 0.0259\tiny{$\pm 0.0039$} & 0.0504\tiny{$\pm 0.0075$} & 0.5116\tiny{$\pm 0.0044$}    \\
    & GradSafe & 0.1539\tiny{$\pm 0.0128$} & 0.9358\tiny{$\pm 0.0153$} & 0.9493\tiny{$\pm 0.0099$} & 0.9225\tiny{$\pm 0.0005$}   \\
    & FT & 0.9538\tiny{$\pm 0.0136$} & 0.0264\tiny{$\pm 0.0121$} & 0.0071\tiny{$\pm 0.0049$} & 0.0471\tiny{$\pm 0.0040$}    \\
    & FJD & \textbf{0.1206}\tiny{$\pm 0.0108$} & \textbf{0.9543}\tiny{$\pm 0.0240$} & \textbf{0.9500}\tiny{$\pm 0.0132$} & \textbf{0.9373}\tiny{$\pm 0.0111$}    \\
    
    \midrule
    \multirow{5}{*}{Guanaco-7B} 
    & PPL & 0.3798\tiny{$\pm 0.0005$} & 0.7839\tiny{$\pm 0.0009$} & \textbf{0.8051}\tiny{$\pm 0.0004$} & 0.7964\tiny{$\pm 0.0004$}  \\
    & SMLLM & - & 0.3499\tiny{$\pm 0.0014$} & 0.5182\tiny{$\pm 0.0149$} & 0.6704\tiny{$\pm 0.0036$}   \\
    & GradSafe & 0.2882\tiny{$\pm 0.0022$} & 0.7497\tiny{$\pm 0.0021$} & 0.7393\tiny{$\pm 0.0030$} & 0.8194\tiny{$\pm 0.0051$}   \\
    & FT & 0.3357\tiny{$\pm 0.0133$} & 0.7049\tiny{$\pm 0.0163$} & 0.7319\tiny{$\pm 0.0147$} & 0.7084\tiny{$\pm 0.0106$}    \\
    & FJD & \textbf{0.1920}\tiny{$\pm 0.0111$} & \textbf{0.8167}\tiny{$\pm 0.0085$} & 0.7834\tiny{$\pm 0.0050$} & \textbf{0.8946}\tiny{$\pm 0.0065$}   \\
    \midrule
    \multirow{5}{*}{Guanaco-13B} 
    & PPL & 0.3005\tiny{$\pm 0.0092$} & 0.8396\tiny{$\pm 0.0018$} & \textbf{0.8063}\tiny{$\pm 0.0037$} & 0.7703\tiny{$\pm 0.0005$}    \\
    & SMLLM & - & 0.0945\tiny{$\pm 0.0093$} & 0.1726\tiny{$\pm 0.0155$} & 0.5583\tiny{$\pm 0.0038$}   \\
    & GradSafe & \textbf{0.2882}\tiny{$\pm 0.0022$} & 0.7497\tiny{$\pm 0.0021$} & 0.7393\tiny{$\pm 0.0030$} & 0.7398\tiny{$\pm 0.0063$}   \\
    & FT & 0.4167\tiny{$\pm 0.0236$} & 0.8438\tiny{$\pm 0.0278$} & 0.7254\tiny{$\pm 0.0135$} & \textbf{0.7710}\tiny{$\pm 0.0172$}   \\
    & FJD & 0.4413\tiny{$\pm 0.0251$} & \textbf{0.8679}\tiny{$\pm 0.0295$} & 0.7309\tiny{$\pm 0.0175$} & 0.7470\tiny{$\pm 0.0135$}   \\
    
    \bottomrule
  \end{tabular} 

\end{table*}

\begin{table*}[t]
  \caption{Detection results (FPR, TPR, F1 and AUC) of jailbreak prompt under AdvPrompter. FJD outperforms baseline methods on almost all the LLMs.}
  \label{pro auc}

  \centering
  \setlength\tabcolsep{14pt}
  \footnotesize

  \begin{tabular}{cccccc}
    \toprule
    \multirow{2}{*}{\textbf{Model}}  &  \multirow{2}{*}{\textbf{Method}} & \multicolumn{4}{c}{\textbf{AdvPrompter}} \\
    \cmidrule(r){3-6}

    & & \textbf{FPR$\downarrow$}& \textbf{TPR $\uparrow$}& \textbf{F1$\uparrow$}& \textbf{AUC$\uparrow$}\\
    
    \midrule
    \multirow{5}{*}{Vicuna-7B} 
    & PPL& 0.3816\tiny{$\pm 0.0361$} & 0.7273\tiny{$\pm 0.0311$} & 0.5197\tiny{$\pm 0.0927$} & 0.6891\tiny{$\pm 0.0049$} \\
    & SMLLM & - & 0.5036\tiny{$\pm 0.0051$} & 0.6699\tiny{$\pm 0.0045$} & 0.7518\tiny{$\pm 0.0026$} \\
    & GradSafe & \textbf{0.1710}\tiny{$\pm 0.0250$} & 0.8245\tiny{$\pm 0.0166$} & \textbf{0.7571}\tiny{$\pm 0.0254$} & 0.8823\tiny{$\pm 0.0056$} \\
    & FT & 0.1920\tiny{$\pm 0.0057$} & 0.7289\tiny{$\pm 0.0293$} & 0.6071\tiny{$\pm 0.0192$} & 0.8471\tiny{$\pm 0.0142$} \\
    & FJD & 0.1949\tiny{$\pm 0.0141$} & \textbf{0.8763}\tiny{$\pm 0.0153$} & 0.6850\tiny{$\pm 0.0175$} & \textbf{0.9041}\tiny{$\pm 0.0072$} \\
    \midrule
    \multirow{5}{*}{Vicuna-13B} 
    & PPL  & 0.3661\tiny{$\pm 0.0140$} & 0.5606\tiny{$\pm 0.0107$} & 0.3252\tiny{$\pm 0.0741$} & 0.5933\tiny{$\pm 0.0038$}\\
    & SMLLM & - & 0.4630\tiny{$\pm 0.0080$} & 0.6287\tiny{$\pm 0.0078$} & 0.7315\tiny{$\pm 0.0040$} \\
    & GradSafe & 0.3861\tiny{$\pm 0.0114$} & 0.6431\tiny{$\pm 0.0277$} & 0.6988\tiny{$\pm 0.0167$} & 0.6641\tiny{$\pm 0.0133$}   \\
    & FT & \textbf{0.1725}\tiny{$\pm 0.0098$} & \textbf{0.8227}\tiny{$\pm 0.0170$} & \textbf{0.7762}\tiny{$\pm 0.0082$} & \textbf{0.9021}\tiny{$\pm 0.0071$} \\
    & FJD & 0.3120\tiny{$\pm 0.0149$} & 0.7045\tiny{$\pm 0.0249$} & 0.6046\tiny{$\pm 0.0148$} & 0.7218\tiny{$\pm 0.0180$} \\
    
    \midrule
    \multirow{5}{*}{Guanaco-7B} 
    & PPL  & 0.3707\tiny{$\pm 0.0129$} & 0.5292\tiny{$\pm 0.0072$} & 0.5274\tiny{$\pm 0.0581$} & 0.5542\tiny{$\pm 0.0046$} \\
    & SMLLM & - & 0.3721\tiny{$\pm 0.0264$} & 0.5419\tiny{$\pm 0.0279$} & 0.6861\tiny{$\pm 0.0132$} \\
    & GradSafe & \textbf{0.2975}\tiny{$\pm 0.0141$} & \textbf{0.7520}\tiny{$\pm 0.0036$} & \textbf{0.6718}\tiny{$\pm 0.0066$} & \textbf{0.8007}\tiny{$\pm 0.0059$} \\
    & FT & 0.6132\tiny{$\pm 0.0403$} & 0.4514\tiny{$\pm 0.0502$} & 0.3636\tiny{$\pm 0.0226$} & 0.3327\tiny{$\pm 0.0048$} \\
    & FJD & 0.4050\tiny{$\pm 0.0093$} & 0.6398\tiny{$\pm 0.0197$} & 0.5606\tiny{$\pm 0.0079$} & 0.7276\tiny{$\pm 0.0050$} \\
    \midrule
    \multirow{5}{*}{Guanaco-13B} 
    & PPL & 0.6667\tiny{$\pm 0.0067$} & \textbf{0.7500}\tiny{$\pm 0.0142$} & 0.0245\tiny{$\pm 0.0095$} & 0.3373\tiny{$\pm 0.0015$} \\
    & SMLLM & - & 0.7333\tiny{$\pm 0.0094$} & \textbf{0.8426}\tiny{$\pm 0.0065$} & \textbf{0.8667}\tiny{$\pm 0.0047$}\\
    & GradSafe & 0.2119\tiny{$\pm 0.0042$} & 0.6623\tiny{$\pm 0.0091$} & 0.7553\tiny{$\pm 0.0053$} & 0.7852\tiny{$\pm 0.0073$}   \\
    & FT & 0.3712\tiny{$\pm 0.0134$} & 0.5500\tiny{$\pm 0.0187$} & 0.2571\tiny{$\pm 0.0054$} & 0.6656\tiny{$\pm 0.0042$} \\
    & FJD & \textbf{0.2032}\tiny{$\pm 0.0192$} & 0.6510\tiny{$\pm 0.0151$} & 0.5023\tiny{$\pm 0.0018$} & 0.7985\tiny{$\pm 0.0030$}\\
    
    \bottomrule
  \end{tabular}
\end{table*}

\begin{table*}[t]
  \caption{Detection results (FPR, TPR, F1 and AUC) of jailbreak prompt under GCG. FJD outperforms baseline methods on Llama2 and achieves comparable performance to PPL with other LLMs.}
  \label{gcg auc}

  \centering
  \setlength\tabcolsep{14pt}
  \footnotesize

  \begin{tabular}{cccccc}
    \toprule
    \multirow{2}{*}{\textbf{Model}}  &  \multirow{2}{*}{\textbf{Method}} & \multicolumn{4}{c}{\textbf{GCG}} \\
    \cmidrule(r){3-6}

    & & \textbf{FPR$\downarrow$}& \textbf{TPR $\uparrow$}& \textbf{F1$\uparrow$}&\textbf{AUC$\uparrow$}\\
    
    \midrule
    \multirow{5}{*}{Llama2-7B} 
    & PPL  & 0.0624\tiny{$\pm 0.0084$} & 0.9756\tiny{$\pm 0.0054$} & 0.8506\tiny{$\pm 0.0543$} & 0.9717\tiny{$\pm 0.0004$}  \\
    & SMLLM & - & 0.8707\tiny{$\pm 0.0041$} & 0.9308\tiny{$\pm 0.0023$} & 0.9423\tiny{$\pm 0.0027$}     \\
    & GradSafe & 0.2306\tiny{$\pm 0.0204$} & 0.8148\tiny{$\pm 0.0354$} & 0.8756\tiny{$\pm 0.0212$} & 0.8943\tiny{$\pm 0.0035$} \\
    & FT & \textbf{0.0188}\tiny{$\pm 0.0153$} & 0.9738\tiny{$\pm 0.0032$}  & 0.9835\tiny{$\pm 0.0008$} & 0.9939\tiny{$\pm 0.0005$}\\
    & FJD & 0.0244\tiny{$\pm 0.0092$} & \textbf{0.9905}\tiny{$\pm 0.0082$}  & \textbf{0.9912}\tiny{$\pm 0.0041$} & \textbf{0.9990}\tiny{$\pm 0.0002$}\\
    \midrule
    \multirow{5}{*}{Llama2-13B} 
    & PPL & 0.0670\tiny{$\pm 0.0011$} & 0.9465\tiny{$\pm 0.0003$} & 0.9605\tiny{$\pm 0.0054$} & 0.9625\tiny{$\pm 0.0001$}  \\
    & SMLLM & -& 0.9585\tiny{$\pm 0.0099$} & 0.9788\tiny{$\pm 0.0067$} & 0.9798\tiny{$\pm 0.0027$}  \\
    & GradSafe & 0.3720\tiny{$\pm 0.0102$} & 0.7188\tiny{$\pm 0.0015$} & 0.7640\tiny{$\pm 0.0010$} & 0.7280\tiny{$\pm 0.0076$} \\
    & FT & 0.1476\tiny{$\pm 0.0098$} & 0.9537\tiny{$\pm 0.0050$} & 0.9440\tiny{$\pm 0.0013$} & 0.9558\tiny{$\pm 0.0031$} \\
    & FJD & \textbf{0.0592}\tiny{$\pm 0.0043$} & \textbf{0.9750}\tiny{$\pm 0.0024$} & \textbf{0.9651}\tiny{$\pm 0.0018$} & \textbf{0.9725}\tiny{$\pm 0.0010$}  \\
    
    \midrule
    \multirow{5}{*}{Vicuna-7B} 
    & PPL & \textbf{0.0382}\tiny{$\pm 0.0055$} & \textbf{0.9717}\tiny{$\pm 0.0003$} & \textbf{0.9776}\tiny{$\pm 0.0038$} &\textbf{0.9860}\tiny{$\pm 0.0002$}  \\
    & SMLLM & - & 0.8964\tiny{$\pm 0.0110$} & 0.9454\tiny{$\pm 0.0092$} & 0.9575\tiny{$\pm 0.0071$}    \\
    & GradSafe & 0.2334\tiny{$\pm 0.0249$} & 0.6428\tiny{$\pm 0.0326$} & 0.7305\tiny{$\pm 0.0194$} & 0.7575\tiny{$\pm 0.0117$}   \\
    & FT & 0.8986\tiny{$\pm 0.0163$} & 0.0827\tiny{$\pm 0.0236$} & 0.0673\tiny{$\pm 0.0087$} & 0.0300\tiny{$\pm 0.0018$}   \\
    & FJD & 0.2783\tiny{$\pm 0.0292$} & 0.6210\tiny{$\pm 0.0178$} & 0.7031\tiny{$\pm 0.0083$} & 0.7250\tiny{$\pm 0.0044$}   \\
    \midrule
    \multirow{5}{*}{Vicuna-13B} 
    & PPL & \textbf{0.0447}\tiny{$\pm 0.0043$} & \textbf{0.9892}\tiny{$\pm 0.0002$} & \textbf{0.9899}\tiny{$\pm 0.0023$} & \textbf{0.9851}\tiny{$\pm 0.0009$}  \\
    & SMLLM & - & 0.8974\tiny{$\pm 0.0036$} & 0.9459\tiny{$\pm 0.0030$} & 0.9550\tiny{$\pm 0.0032$}  \\
    & GradSafe & 0.1488\tiny{$\pm 0.0267$} & 0.6447\tiny{$\pm 0.0360$} & 0.7788\tiny{$\pm 0.0278$} & 0.7621\tiny{$\pm 0.0090$}   \\
    & FT & 0.3611\tiny{$\pm 0.0066$} & 0.5687\tiny{$\pm 0.0029$} & 0.6897\tiny{$\pm 0.0020$} & 0.5203\tiny{$\pm 0.0036$}   \\
    & FJD & 0.1874\tiny{$\pm 0.0271$} & 0.6581\tiny{$\pm 0.0283$} & 0.7539\tiny{$\pm 0.0136$} & 0.7829\tiny{$\pm 0.0128$}  \\
    
    \midrule
    \multirow{5}{*}{Guanaco-7B} 
    & PPL  & \textbf{0.0503}\tiny{$\pm 0.0059$} & \textbf{0.9803}\tiny{$\pm 0.0009$} & \textbf{0.9837}\tiny{$\pm 0.0034$} & \textbf{0.9833}\tiny{$\pm 0.0001$}  \\
    & SMLLM & - & 0.7767\tiny{$\pm 0.0083$} & 0.8743\tiny{$\pm 0.0053$} & 0.8811\tiny{$\pm 0.0029$}  \\
    & GradSafe & 0.4704\tiny{$\pm 0.0108$} & 0.7712\tiny{$\pm 0.0068$} & 0.6695\tiny{$\pm 0.0070$} & 0.7501\tiny{$\pm 0.0019$}  \\
    & FT & 0.0848\tiny{$\pm 0.0063$} & 0.9145\tiny{$\pm 0.0043$} & 0.9316\tiny{$\pm 0.0027$} & 0.9640\tiny{$\pm 0.0008$}  \\
    & FJD & 0.1119\tiny{$\pm 0.0095$} & 0.9015\tiny{$\pm 0.0086$} & 0.9129\tiny{$\pm 0.0060$} & 0.9515\tiny{$\pm 0.0040$}  \\
    \midrule
    \multirow{5}{*}{Guanaco-13B} 
    & PPL & \textbf{0.0615}\tiny{$\pm 0.0048$} & \textbf{0.9758}\tiny{$\pm 0.0045$} & \textbf{0.9825}\tiny{$\pm 0.0037$} & \textbf{0.9779}\tiny{$\pm 0.0003$}  \\
    & SMLLM & - & 0.8352\tiny{$\pm 0.0117$} & 0.9102\tiny{$\pm 0.0070$} & 0.9150\tiny{$\pm 0.0077$}  \\
    & GradSafe & 0.1592\tiny{$\pm 0.0093$} & 0.8539\tiny{$\pm 0.0101$} & 0.7518\tiny{$\pm 0.0066$} & 0.8364\tiny{$\pm 0.0084$}  \\
    & FT & 0.3056\tiny{$\pm 0.0293$} & 0.5825\tiny{$\pm 0.0180$} & 0.7066\tiny{$\pm 0.0129$} & 0.6317\tiny{$\pm 0.0042$}   \\
    & FJD & 0.2587\tiny{$\pm 0.0369$} & 0.6560\tiny{$\pm 0.0293$} & 0.7648\tiny{$\pm 0.0182$} & 0.7118\tiny{$\pm 0.0041$}  \\

    \bottomrule
  \end{tabular} 

\end{table*}

\begin{table*}[t]
  \caption{Detection results (FPR, TPR, F1 and AUC) of jailbreak prompt under MAC. FJD outperforms baseline methods on Llama2 and achieves comparable performance to PPL with other LLMs.}
  \label{mac auc}

  \centering
  \setlength\tabcolsep{14pt}
  \footnotesize

  \begin{tabular}{cccccc}
    \toprule
    \multirow{2}{*}{\textbf{Model}}  &  \multirow{2}{*}{\textbf{Method}} & \multicolumn{4}{c}{\textbf{MAC}} \\
    \cmidrule(r){3-6}

    & & \textbf{FPR$\downarrow$}& \textbf{TPR$\uparrow$}& \textbf{F1$\uparrow$}&\textbf{AUC$\uparrow$}\\
    
    \midrule
    \multirow{5}{*}{Llama2-7B} 
    & PPL  & 0.0391\tiny{$\pm 0.0016$} & 0.9404\tiny{$\pm 0.0208$} & 0.3192\tiny{$\pm 0.0921$} & 0.9816\tiny{$\pm 0.0001$} \\
    & SMLLM & - & 0.6482\tiny{$\pm 0.0128$} & 0.7866\tiny{$\pm 0.0123$} & 0.9091\tiny{$\pm 0.0064$}  \\
    & GradSafe  & 0.1001\tiny{$\pm 0.0083$} & 0.9136\tiny{$\pm 0.0077$} & \textbf{0.9209}\tiny{$\pm 0.0041$} & 0.9565\tiny{$\pm 0.0067$} \\
    & FT & 0.0516\tiny{$\pm 0.0032$} & \textbf{0.9335}\tiny{$\pm 0.0071$} & 0.6156\tiny{$\pm 0.0267$} & 0.9815\tiny{$\pm 0.0022$} \\
    & FJD & \textbf{0.0325}\tiny{$\pm 0.0030$} & 0.9307\tiny{$\pm 0.0073$} & 0.9093\tiny{$\pm 0.0037$} & \textbf{0.9839}\tiny{$\pm 0.0024$}  \\
    \midrule
    \multirow{5}{*}{Llama2-13B} 
    & PPL & 0.0411\tiny{$\pm 0.0011$} & 0.9091\tiny{$\pm 0.077$} & 0.2179\tiny{$\pm 0.0721$} & 0.9882\tiny{$\pm 0.0003$} \\
    & SMLLM & - & 0.8667\tiny{$\pm 0.0091$} & \textbf{0.9286}\tiny{$\pm 0.0058$} & 0.9333\tiny{$\pm 0.0021$} \\
    & GradSafe & 0.1813\tiny{$\pm 0.0048$} & 0.9231\tiny{$\pm 0.0362$} & 0.8471\tiny{$\pm 0.0193$} & 0.9398\tiny{$\pm 0.0059$} \\
    & FT & 0.0722\tiny{$\pm 0.0037$} & 0.9636\tiny{$\pm 0.0045$} & 0.5345\tiny{$\pm 0.0165$} & 0.9833\tiny{$\pm 0.0048$} \\
    & FJD & \textbf{0.0397}\tiny{$\pm 0.0033$} & \textbf{0.9999}\tiny{$\pm 0.0001$} & 0.8997\tiny{$\pm 0.0207$} & \textbf{0.9964}\tiny{$\pm 0.0030$} \\
    
    \midrule
    \multirow{5}{*}{Vicuna-7B} 
    & PPL & \textbf{0.0419}\tiny{$\pm 0.0092$} & \textbf{0.9849}\tiny{$\pm 0.0003$} & \textbf{0.9218}\tiny{$\pm 0.0333$} & \textbf{0.9853}\tiny{$\pm 0.0005$} \\
    & SMLLM & - & 0.7673\tiny{$\pm 0.0130$} & 0.8683\tiny{$\pm 0.0083$} & 0.8837\tiny{$\pm 0.0065$} \\
    & GradSafe & 0.0873\tiny{$\pm 0.0343$} & 0.8740\tiny{$\pm 0.0306$} & 0.9114\tiny{$\pm 0.0102$} & 0.9686\tiny{$\pm 0.0010$}     \\
    & FT & 0.7261\tiny{$\pm 0.0040$} & 0.4593\tiny{$\pm 0.0305$} & 0.5237\tiny{$\pm 0.0342$} & 0.2911\tiny{$\pm 0.0044$} \\
    & FJD & 0.1964\tiny{$\pm 0.0019$} & 0.8293\tiny{$\pm 0.095$} & 0.8561\tiny{$\pm 0.0071$} & 0.8703\tiny{$\pm 0.0101$}  \\
    \midrule
    \multirow{5}{*}{Vicuna-13B} 
    & PPL  & \textbf{0.0279}\tiny{$\pm 0.0003$} & \textbf{0.9813}\tiny{$\pm 0.0004$} & \textbf{0.9430}\tiny{$\pm 0.0249$} & \textbf{0.9902}\tiny{$\pm 0.0002$} \\
    & SMLLM & - & 0.9462\tiny{$\pm 0.0044$} & 0.9723\tiny{$\pm 0.0024$} & 0.9730\tiny{$\pm 0.0022$} \\
    & GradSafe & 0.2726\tiny{$\pm 0.0066$} & 0.6903\tiny{$\pm 0.0062$} & 0.7743\tiny{$\pm 0.0026$} & 0.7785\tiny{$\pm 0.0021$}   \\
    & FT & 0.7824\tiny{$\pm 0.0284$} & 0.6021\tiny{$\pm 0.0084$} & 0.6450\tiny{$\pm 0.0059$} & 0.3173\tiny{$\pm 0.0072$} \\
    & FJD & 0.2154\tiny{$\pm 0.074$} & 0.7847\tiny{$\pm 0.0092$} & 0.8250\tiny{$\pm 0.0079$} & 0.8091\tiny{$\pm 0.0129$} \\
    
    \midrule
    \multirow{5}{*}{Guanaco-7B} 
    & PPL  & \textbf{0.0514}\tiny{$\pm 0.0073$} & \textbf{0.9703}\tiny{$\pm 0.0005$} & \textbf{0.9385}\tiny{$\pm 0.0267$} & \textbf{0.9867}\tiny{$\pm 0.0006$} \\
    & SMLLM & - & 0.8143\tiny{$\pm 0.0010$} & 0.8976\tiny{$\pm 0.0006$} & 0.9071\tiny{$\pm 0.0005$} \\
    & GradSafe & 0.3212\tiny{$\pm 0.0040$} & 0.6049\tiny{$\pm 0.0039$} & 0.7031\tiny{$\pm 0.0018$} & 0.6662\tiny{$\pm 0.0019$} \\
    & FT & 0.2118\tiny{$\pm 0.0147$} & 0.7527\tiny{$\pm 0.0100$} & 0.8233\tiny{$\pm 0.0056$} & 0.8076\tiny{$\pm 0.0083$} \\
    & FJD & 0.1328\tiny{$\pm 0.0117$} & 0.8584\tiny{$\pm 0.0068$} & 0.9006\tiny{$\pm 0.0041$} & 0.9378\tiny{$\pm 0.0029$} \\
    \midrule
    \multirow{5}{*}{Guanaco-13B} 
    & PPL & \textbf{0.0255}\tiny{$\pm 0.0044$} & \textbf{0.9804}\tiny{$\pm 0.0002$} & 0.5476\tiny{$\pm 0.1103$} & \textbf{0.9895}\tiny{$\pm 0.0001$} \\
    & SMLLM & - & 0.8798\tiny{$\pm 0.0077$} & \textbf{0.9360}\tiny{$\pm 0.0044$} & 0.9399\tiny{$\pm 0.0039$}\\
    & GradSafe & 0.2478\tiny{$\pm 0.0102$} & 0.7758\tiny{$\pm 0.0255$} & 0.6437\tiny{$\pm 0.0177$} & 0.8271\tiny{$\pm 0.0072$}  \\
    & FT & 0.9889\tiny{$\pm 0.0063$} & 0.9020\tiny{$\pm 0.0328$} & 0.2591\tiny{$\pm 0.0071$} & 0.1424\tiny{$\pm 0.0044$} \\
    & FJD & 0.2295\tiny{$\pm 0.0063$} & 0.7686\tiny{$\pm 0.0328$} & 0.5176\tiny{$\pm 0.0071$} & 0.8490\tiny{$\pm 0.0044$} \\
    
    \bottomrule
  \end{tabular} 

\end{table*}

\section{Jailbreak Detection under Attacks with Mismatched Generalization}\label{app:mg}

In order to fully evaluate the performance of FJD under attacks via mismatched generalization, we supplement Cipher experiments on Llama2 7B/13B, Vicuna 7B/13B and Guanaco 7B/13B in Tab.~\ref{cipher more}. We supplement PAIR experiments on Vicuna 7B/13B and Llama2 7B/13B. In Tab.~\ref{pair more} illustrates the detection results (AUC) of jailbreak prompt and shows the effective detection of Jailbreak Prompts by FJD under PAIR attack. For the two jailbreak attacks, FJD can effectively detect these on all models.

\section{Jailbreak Detection under Hand-crafted Attacks}\label{app:hand}
We concurrently assess the detection efficacy of FJD on 28 manual attack methods in Hand-Crafted~\citep{chen2024red} method on Llama2 7B/13B (Tab.~\ref{hand l7 more}, ~\ref{hand l13 more}), Vicuna 7B/13B (Tab.~\ref{hand v7 more}, ~\ref{hand v13 more}) and Guanaco 7B/13B (Tab.~\ref{hand g7 more}, ~\ref{hand g13 more}). Both attack methods are human-readable, and FJD achieves the best performance on competing objectives and mismatched generalization. We hypothesize that this is attributed to the low perplexity of jailbreak prompts created by hand-crafted or semantically meaningful jailbreaks. Furthermore, benign prompts also exhibit relatively high perplexity, leading to PPL essentially performing reverse detection.

\section{Jailbreak Detection under Transferable Jailbreak Attack}\label{app:transfer}

We also provide complete jailbreak detection results under transferable attacks. This experiment employs Vicuna 7B, Llama2 7B and Guanaco 7B as the source models and aggregates jailbreak prompts acquired from GCG and AutoDAN. We systematically merge Vicuna 7B, Llama2 7B and Guanaco 7B to produce transferable jailbreak prompts using the transferable attack method within GCG. Then, we evaluate Vicuna 7B/13B, Llama2 7B/13B and Guanaco 7B/13B as the target models. In Tab.~\ref{transfer 13b} shows that, for the comprehensive migration of a successful jailbreak prompt generated on a single model, FJD demonstrates a more effective detection capability. In the case of jailbreak prompts generated by GCG transferable attack, FJD also demonstrates competitive results compared to PPL, which almost requires no extra model inference.

\begin{table*}[t]
  \caption{Detection results (FPR, TPR, F1 and AUC) of jailbreak prompt under Cipher. FJD outperforms baseline methods on almost all the LLMs.}
  \label{cipher more}

  \centering
  \setlength\tabcolsep{14pt}
  \footnotesize

  \begin{tabular}{cccccc}
    \toprule
    \multirow{2}{*}{\textbf{Model}}  &  \multirow{2}{*}{\textbf{Method}} & \multicolumn{4}{c}{\textbf{Cipher}} \\
    \cmidrule(r){3-6}

    & & \textbf{FPR$\downarrow$}& \textbf{TPR$\uparrow$}& \textbf{F1$\uparrow$}&\textbf{AUC$\uparrow$}\\
    
    \midrule
    \multirow{5}{*}{Llama2-7B} 
    & PPL & 0.9672\tiny{$\pm 0.0013$} & 0.0038\tiny{$\pm 0.0008$} & 0.0069\tiny{$\pm 0.0005$} & 0.0070\tiny{$\pm 0.0005$} \\
    & SMLLM & - & 0.0101\tiny{$\pm 0.0048$} & 0.0200\tiny{$\pm 0.0094$} & 0.5034\tiny{$\pm 0.0024$}  \\
    & GradSafe & 0.2070\tiny{$\pm 0.0092$} & 0.6345\tiny{$\pm 0.0096$} & 0.5477\tiny{$\pm 0.0137$} & 0.7862\tiny{$\pm 0.0045$}   \\
    & FT & 0.0629\tiny{$\pm 0.0067$} & 0.9812\tiny{$\pm 0.0147$} & 0.8730\tiny{$\pm 0.0198$} & 0.9636\tiny{$\pm 0.0025$} \\
    & FJD & \textbf{0.0386}\tiny{$\pm 0.0077$} & \textbf{0.9845}\tiny{$\pm 0.0096$} & \textbf{0.9257}\tiny{$\pm 0.0203$} & \textbf{0.9896}\tiny{$\pm 0.0014$}  \\
    \midrule
    \multirow{5}{*}{Llama2-13B} 
    & PPL & 0.9978\tiny{$\pm 0.0065$} & 0.0089\tiny{$\pm 0.0003$} & 0.0076\tiny{$\pm 0.0002$} & 0.0221\tiny{$\pm 0.0011$} \\
    & SMLLM & - & 0.8192\tiny{$\pm 0.0211$} & 0.8211\tiny{$\pm 0.0096$} & 0.9096\tiny{$\pm 0.0105$} \\
    & GradSafe & 0.1513\tiny{$\pm 0.0098$} & 0.7831\tiny{$\pm 0.0237$} & 0.6340\tiny{$\pm 0.0198$} & 0.8723\tiny{$\pm 0.0073$} \\
    & FT & 0.0493\tiny{$\pm 0.0069$} & 0.9839\tiny{$\pm 0.0126$} & 0.8901\tiny{$\pm 0.0135$} & 0.9837\tiny{$\pm 0.0031$} \\
    & FJD & \textbf{0.0114}\tiny{$\pm 0.0037$} & \textbf{0.9869}\tiny{$\pm 0.0102$} & \textbf{0.9658}\tiny{$\pm 0.0109$} & \textbf{0.9909}\tiny{$\pm 0.0091$} \\
    
    \midrule
    \multirow{5}{*}{Vicuna-7B} 
    & PPL & 0.9876\tiny{$\pm 0.0051$} & 0.0512\tiny{$\pm 0.0039$} & 0.0043\tiny{$\pm 0.0006$} & 0.0266\tiny{$\pm 0.0004$} \\
    & SMLLM & - & 0.0465\tiny{$\pm 0.0019$} & 0.0889\tiny{$\pm 0.0034$} & 0.5233\tiny{$\pm 0.0009$} \\
    & GradSafe & 0.4190\tiny{$\pm 0.0199$} & 0.7549\tiny{$\pm 0.0303$} & 0.7284\tiny{$\pm 0.0136$} & 0.7094\tiny{$\pm 0.0201$}   \\
    & FT & 0.2731\tiny{$\pm 0.0267$} & 0.7329\tiny{$\pm 0.0110$} & 0.8150\tiny{$\pm 0.0051$} & 0.7966\tiny{$\pm 0.0055$} \\
    & FJD & \textbf{0.1960}\tiny{$\pm 0.0189$} & \textbf{0.8362}\tiny{$\pm 0.0164$} & \textbf{0.8474}\tiny{$\pm 0.0053$} & \textbf{0.8633}\tiny{$\pm 0.0033$}  \\
    \midrule
    \multirow{5}{*}{Vicuna-13B} 
    & PPL  & 0.9913\tiny{$\pm 0.0110$} & 0.0477\tiny{$\pm 0.0015$} & 0.0036\tiny{$\pm 0.0002$} & 0.0259\tiny{$\pm 0.0005$} \\
    & SMLLM & - & 0.0690\tiny{$\pm 0.0050$} & 0.0110\tiny{$\pm 0.0084$} & 0.5344\tiny{$\pm 0.0025$} \\
    & GradSafe & 0.1894\tiny{$\pm 0.0041$} & 0.6683\tiny{$\pm 0.0111$} & 0.7783\tiny{$\pm 0.0073$} & 0.7972\tiny{$\pm 0.0036$}   \\
    & FT & 0.7262\tiny{$\pm 0.0125$} & 0.6528\tiny{$\pm 0.0271$} & 0.6712\tiny{$\pm 0.0245$} & 0.3030\tiny{$\pm 0.0150$} \\
    & FJD & \textbf{0.1405}\tiny{$\pm 0.0156$} & \textbf{0.9918}\tiny{$\pm 0.0046$} & \textbf{0.9680}\tiny{$\pm 0.0047$} & \textbf{0.8876}\tiny{$\pm 0.0170$} \\

    \midrule
    \multirow{5}{*}{Guanaco-7B} 
    & PPL & 0.9803\tiny{$\pm 0.0095$} & 0.0396\tiny{$\pm 0.0003$} & 0.0013\tiny{$\pm 0.0003$} & 0.0248\tiny{$\pm 0.0005$} \\
    & SMLLM & - & 0.0919\tiny{$\pm 0.0052$} & 0.1683\tiny{$\pm 0.0087$} & 0.5460\tiny{$\pm 0.0026$} \\
    & GradSafe & 0.3391\tiny{$\pm 0.0197$} & 0.6607\tiny{$\pm 0.0040$} & 0.7569\tiny{$\pm 0.0029$} & 0.8112\tiny{$\pm 0.0088$}   \\
    & FT & 0.9729\tiny{$\pm 0.0190$} & 0.7528\tiny{$\pm 0.0215$} & 0.2699\tiny{$\pm 0.0146$} & 0.4905\tiny{$\pm 0.0173$} \\
    & FJD & \textbf{0.2610}\tiny{$\pm 0.0277$} & \textbf{0.8122}\tiny{$\pm 0.0243$} & \textbf{0.8307}\tiny{$\pm 0.0120$} & \textbf{0.8299}\tiny{$\pm 0.0043$}  \\

    \midrule
    \multirow{5}{*}{Guanaco-13B} 
    & PPL  & 0.9782\tiny{$\pm 0.0071$} & 0.0374\tiny{$\pm 0.0005$} & 0.0051\tiny{$\pm 0.0002$} & 0.0254\tiny{$\pm 0.0008$} \\
    & SMLLM & - & 0.0964\tiny{$\pm 0.0039$} & 0.1724\tiny{$\pm 0.0066$} & 0.5482\tiny{$\pm 0.0020$} \\
    & GradSafe & 0.3418\tiny{$\pm 0.0116$} & 0.7401\tiny{$\pm 0.0227$} & 0.7425\tiny{$\pm 0.0050$} & 0.7691\tiny{$\pm 0.0105$}  \\
    & FT & 0.6230\tiny{$\pm 0.0250$} & 0.7723\tiny{$\pm 0.0236$} & 0.7624\tiny{$\pm 0.0237$} & 0.4724\tiny{$\pm 0.0148$} \\
    & FJD & \textbf{0.2825}\tiny{$\pm 0.0299$} & \textbf{0.8415}\tiny{$\pm 0.0235$} & \textbf{0.8810}\tiny{$\pm 0.0223$} & \textbf{0.8216}\tiny{$\pm 0.0191$} \\
    
    \bottomrule
  \end{tabular}

\end{table*}

\begin{table*}[t]
  \caption{Detection results (FPR, TPR, F1 and AUC) of jailbreak prompt under PAIR. FJD outperforms baseline methods on almost all the LLMs.}
  \label{pair more}

  \centering
  \setlength\tabcolsep{14pt}
  \footnotesize

  \begin{tabular}{cccccc}
    \toprule
    \multirow{2}{*}{\textbf{Model}}  &  \multirow{2}{*}{\textbf{Method}} & \multicolumn{4}{c}{\textbf{PAIR}} \\
    \cmidrule(r){3-6}

    & & \textbf{FPR$\downarrow$}& \textbf{TPR$\uparrow$}& \textbf{F1$\uparrow$}&\textbf{AUC$\uparrow$}\\
    
    \midrule
    \multirow{5}{*}{Llama2-7B} 
    & PPL & 0.7897\tiny{$\pm 0.0144$} & 0.0382\tiny{$\pm 0.0008$} & 0.0823\tiny{$\pm 0.0250$} & 0.2715\tiny{$\pm 0.0061$} \\
    & SMLLM & - & 0.7423\tiny{$\pm 0.0158$} & \textbf{0.8502}\tiny{$\pm 0.0110$} & 0.8625\tiny{$\pm 0.0019$}  \\
    & GradSafe & 0.0681\tiny{$\pm 0.0097$} & 0.9625\tiny{$\pm 0.0076$} & 0.7952\tiny{$\pm 0.0121$} & 0.9697\tiny{$\pm 0.0056$} \\
    & FT & 0.0937\tiny{$\pm 0.0040$} & \textbf{0.9750}\tiny{$\pm 0.0125$} & 0.7040\tiny{$\pm 0.0093$} & 0.9470\tiny{$\pm 0.0028$} \\
    & FJD & \textbf{0.0516}\tiny{$\pm 0.0212$} & 0.9687\tiny{$\pm 0.0087$} & 0.8042\tiny{$\pm 0.0059$} & \textbf{0.9761}\tiny{$\pm 0.0009$}  \\
    \midrule
    \multirow{5}{*}{Llama2-13B} 
    & PPL & 0.9367\tiny{$\pm 0.0033$} & 0.0067\tiny{$\pm 0.0009$} & 0.0088\tiny{$\pm 0.0007$} & 0.1140\tiny{$\pm 0.0142$} \\
    & SMLLM & - & 0.8889\tiny{$\pm 0.0079$} & 0.9394\tiny{$\pm 0.0043$} & 0.9244\tiny{$\pm 0.0024$} \\
    & GradSafe & 0.1161\tiny{$\pm 0.0031$} & 0.9998\tiny{$\pm 0.0002$} & 0.8797\tiny{$\pm 0.0195$} & 0.9185\tiny{$\pm 0.0029$} \\
    & FT & 0.1674\tiny{$\pm 0.0039$} & 0.9667\tiny{$\pm 0.0082$} & 0.9586\tiny{$\pm 0.0030$} & 0.9153\tiny{$\pm 0.0039$} \\
    & FJD & \textbf{0.1024}\tiny{$\pm 0.0011$} & \textbf{1.0000}\tiny{$\pm 0.0000$} & \textbf{0.9732}\tiny{$\pm 0.0021$} & \textbf{0.9264}\tiny{$\pm 0.0013$} \\
    
    \midrule
    \multirow{5}{*}{Vicuna-7B} 
    & PPL & 0.8886\tiny{$\pm 0.0032$} & 0.1222\tiny{$\pm 0.0006$} & 0.2256\tiny{$\pm 0.0167$} & 0.3245\tiny{$\pm 0.0024$} \\
    & SMLLM & - & 0.7622\tiny{$\pm 0.0074$} & \textbf{0.8615}\tiny{$\pm 0.0135$} & 0.8738\tiny{$\pm 0.0082$} \\
    & GradSafe & 0.2174\tiny{$\pm 0.0207$} & 0.8169\tiny{$\pm 0.0122$} & 0.7998\tiny{$\pm 0.0159$} & 0.8987\tiny{$\pm 0.0024$}     \\
    & FT & 0.4738\tiny{$\pm 0.0081$} & 0.5999\tiny{$\pm 0.0167$} & 0.4770\tiny{$\pm 0.0127$} & 0.5526\tiny{$\pm 0.0054$} \\
    & FJD & \textbf{0.1452}\tiny{$\pm 0.0094$} & \textbf{0.8702}\tiny{$\pm 0.0120$} & 0.8079\tiny{$\pm 0.0128$} & \textbf{0.9025}\tiny{$\pm 0.0027$} \\ 
    \midrule
    \multirow{5}{*}{Vicuna-13B} 
    & PPL  & 0.4701\tiny{$\pm 0.0471$} & 0.3333\tiny{$\pm 0.0114$} & 0.0991\tiny{$\pm 0.0232$} & 0.2272\tiny{$\pm 0.0010$} \\
    & SMLLM & - & 0.9167\tiny{$\pm 0.0035$} & \textbf{0.9562}\tiny{$\pm 0.0190$} & 0.9583\tiny{$\pm 0.0172$} \\
    & GradSafe & 0.2007\tiny{$\pm 0.0332$} & 0.8428\tiny{$\pm 0.0424$} & 0.7163\tiny{$\pm 0.0262$} & 0.8068\tiny{$\pm 0.0098$}   \\
    & FT & 0.5120\tiny{$\pm 0.0050$} & 0.7762\tiny{$\pm 0.0149$} & 0.0539\tiny{$\pm 0.0088$} & 0.5285\tiny{$\pm 0.0077$} \\
    & FJD & \textbf{0.0332}\tiny{$\pm 0.0023$} & \textbf{0.9895}\tiny{$\pm 0.0100$} & 0.9358\tiny{$\pm 0.0109$} & \textbf{0.9957}\tiny{$\pm 0.0009$} \\

    \bottomrule
  \end{tabular} 
\end{table*}

\begin{table*}[t]
  \caption{Detection results (AUC) of jailbreak prompt under Hand-crafted attacks on Llama2 7B. FJD outperforms baseline methods on almost all attacks and LLMs.}
  \label{hand l7 more}

  \centering
  \setlength\tabcolsep{9pt}

  \footnotesize

  \begin{tabular}{cccccc}
    \toprule
    \textbf{Attack on Llama2-7B}  & \textbf{PPL} & \textbf{SMLLM} & \textbf{GradSafe} & \textbf{FT} & \textbf{FJD}  \\
    
    \midrule
    aim &  0.5228\tiny{$\pm 0.0004$} & 0.6283\tiny{$\pm 0.0027$} & 0.9892\tiny{$\pm 0.0012$} & 0.9608\tiny{$\pm 0.0078$} & 0.9956\tiny{$\pm 0.0031$} \\
    dev\_mode\_v2 & 0.4289\tiny{$\pm 0.0015$} & 0.5050\tiny{$\pm 0.0012$} & 0.9746\tiny{$\pm 0.0009$} & 0.9812\tiny{$\pm 0.0023$} & 0.9985\tiny{$\pm 0.0018$}   \\
    dev\_mode\_ranti & 0.5485\tiny{$\pm 0.0003$} & 0.5219\tiny{$\pm 0.0015$} & 0.9825\tiny{$\pm 0.0059$} & 0.9829\tiny{$\pm 0.0016$} &  0.9995\tiny{$\pm 0.0007$}  \\
    distractors & 0.6897\tiny{$\pm 0.0042$} & 0.9514\tiny{$\pm 0.0354$} & 0.8236\tiny{$\pm 0.0090$} & 0.8510\tiny{$\pm 0.0045$} &  0.9024\tiny{$\pm 0.0289$}  \\
    distractors\_negated & 0.9718\tiny{$\pm 0.0003$} & 0.9991\tiny{$\pm 0.0002$} & 0.8978\tiny{$\pm 0.0016$} & 0.7267\tiny{$\pm 0.0135$} &  0.8167\tiny{$\pm 0.0161$}  \\
    evil\_confidant & 0.8422\tiny{$\pm 0.0017$} & 0.5632\tiny{$\pm 0.0065$} & 0.9989\tiny{$\pm 0.0015$} & 0.9998\tiny{$\pm 0.0004$} & 0.9973\tiny{$\pm 0.0022$}   \\
    poems & 0.9377\tiny{$\pm 0.0029$} & 0.9087\tiny{$\pm 0.0022$} & 0.9241\tiny{$\pm 0.0015$} & 0.8584\tiny{$\pm 0.0032$} & 0.9406\tiny{$\pm 0.0028$}   \\
    prefix\_injection\_1 & 0.9578\tiny{$\pm 0.0013$} & 0.8571\tiny{$\pm 0.0111$} & 0.9091\tiny{$\pm 0.0085$} & 0.8962\tiny{$\pm 0.0069$} & 0.9546\tiny{$\pm 0.0109$}   \\
    prefix\_injection\_2 & 0.1477\tiny{$\pm 0.0016$} & 0.7381\tiny{$\pm 0.0168$} & 0.9231\tiny{$\pm 0.0016$} & 0.9714\tiny{$\pm 0.0035$} &  0.9926\tiny{$\pm 0.0040$}  \\
    prefix\_injection\_hello & 0.8529\tiny{$\pm 0.0170$} & 0.9258\tiny{$\pm 0.0121$} & 0.8889\tiny{$\pm 0.0104$} & 0.9467\tiny{$\pm 0.0035$} & 0.9851\tiny{$\pm 0.0057$}   \\
    refusal\_suppression & 0.0073\tiny{$\pm 0.0005$} & 0.5552\tiny{$\pm 0.0231$} & 0.9832\tiny{$\pm 0.0008$} & 0.9043\tiny{$\pm 0.0024$} &  0.9809\tiny{$\pm 0.0007$}  \\
    refusal\_suppression\_inv & 0.0094\tiny{$\pm 0.0008$} & 0.5619\tiny{$\pm 0.0210$} & 0.9919\tiny{$\pm 0.0017$} & 0.9722\tiny{$\pm 0.0036$} & 0.9956\tiny{$\pm 0.0030$}   \\
    style\_injection\_short & 0.0068\tiny{$\pm 0.0001$} & 0.5519\tiny{$\pm 0.0026$} & 0.9232\tiny{$\pm 0.0085$} & 0.9652\tiny{$\pm 0.0029$} &  0.9724\tiny{$\pm 0.0057$}  \\
    \midrule
    Average of CO & 0.5326\tiny{$\pm 0.0025$} & 0.7129\tiny{$\pm 0.0105$} & 0.9392\tiny{$\pm 0.0041$} & 0.9244\tiny{$\pm 0.0043$}  & \textbf{0.9640}\tiny{$\pm 0.0067$}   \\
    \midrule
    auto\_payload\_splitting & 0.9290\tiny{$\pm 0.0005$} & 0.5670\tiny{$\pm 0.0053$} & 0.9853\tiny{$\pm 0.0007$} & 0.6133\tiny{$\pm 0.0133$} &  0.8081\tiny{$\pm 0.0114$}  \\
    base64 & 0.9205\tiny{$\pm 0.0003$} & 0.5313\tiny{$\pm 0.0059$} & 0.9643\tiny{$\pm 0.0047$} & 0.9939\tiny{$\pm 0.0009$} & 0.9884\tiny{$\pm 0.0039$}   \\
    base64\_raw & 0.9191\tiny{$\pm 0.0004$} & 0.5063\tiny{$\pm 0.0017$} & 0.9638\tiny{$\pm 0.0018$} & 0.9826\tiny{$\pm 0.0046$} & 0.9305\tiny{$\pm 0.0076$}   \\
    base64\_input\_only & 0.9281\tiny{$\pm 0.0006$} & 0.8996\tiny{$\pm 0.0062$} & 0.9376\tiny{$\pm 0.0092$} & 0.9939\tiny{$\pm 0.0020$} &  0.9954\tiny{$\pm 0.0008$}  \\
    base64\_output\_only & 0.9240\tiny{$\pm 0.0031$} & 0.7796\tiny{$\pm 0.0274$} & 0.9504\tiny{$\pm 0.0049$} & 0.7333\tiny{$\pm 0.0027$} & 0.9794\tiny{$\pm 0.0078$}   \\
    combination\_1 & 0.0031\tiny{$\pm 0.0001$} & 0.5050\tiny{$\pm 0.0033$} & 0.6328\tiny{$\pm 0.0189$} & 0.9918\tiny{$\pm 0.0040$} &  0.9770\tiny{$\pm 0.0072$}  \\
    combination\_2 & 0.0031\tiny{$\pm 0.0001$} & 0.5379\tiny{$\pm 0.0028$} & 0.6300\tiny{$\pm 0.0043$} & 0.9929\tiny{$\pm 0.0022$} & 0.9786\tiny{$\pm 0.0018$}   \\
    combination\_3 & 0.0053\tiny{$\pm 0.0001$} & 0.5682\tiny{$\pm 0.0030$} & 0.6734\tiny{$\pm 0.0147$} & 0.9916\tiny{$\pm 0.0026$} & 0.9869\tiny{$\pm 0.0037$}   \\
    disemvowel & 0.9895\tiny{$\pm 0.0004$} & 0.9792\tiny{$\pm 0.0295$} & 0.9398\tiny{$\pm 0.0015$} & 0.9908\tiny{$\pm 0.0047$} & 0.9262\tiny{$\pm 0.0152$} \\
    few\_shot\_json & 0.0104\tiny{$\pm 0.0007$} & 0.5218\tiny{$\pm 0.0024$} & 0.8938\tiny{$\pm 0.0010$} & 0.9385\tiny{$\pm 0.0093$} & 0.9872\tiny{$\pm 0.0024$}   \\
    leetspeak & 0.9797\tiny{$\pm 0.0011$} & 0.9111\tiny{$\pm 0.0240$} & 0.9258\tiny{$\pm 0.0064$} & 0.8975\tiny{$\pm 0.0023$} &  0.9314\tiny{$\pm 0.0086$}  \\
    rot13 & 0.9993\tiny{$\pm 0.0002$} & 0.9958\tiny{$\pm 0.0059$} & 0.9325\tiny{$\pm 0.0073$} & 0.9778\tiny{$\pm 0.0002$} & 0.9823\tiny{$\pm 0.0025$}   \\
    style\_injection\_json & 0.9176\tiny{$\pm 0.0101$} & 0.9457\tiny{$\pm 0.0128$} & 0.9120\tiny{$\pm 0.0061$} & 0.9693\tiny{$\pm 0.0032$}  & 0.9940\tiny{$\pm 0.0043$}   \\
    wikipedia & 0.8210\tiny{$\pm 0.0011$} & 0.9167\tiny{$\pm 0.0118$} & 0.8980\tiny{$\pm 0.0020$} & 0.8525\tiny{$\pm 0.0267$} & 0.8629\tiny{$\pm 0.0296$}   \\
    wikipedia\_with\_title & 0.9315\tiny{$\pm 0.0025$} & 0.9593\tiny{$\pm 0.0239$} & 0.9252\tiny{$\pm 0.0031$} & 0.9233\tiny{$\pm 0.0036$} & 0.9946\tiny{$\pm 0.0015$}   \\
    \midrule
    Average of MG & 0.6854\tiny{$\pm 0.0014$} & 0.7146\tiny{$\pm 0.0111$} & 0.8777\tiny{$\pm 0.0058$} & 0.9229\tiny{$\pm 0.0055$} & \textbf{0.9549}\tiny{$\pm 0.0072$}   \\

    \bottomrule
  \end{tabular}

\end{table*}

\begin{table*}[t]
  \caption{Detection results (AUC) of jailbreak prompt under Hand-crafted attacks on Llama2 13B. FJD outperforms baseline methods on almost all attacks and LLMs.}
  \label{hand l13 more}
  
  \centering
  \setlength\tabcolsep{9pt}
  \footnotesize

  \begin{tabular}{cccccc}
    \toprule
    \textbf{Attack on Llama2-13B}  & \textbf{PPL} & \textbf{SMLLM} & \textbf{GradSafe} & \textbf{FT} & \textbf{FJD}  \\
    
    \midrule
    aim &  0.5244\tiny{$\pm 0.0005$} & 0.7185\tiny{$\pm 0.0029$} & 0.9886\tiny{$\pm 0.0010$} & 0.6650\tiny{$\pm 0.0297$} & 0.9997\tiny{$\pm 0.0002$} \\
    dev\_mode\_v2 & 0.4292\tiny{$\pm 0.0003$} & 0.6128\tiny{$\pm 0.0019$} & 0.9943\tiny{$\pm 0.0009$} & 0.9774\tiny{$\pm 0.0015$} & 0.9974\tiny{$\pm 0.0005$}   \\
    dev\_mode\_ranti & 0.5485\tiny{$\pm 0.0010$} & 0.6379\tiny{$\pm 0.0021$} & 0.9728\tiny{$\pm 0.0026$} & 0.6893\tiny{$\pm 0.0094$} & 0.9826\tiny{$\pm 0.0011$}   \\
    distractors & 0.6906\tiny{$\pm 0.0040$} & 0.8955\tiny{$\pm 0.0362$} & 0.8627\tiny{$\pm 0.0215$} & 0.8397\tiny{$\pm 0.093$} & 0.8469\tiny{$\pm 0.0144$}   \\
    distractors\_negated & 0.9680\tiny{$\pm 0.0034$} & 0.9523\tiny{$\pm 0.0122$} & 0.8934\tiny{$\pm 0.0015$} & 0.8244\tiny{$\pm 0.0087$} & 0.8947\tiny{$\pm 0.0074$}   \\
    evil\_confidant & 0.8415\tiny{$\pm 0.0015$} & 0.5657\tiny{$\pm 0.0069$} & 0.9643\tiny{$\pm 0.0021$} & 0.8843\tiny{$\pm 0.0023$} & 0.9665\tiny{$\pm 0.0030$}   \\
    poems & 0.9225\tiny{$\pm 0.0007$} & 0.9478\tiny{$\pm 0.0048$} & 0.9773\tiny{$\pm 0.0066$} & 0.9486\tiny{$\pm 0.0047$} & 0.9631\tiny{$\pm 0.0056$}   \\
    prefix\_injection\_1 & 0.9733\tiny{$\pm 0.0003$} & 0.7312\tiny{$\pm 0.0099$} & 0.9675\tiny{$\pm 0.0018$} & 0.9536\tiny{$\pm 0.0081$} & 0.9792\tiny{$\pm 0.0017$}   \\
    prefix\_injection\_2 & 0.1042\tiny{$\pm 0.0104$} & 0.7039\tiny{$\pm 0.0152$} & 0.9893\tiny{$\pm 0.0016$} & 0.9063\tiny{$\pm 0.0055$} & 0.9996\tiny{$\pm 0.0005$}   \\
    prefix\_injection\_hello & 0.8237\tiny{$\pm 0.0075$} & 0.8837\tiny{$\pm 0.0129$} & 0.9963\tiny{$\pm 0.0012$} & 0.7619\tiny{$\pm 0.0161$} & 0.9990\tiny{$\pm 0.0009$}   \\
    refusal\_suppression & 0.0035\tiny{$\pm 0.0003$} & 0.5121\tiny{$\pm 0.0177$} & 0.9252\tiny{$\pm 0.0023$} & 0.6059\tiny{$\pm 0.0108$} & 0.9352\tiny{$\pm 0.0054$}   \\
    refusal\_suppression\_inv & 0.0051\tiny{$\pm 0.0004$} & 0.6284\tiny{$\pm 0.0173$} & 0.9776\tiny{$\pm 0.0033$} & 0.8568\tiny{$\pm 0.0094$} & 0.9987\tiny{$\pm 0.0016$}   \\
    style\_injection\_short & 0.0027\tiny{$\pm 0.0002$} & 0.5610\tiny{$\pm 0.0033$} & 0.9949\tiny{$\pm 0.0008$} & 0.8564\tiny{$\pm 0.0179$} & 0.9826\tiny{$\pm 0.0150$}   \\
    \midrule
    Average of CO & 0.5259\tiny{$\pm 0.0023$} & 0.7193\tiny{$\pm 0.0110$} & 0.9619\tiny{$\pm 0.0036$} & 0.8284\tiny{$\pm 0.0167$} & \textbf{0.9650}\tiny{$\pm 0.0044$}   \\
    \midrule
    auto\_payload\_splitting & 0.9290\tiny{$\pm 0.0011$} & 0.9454\tiny{$\pm 0.0048$} & 0.9780\tiny{$\pm 0.0017$} & 0.6326\tiny{$\pm 0.0327$} & 0.9863\tiny{$\pm 0.0106$}   \\
    base64 & 0.9264\tiny{$\pm 0.0009$} & 0.7655\tiny{$\pm 0.0121$} & 0.9412\tiny{$\pm 0.0048$} & 0.8416\tiny{$\pm 0.0109$} & 0.9428\tiny{$\pm 0.0070$}   \\
    base64\_raw & 0.9201\tiny{$\pm 0.0005$} & 0.6926\tiny{$\pm 0.0061$} & 0.7832\tiny{$\pm 0.0116$} & 0.4950\tiny{$\pm 0.0067$} & 0.9578\tiny{$\pm 0.0049$}   \\
    base64\_input\_only & 0.9264\tiny{$\pm 0.0008$} & 0.7290\tiny{$\pm 0.0055$} & 0.9419\tiny{$\pm 0.0096$} & 0.8813\tiny{$\pm 0.0081$} & 0.9482\tiny{$\pm 0.0058$}   \\
    base64\_output\_only & 0.8980\tiny{$\pm 0.0065$} & 0.9045\tiny{$\pm 0.0115$} & 0.8943\tiny{$\pm 0.0054$} & 0.7232\tiny{$\pm 0.0065$} & 0.9486\tiny{$\pm 0.0063$}   \\
    combination\_1 & 0.0031\tiny{$\pm 0.0001$} & 0.5151\tiny{$\pm 0.0023$} & 0.5120\tiny{$\pm 0.0023$} & 0.4738\tiny{$\pm 0.0152$} & 0.8133\tiny{$\pm 0.0230$}   \\
    combination\_2 & 0.0032\tiny{$\pm 0.0003$} & 0.5284\tiny{$\pm 0.0027$} & 0.5082\tiny{$\pm 0.0114$} & 0.4864\tiny{$\pm 0.0137$} & 0.8896\tiny{$\pm 0.0178$}   \\
    combination\_3 & 0.0051\tiny{$\pm 0.0003$} & 0.5168\tiny{$\pm 0.0030$} & 0.6146\tiny{$\pm 0.0241$} & 0.5668\tiny{$\pm 0.0124$} & 0.9989\tiny{$\pm 0.0003$}   \\
    disemvowel & 0.9894\tiny{$\pm 0.0007$} & 0.5889\tiny{$\pm 0.0048$} & 0.9041\tiny{$\pm 0.0014$} & 0.8387\tiny{$\pm 0.0156$} & 0.8430\tiny{$\pm 0.0162$}   \\
    few\_shot\_json & 0.0041\tiny{$\pm 0.0002$} & 0.5635\tiny{$\pm 0.0022$} & 0.9942\tiny{$\pm 0.0051$} & 0.9260\tiny{$\pm 0.0159$} & 0.9953\tiny{$\pm 0.0024$}   \\
    leetspeak & 0.9815\tiny{$\pm 0.0005$} & 0.9114\tiny{$\pm 0.0040$} & 0.9641\tiny{$\pm 0.0080$} & 0.9341\tiny{$\pm 0.0140$} &  0.9771\tiny{$\pm 0.0049$}  \\
    rot13 & 0.9896\tiny{$\pm 0.0003$} & 0.9374\tiny{$\pm 0.0078$} & 0.8500\tiny{$\pm 0.0056$} & 0.9146\tiny{$\pm 0.0118$} & 0.9618\tiny{$\pm 0.0148$}   \\
    style\_injection\_json & 0.9067\tiny{$\pm 0.0036$} & 0.8610\tiny{$\pm 0.0159$} & 0.8962\tiny{$\pm 0.0076$} & 0.7919\tiny{$\pm 0.0135$} & 0.9598\tiny{$\pm 0.0030$}   \\
    wikipedia & 0.8089\tiny{$\pm 0.0067$} & 0.9480\tiny{$\pm 0.0177$} & 0.9697\tiny{$\pm 0.0031$} & 0.9134\tiny{$\pm 0.0153$} & 0.9444\tiny{$\pm 0.0108$}   \\
    wikipedia\_with\_title & 0.8890\tiny{$\pm 0.0019$} & 0.9725\tiny{$\pm 0.0212$} & 0.9994\tiny{$\pm 0.0005$} & 0.9155\tiny{$\pm 0.0245$} & 0.9998\tiny{$\pm 0.0002$}   \\
    \midrule
    Average & 0.6787\tiny{$\pm 0.0016$} & 0.7587\tiny{$\pm 0.0081$} & 0.8501\tiny{$\pm 0.0068$} & 0.7557\tiny{$\pm 0.0145$} & \textbf{0.9444}\tiny{$\pm 0.0085$}   \\
    
    \bottomrule
  \end{tabular}

\end{table*}

\begin{table*}[t]
  \caption{Detection results (AUC) of jailbreak prompt under Hand-crafted attacks on Vicuna 7B. FJD outperforms baseline methods on almost all attacks and LLMs.}
  \label{hand v7 more}
  
  \centering
  \setlength\tabcolsep{9pt}
  \footnotesize

  \begin{tabular}{cccccc}
    \toprule
    \textbf{Attack on Vicuna-7B}  & \textbf{PPL} & \textbf{SMLLM} & \textbf{GradSafe} & \textbf{FT} & \textbf{FJD}  \\
    
    \midrule
    aim &  0.5250\tiny{$\pm 0.0004$} & 0.5077\tiny{$\pm 0.0036$} & 0.6688\tiny{$\pm 0.0083$} & 0.2783\tiny{$\pm 0.0167$} & 0.8976\tiny{$\pm 0.0074$} \\
    dev\_mode\_v2 & 0.4342\tiny{$\pm 0.0006$} & 0.5424\tiny{$\pm 0.0064$} & 0.8558\tiny{$\pm 0.0025$} & 0.2140\tiny{$\pm 0.0131$} & 0.8393\tiny{$\pm 0.0075$}   \\
    dev\_mode\_ranti & 0.5498\tiny{$\pm 0.0004$} & 0.5181\tiny{$\pm 0.0026$} & 0.8567\tiny{$\pm 0.0087$} & 0.5766\tiny{$\pm 0.0304$} & 0.8763\tiny{$\pm 0.0106$}   \\
    distractors & 0.6794\tiny{$\pm 0.0007$} & 0.5944\tiny{$\pm 0.0052$} & 0.7558\tiny{$\pm 0.0066$} & 0.6616\tiny{$\pm 0.0160$} & 0.8969\tiny{$\pm 0.0201$}   \\
    distractors\_negated & 0.9643\tiny{$\pm 0.0001$} & 0.7833\tiny{$\pm 0.0103$} & 0.7646\tiny{$\pm 0.0086$} & 0.6123\tiny{$\pm 0.0150$} & 0.7121\tiny{$\pm 0.0174$}   \\
    evil\_confidant & 0.8432\tiny{$\pm 0.0004$} & 0.5042\tiny{$\pm 0.0029$} & 0.7116\tiny{$\pm 0.0139$} & 0.0989\tiny{$\pm 0.0108$} & 0.8586\tiny{$\pm 0.0073$}   \\
    poems & 0.9260\tiny{$\pm 0.0004$} & 0.6472\tiny{$\pm 0.0071$} & 0.7783\tiny{$\pm 0.0053$} & 0.6799\tiny{$\pm 0.0105$} & 0.7953\tiny{$\pm 0.0199$}   \\
    prefix\_injection\_1 & 0.9697\tiny{$\pm 0.0002$} & 0.8875\tiny{$\pm 0.0029$} & 0.7911\tiny{$\pm 0.0035$} & 0.1724\tiny{$\pm 0.0203$} & 0.7741\tiny{$\pm 0.0084$}   \\
    prefix\_injection\_2 & 0.1291\tiny{$\pm 0.0043$} & 0.5218\tiny{$\pm 0.0074$} & 0.8254\tiny{$\pm 0.0044$} & 0.0269\tiny{$\pm 0.0071$} & 0.6244\tiny{$\pm 0.0191$}   \\
    prefix\_injection\_hello & 0.8513\tiny{$\pm 0.0015$} & 0.6972\tiny{$\pm 0.0055$} & 0.7377\tiny{$\pm 0.0076$} & 0.3405\tiny{$\pm 0.0149$} & 0.5606\tiny{$\pm 0.0132$}   \\
    refusal\_suppression & 0.0076\tiny{$\pm 0.0001$} & 0.9090\tiny{$\pm 0.0043$} & 0.8881\tiny{$\pm 0.0032$} & 0.6787\tiny{$\pm 0.0176$} & 0.8965\tiny{$\pm 0.0174$}   \\
    refusal\_suppression\_inv & 0.0082\tiny{$\pm 0.0001$} & 0.9465\tiny{$\pm 0.0080$} & 0.8174\tiny{$\pm 0.0037$} & 0.5201\tiny{$\pm 0.0192$} & 0.8635\tiny{$\pm 0.0160$}   \\
    style\_injection\_short & 0.0068\tiny{$\pm 0.0001$} & 0.5417\tiny{$\pm 0.0061$} & 0.7893\tiny{$\pm 0.0035$} & 0.7456\tiny{$\pm 0.0114$} & 0.8670\tiny{$\pm 0.0122$}   \\
    \midrule
    Average of CO & 0.5304\tiny{$\pm 0.0007$} & 0.6616\tiny{$\pm 0.0056$} & 0.7877\tiny{$\pm 0.0061$} & 0.4312\tiny{$\pm 0.0156$} & \textbf{0.8048}\tiny{$\pm 0.0135$}   \\
    \midrule
    auto\_payload\_splitting & 0.9604\tiny{$\pm 0.0002$} & 0.6726\tiny{$\pm 0.0085$} & 0.8068\tiny{$\pm 0.0023$} & 0.5218\tiny{$\pm 0.0159$} & 0.7296\tiny{$\pm 0.0153$}   \\
    base64 & 0.9206\tiny{$\pm 0.0013$} & 0.7671\tiny{$\pm 0.0045$} & 0.8002\tiny{$\pm 0.0034$} & 0.8508\tiny{$\pm 0.0095$} & 0.9133\tiny{$\pm 0.0028$}   \\
    base64\_raw & 0.9172\tiny{$\pm 0.0010$} & 0.5937\tiny{$\pm 0.0058$} & 0.8051\tiny{$\pm 0.0063$} & 0.7521\tiny{$\pm 0.0068$} & 0.8064\tiny{$\pm 0.0149$}   \\
    base64\_input\_only & 0.9264\tiny{$\pm 0.0001$} & 0.8646\tiny{$\pm 0.0079$} & 0.9016\tiny{$\pm 0.0035$} & 0.7544\tiny{$\pm 0.0151$} & 0.8542\tiny{$\pm 0.0293$}   \\
    base64\_output\_only & 0.8792\tiny{$\pm 0.0008$} &0.7806\tiny{$\pm 0.0149$} & 0.8797\tiny{$\pm 0.0040$} & 0.7957\tiny{$\pm 0.0179$} & 0.8762\tiny{$\pm 0.0232$}   \\
    combination\_1 & 0.0033\tiny{$\pm 0.0001$} & 0.5281\tiny{$\pm 0.0047$} & 0.6365\tiny{$\pm 0.0058$} & 0.0930\tiny{$\pm 0.0159$} & 0.7703\tiny{$\pm 0.0124$}   \\
    combination\_2 & 0.0032\tiny{$\pm 0.0001$} & 0.5293\tiny{$\pm 0.0083$} & 0.6847\tiny{$\pm 0.0028$} & 0.0519\tiny{$\pm 0.0110$} & 0.7570\tiny{$\pm 0.0116$}   \\
    combination\_3 & 0.0053\tiny{$\pm 0.0001$} & 0.5022\tiny{$\pm 0.0008$} & 0.6520\tiny{$\pm 0.0135$} & 0.1705\tiny{$\pm 0.0155$} & 0.7713\tiny{$\pm 0.0220$}   \\
    disemvowel & 0.9895\tiny{$\pm 0.0004$} & 0.8174\tiny{$\pm 0.0121$} & 0.8583\tiny{$\pm 0.0038$} & 0.5317\tiny{$\pm 0.0189$} & 0.7747\tiny{$\pm 0.0180$}   \\
    few\_shot\_json & 0.0035\tiny{$\pm 0.0003$} & 0.8521\tiny{$\pm 0.0061$} & 0.7425\tiny{$\pm 0.0049$} & 0.7443\tiny{$\pm 0.0170$} & 0.7556\tiny{$\pm 0.0128$}   \\
    leetspeak & 0.9784\tiny{$\pm 0.0010$} & 0.5563\tiny{$\pm 0.0017$} & 0.8740\tiny{$\pm 0.0022$} & 0.6685\tiny{$\pm 0.0157$} & 0.8160\tiny{$\pm 0.0250$}   \\
    rot13 & 0.9994\tiny{$\pm 0.0002$} & 0.7938\tiny{$\pm 0.0090$} & 0.8020\tiny{$\pm 0.0082$} & 0.7560\tiny{$\pm 0.0177$} & 0.8446\tiny{$\pm 0.0142$}   \\
    style\_injection\_json & 0.9176\tiny{$\pm 0.0101$} & 0.6125\tiny{$\pm 0.0045$} & 0.7889\tiny{$\pm 0.0100$} & 0.4890\tiny{$\pm 0.0106$} & 0.7238\tiny{$\pm 0.0100$}   \\
    wikipedia & 0.8281\tiny{$\pm 0.0026$} &0.9868\tiny{$\pm 0.0043$}  & 0.7781\tiny{$\pm 0.0003$} & 0.7454\tiny{$\pm 0.0162$} & 0.7851\tiny{$\pm 0.0074$}   \\
    wikipedia\_with\_title & 0.9084\tiny{$\pm 0.0005$} & 0.8750\tiny{$\pm 0.0112$} & 0.7860\tiny{$\pm 0.0020$} & 0.5131\tiny{$\pm 0.0137$}& 0.7279\tiny{$\pm 0.0205$}   \\
    \midrule
    Average of MG & 0.6827\tiny{$\pm 0.0013$} & 0.7155\tiny{$\pm 0.0070$} & 0.7864\tiny{$\pm 0.0049$} & 0.5625\tiny{$\pm 0.0145$} & \textbf{0.7937}\tiny{$\pm 0.0160$}   \\
    
    \bottomrule
  \end{tabular} 

\end{table*}

\begin{table*}[t]
  \caption{Detection results (AUC) of jailbreak prompt under Hand-crafted attacks on Vicuna 13B. FJD outperforms baseline methods on almost all attacks and LLMs.}
  \label{hand v13 more}

  \centering
  \setlength\tabcolsep{9pt}
  \footnotesize

  \begin{tabular}{ccccccc}
    \toprule
    \textbf{Attack on Vicuna-13B}  & \textbf{PPL} & \textbf{SMLLM} & \textbf{GradSafe} & \textbf{FT} & \textbf{FJD}  \\
    
    \midrule
    aim &  0.5254\tiny{$\pm 0.0009$} & 0.5014\tiny{$\pm 0.0010$} & 0.9409\tiny{$\pm 0.0032$} & 0.2218\tiny{$\pm 0.0128$} & 0.9458\tiny{$\pm 0.0133$} \\
    dev\_mode\_v2 & 0.4302\tiny{$\pm 0.0002$} & 0.8333\tiny{$\pm 0.0059$} & 0.8126\tiny{$\pm 0.0052$} & 0.4567\tiny{$\pm 0.0186$} & 0.9491\tiny{$\pm 0.0121$}   \\
    dev\_mode\_ranti & 0.5484\tiny{$\pm 0.0001$} & 0.6340\tiny{$\pm 0.0065$} & 0.8086\tiny{$\pm 0.0031$} & 0.5842\tiny{$\pm 0.0049$} & 0.9303\tiny{$\pm 0.0022$}   \\
    distractors & 0.6832\tiny{$\pm 0.0007$} & 0.7452\tiny{$\pm 0.0242$} & 0.7118\tiny{$\pm 0.0060$} & 0.6271\tiny{$\pm 0.0057$} & 0.9699\tiny{$\pm 0.0024$}   \\
    distractors\_negated & 0.9624\tiny{$\pm 0.0005$} & 0.9899\tiny{$\pm 0.0072$} & 0.9864\tiny{$\pm 0.0037$} & 0.6944\tiny{$\pm 0.0168$} & 0.9251\tiny{$\pm 0.0120$}   \\
    evil\_confidant & 0.8418\tiny{$\pm 0.0005$} & 0.5094\tiny{$\pm 0.0010$} & 0.6169\tiny{$\pm 0.0034$} & 0.4899\tiny{$\pm 0.0343$} &  0.9527\tiny{$\pm 0.0124$}  \\
    poems & 0.9250\tiny{$\pm 0.0004$} & 0.9513\tiny{$\pm 0.0053$} & 0.7733\tiny{$\pm 0.0081$} & 0.6919\tiny{$\pm 0.0266$} & 0.9984\tiny{$\pm 0.0139$}   \\
    prefix\_injection\_1 & 0.9605\tiny{$\pm 0.0015$} & 0.9403\tiny{$\pm 0.0156$} & 0.9126\tiny{$\pm 0.0018$} & 0.5745\tiny{$\pm 0.0166$} & 0.9278\tiny{$\pm 0.0081$}   \\
    prefix\_injection\_2 & 0.1292\tiny{$\pm 0.0011$} & 0.5731\tiny{$\pm 0.0063$} & 0.6094\tiny{$\pm 0.0165$} & 0.2526\tiny{$\pm 0.0076$} & 0.9244\tiny{$\pm 0.0065$}   \\
    prefix\_injection\_hello & 0.8464\tiny{$\pm 0.0009$} & 0.9760\tiny{$\pm 0.0006$} & 0.5527\tiny{$\pm 0.0069$} & 0.4665\tiny{$\pm 0.0172$} & 0.9114\tiny{$\pm 0.0066$}   \\
    refusal\_suppression & 0.0068\tiny{$\pm 0.0003$} & 0.5726\tiny{$\pm 0.0049$} & 0.8108\tiny{$\pm 0.0032$} & 0.6829\tiny{$\pm 0.0214$} & 0.9590\tiny{$\pm 0.0125$}   \\
    refusal\_suppression\_inv & 0.0063\tiny{$\pm 0.0002$} & 0.9825\tiny{$\pm 0.0070$} & 0.8392\tiny{$\pm 0.0087$} & 0.6891\tiny{$\pm 0.0125$} & 0.9529\tiny{$\pm 0.0073$}   \\
    style\_injection\_short & 0.0070\tiny{$\pm 0.0001$} & 0.5058\tiny{$\pm 0.0123$} & 0.9822\tiny{$\pm 0.0021$} & 0.7312\tiny{$\pm 0.0204$} & 0.9951\tiny{$\pm 0.0059$}   \\
    \midrule
    Average of CO & 0.5287\tiny{$\pm 0.0006$} & 0.7473\tiny{$\pm 0.0075$} & 0.7967\tiny{$\pm 0.0055$} & 0.5510\tiny{$\pm 0.0166$} & \textbf{0.9494}\tiny{$\pm 0.0089$}   \\
    \midrule
    auto\_payload\_splitting & 0.9612\tiny{$\pm 0.0008$} & 0.6709\tiny{$\pm 0.0107$}  & 0.5258\tiny{$\pm 0.0065$} & 0.4448\tiny{$\pm 0.0260$} & 0.9477\tiny{$\pm 0.0036$}   \\
    base64 & 0.9200\tiny{$\pm 0.0001$} & 0.5232\tiny{$\pm 0.0030$} & 0.5501\tiny{$\pm 0.0070$} & 0.7413\tiny{$\pm 0.0061$} & 0.9431\tiny{$\pm 0.0205$}   \\
    base64\_raw & 0.9218\tiny{$\pm 0.0004$} & 0.7395\tiny{$\pm 0.0126$} & 0.5155\tiny{$\pm 0.0057$} & 0.7450\tiny{$\pm 0.0111$} & 0.9713\tiny{$\pm 0.0151$}   \\
    base64\_input\_only & 0.9271\tiny{$\pm 0.0002$} & 0.7448\tiny{$\pm 0.0085$} & 0.6481\tiny{$\pm 0.0078$} & 0.6932\tiny{$\pm 0.0300$} &  0.9548\tiny{$\pm 0.0116$}  \\
    base64\_output\_only & 0.8879\tiny{$\pm 0.0030$} & 0.6027\tiny{$\pm 0.0117$} & 0.9589\tiny{$\pm 0.0009$} & 0.7283\tiny{$\pm 0.0272$} &  0.9204\tiny{$\pm 0.0109$}  \\
    combination\_1 & 0.0031\tiny{$\pm 0.0001$} & 0.5843\tiny{$\pm 0.0045$} & 0.9385\tiny{$\pm 0.0043$} & 0.5631\tiny{$\pm 0.0192$} & 0.9564\tiny{$\pm 0.0084$}    \\
    combination\_2 & 0.0030\tiny{$\pm 0.0001$} & 0.5221\tiny{$\pm 0.0049$} & 0.9425\tiny{$\pm 0.0018$} & 0.5544\tiny{$\pm 0.0071$} & 0.9565\tiny{$\pm 0.0078$}   \\
    combination\_3 & 0.0054\tiny{$\pm 0.0001$} & 0.5508\tiny{$\pm 0.0039$} & 0.9533\tiny{$\pm 0.0025$} & 0.6522\tiny{$\pm 0.0161$} & 0.9691\tiny{$\pm 0.0044$}   \\
    disemvowel & 0.9995\tiny{$\pm 0.0001$} & 0.7070\tiny{$\pm 0.0099$} & 0.9125\tiny{$\pm 0.0087$} & 0.7155\tiny{$\pm 0.0096$} & 0.9903\tiny{$\pm 0.0021$}   \\
    few\_shot\_json & 0.0079\tiny{$\pm 0.0001$} & 0.6630\tiny{$\pm 0.0078$} & 0.9581\tiny{$\pm 0.0011$} & 0.6996\tiny{$\pm 0.0091$} & 0.9707\tiny{$\pm 0.0089$}   \\
    leetspeak & 0.9759\tiny{$\pm 0.0006$} & 0.5747\tiny{$\pm 0.0037$} & 0.9455\tiny{$\pm 0.0005$} & 0.7210\tiny{$\pm 0.0091$} & 0.9257\tiny{$\pm 0.0109$}   \\
    rot13 & 0.9935\tiny{$\pm 0.0006$} & 0.6806\tiny{$\pm 0.0035$} & 0.9882\tiny{$\pm 0.0011$} & 0.7488\tiny{$\pm 0.0124$} & 0.9051\tiny{$\pm 0.0179$}   \\
    style\_injection\_json & 0.9031\tiny{$\pm 0.0017$} & 0.6109\tiny{$\pm 0.0094$} & 0.5256\tiny{$\pm 0.0052$} & 0.6661\tiny{$\pm 0.0209$} & 0.9045\tiny{$\pm 0.0147$}   \\
    wikipedia & 0.7794\tiny{$\pm 0.0011$} & 0.9583\tiny{$\pm 0.0295$} & 0.9967\tiny{$\pm 0.0011$} & 0.7066\tiny{$\pm 0.0262$} & 0.9688\tiny{$\pm 0.0126$}   \\
    wikipedia\_with\_title & 0.9065\tiny{$\pm 0.0008$} & 0.9096\tiny{$\pm 0.0126$} & 0.9186\tiny{$\pm 0.0047$} & 0.5204\tiny{$\pm 0.0216$} & 0.9813\tiny{$\pm 0.0069$}   \\
    \midrule
    Average of MG & 0.6797\tiny{$\pm 0.0007$} & 0.6695\tiny{$\pm 0.0091$} & 0.8185\tiny{$\pm 0.0039$} & 0.6600\tiny{$\pm 0.0168$} & \textbf{0.9510}\tiny{$\pm 0.0104$}   \\
    
    \bottomrule
  \end{tabular} 

\end{table*}

\begin{table*}[t]
  \caption{Detection results (AUC) of jailbreak prompt under Hand-crafted attacks on Guanaco 7B. FJD outperforms baseline methods on almost all attacks and LLMs.}
  \label{hand g7 more}
  
  \centering
  \setlength\tabcolsep{9pt}
  \footnotesize

  \begin{tabular}{cccccc}
    \toprule
    \textbf{Attack on Guanaco-7B}  & \textbf{PPL} & \textbf{SMLLM} & \textbf{GradSafe} & \textbf{FT} & \textbf{FJD}  \\
    
    \midrule
    aim &  0.5258\tiny{$\pm 0.0006$} & 0.8632\tiny{$\pm 0.0043$} & 0.7448\tiny{$\pm 0.0073$} & 0.8635\tiny{$\pm 0.0100$} & 0.9646\tiny{$\pm 0.0082$} \\
    dev\_mode\_v2 & 0.4292\tiny{$\pm 0.0011$} &0.5215\tiny{$\pm 0.0055$}  & 0.8763\tiny{$\pm 0.0057$} & 0.3517\tiny{$\pm 0.0161$} & 0.6243\tiny{$\pm 0.0236$}   \\
    dev\_mode\_ranti & 0.5486\tiny{$\pm 0.0004$} & 0.5757\tiny{$\pm 0.0055$} & 0.5532\tiny{$\pm 0.0142$} & 0.6699\tiny{$\pm 0.0302$} & 0.8346\tiny{$\pm 0.0087$}   \\
    distractors & 0.6778\tiny{$\pm 0.0003$} & 0.5056\tiny{$\pm 0.0026$} & 0.8878\tiny{$\pm 0.0015$} & 0.5649\tiny{$\pm 0.0205$} & 0.7928\tiny{$\pm 0.0230$}   \\
    distractors\_negated & 0.9562\tiny{$\pm 0.0010$} & 0.8285\tiny{$\pm 0.0064$} & 0.8914\tiny{$\pm 0.0011$} & 0.3073\tiny{$\pm 0.0164$} & 0.7874\tiny{$\pm 0.0093$}   \\
    evil\_confidant & 0.8423\tiny{$\pm 0.0002$} & 0.5521\tiny{$\pm 0.0017$} & 0.5760\tiny{$\pm 0.0007$} & 0.3389\tiny{$\pm 0.0149$} & 0.6062\tiny{$\pm 0.0250$}   \\
    poems & 0.9190\tiny{$\pm 0.0015$} & 0.5118\tiny{$\pm 0.0077$} & 0.8449\tiny{$\pm 0.0050$} & 0.4110\tiny{$\pm 0.0172$} & 0.7476\tiny{$\pm 0.0223$}   \\
    prefix\_injection\_1 & 0.9611\tiny{$\pm 0.0007$} & 0.8542\tiny{$\pm 0.0088$} & 0.6972\tiny{$\pm 0.0054$} & 0.9215\tiny{$\pm 0.0058$} & 0.9252\tiny{$\pm 0.0022$}   \\
    prefix\_injection\_2 & 0.1288\tiny{$\pm 0.0004$} & 0.5683\tiny{$\pm 0.0090$} & 0.5532\tiny{$\pm 0.0060$} & 0.9806\tiny{$\pm 0.0047$} & 0.9931\tiny{$\pm 0.0020$}   \\
    prefix\_injection\_hello & 0.8267\tiny{$\pm 0.0003$} & 0.8410\tiny{$\pm 0.0026$} & 0.7944\tiny{$\pm 0.0086$} & 0.6736\tiny{$\pm 0.0105$} & 0.9535\tiny{$\pm 0.0024$}   \\
    refusal\_suppression & 0.0066\tiny{$\pm 0.0002$} & 0.8840\tiny{$\pm 0.0084$} & 0.9035\tiny{$\pm 0.0025$} & 0.4061\tiny{$\pm 0.0321$} & 0.7954\tiny{$\pm 0.0148$}   \\
    refusal\_suppression\_inv & 0.0033\tiny{$\pm 0.0001$} & 0.8764\tiny{$\pm 0.0104$} & 0.8867\tiny{$\pm 0.0070$} & 0.4867\tiny{$\pm 0.0205$} & 0.9269\tiny{$\pm 0.0149$}   \\
    style\_injection\_short & 0.0059\tiny{$\pm 0.0001$} & 0.7611\tiny{$\pm 0.0116$} & 0.9240\tiny{$\pm 0.0028$} & 0.3274\tiny{$\pm 0.0284$} & 0.8508\tiny{$\pm 0.0038$}   \\
    \midrule
    Average of CO & 0.5255\tiny{$\pm 0.0005$} & 0.7033\tiny{$\pm 0.0065$} & 0.7795\tiny{$\pm 0.0052$} & 0.5618\tiny{$\pm 0.0175$} & \textbf{0.8310}\tiny{$\pm 0.0123$}   \\
    \midrule
    auto\_payload\_splitting & 0.9637\tiny{$\pm 0.0004$} & 0.7951\tiny{$\pm 0.0010$} & 0.9538\tiny{$\pm 0.0019$} & 0.4236\tiny{$\pm 0.0058$} & 0.9578\tiny{$\pm 0.0159$}   \\
    base64 & 0.9221\tiny{$\pm 0.0006$} & 0.9431\tiny{$\pm 0.0035$} & 0.6072\tiny{$\pm 0.0098$} & 0.3697\tiny{$\pm 0.0088$} & 0.6328\tiny{$\pm 0.0264$}   \\
    base64\_raw & 0.9190\tiny{$\pm 0.0010$} & 0.8611\tiny{$\pm 0.0071$} & 0.6806\tiny{$\pm 0.0048$} & 0.3287\tiny{$\pm 0.0068$} & 0.9141\tiny{$\pm 0.0190$}   \\
    base64\_input\_only & 0.9281\tiny{$\pm 0.0007$} & 0.9028\tiny{$\pm 0.0069$} & 0.5447\tiny{$\pm 0.0147$} & 0.4175\tiny{$\pm 0.0089$} & 0.7910\tiny{$\pm 0.0184$}   \\
    base64\_output\_only & 0.8838\tiny{$\pm 0.0008$} & 0.7569\tiny{$\pm 0.0113$} & 0.8771\tiny{$\pm 0.0081$} & 0.4180\tiny{$\pm 0.0192$} & 0.8431\tiny{$\pm 0.0134$}   \\
    combination\_1 & 0.0032\tiny{$\pm 0.0001$} & 0.6792\tiny{$\pm 0.0151$} & 0.8659\tiny{$\pm 0.0073$} & 0.9706\tiny{$\pm 0.0066$} & 0.9108\tiny{$\pm 0.0086$}   \\
    combination\_2 & 0.0031\tiny{$\pm 0.0001$} & 0.6854\tiny{$\pm 0.0103$} & 0.8837\tiny{$\pm 0.0014$} & 0.9770\tiny{$\pm 0.0051$} & 0.9874\tiny{$\pm 0.0193$}   \\
    combination\_3 & 0.0052\tiny{$\pm 0.0001$} & 0.8938\tiny{$\pm 0.0168$} & 0.5848\tiny{$\pm 0.0086$} & 0.8303\tiny{$\pm 0.0098$} & 0.9826\tiny{$\pm 0.0095$}   \\
    disemvowel & 0.9884\tiny{$\pm 0.0007$} & 0.8611\tiny{$\pm 0.0039$} & 0.9319\tiny{$\pm 0.0068$} & 0.3832\tiny{$\pm 0.0250$} & 0.9829\tiny{$\pm 0.0231$}   \\
    few\_shot\_json & 0.0017\tiny{$\pm 0.0001$} & 0.7563\tiny{$\pm 0.0051$} & 0.8124\tiny{$\pm 0.0084$} & 0.3417\tiny{$\pm 0.0275$} & 0.7719\tiny{$\pm 0.0134$}   \\
    leetspeak & 0.9793\tiny{$\pm 0.0002$} & 0.7653\tiny{$\pm 0.0087$} & 0.9264\tiny{$\pm 0.0031$} & 0.3738\tiny{$\pm 0.0117$} & 0.8922\tiny{$\pm 0.0133$}   \\
    rot13 & 0.9981\tiny{$\pm 0.0001$} & 0.8368\tiny{$\pm 0.0060$} & 0.8631\tiny{$\pm 0.0047$} & 0.4398\tiny{$\pm 0.0145$} & 0.9018\tiny{$\pm 0.0108$}   \\
    style\_injection\_json & 0.9000\tiny{$\pm 0.0010$} &  0.8368\tiny{$\pm 0.0060$} & 0.9803\tiny{$\pm 0.0012$} & 0.4005\tiny{$\pm 0.0138$} & 0.8547\tiny{$\pm 0.0135$}   \\
    wikipedia & 0.7799\tiny{$\pm 0.0024$} & 0.9271\tiny{$\pm 0.0090$} & 0.9359\tiny{$\pm 0.0007$} & 0.3493\tiny{$\pm 0.0139$} & 0.9474\tiny{$\pm 0.0086$}   \\
    wikipedia\_with\_title & 0.8962\tiny{$\pm 0.0003$} & 0.8472\tiny{$\pm 0.0039$} & 0.9499\tiny{$\pm 0.0015$} & 0.3035\tiny{$\pm 0.0113$} & 0.9526\tiny{$\pm 0.0161$}   \\
    \midrule
    Average of MG & 0.6781\tiny{$\pm 0.0006$} & 0.8232\tiny{$\pm 0.0076$} & 0.8265\tiny{$\pm 0.0055$} & 0.4885\tiny{$\pm 0.0126$} & \textbf{0.8882}\tiny{$\pm 0.0153$}   \\
    
    \bottomrule
  \end{tabular} 

\end{table*}

\begin{table*}[t]
  \caption{Detection results (AUC) of jailbreak prompt under Hand-crafted attacks on Guanaco 13B. FJD outperforms baseline methods on almost all attacks and LLMs.}
  \label{hand g13 more}
  
  \centering
  \setlength\tabcolsep{9pt}
  \footnotesize

  \begin{tabular}{ccccccc}
    \toprule
    \textbf{Attack on Guanaco-13B}  & \textbf{PPL} & \textbf{SMLLM} & \textbf{GradSafe} & \textbf{FT} & \textbf{FJD}  \\
    
    \midrule
    aim &  0.5262\tiny{$\pm 0.0011$} & 0.6211\tiny{$\pm 0.0048$} & 0.9235\tiny{$\pm 0.0082$} & 0.7403\tiny{$\pm 0.0197$} & 0.9063\tiny{$\pm 0.0106$} \\
    dev\_mode\_v2 & 0.4308\tiny{$\pm 0.0001$} & 0.5633\tiny{$\pm 0.0099$} & 0.8662\tiny{$\pm 0.0058$} & 0.7465\tiny{$\pm 0.0190$} & 0.8974\tiny{$\pm 0.0102$}   \\
    dev\_mode\_ranti & 0.5491\tiny{$\pm 0.0011$} & 0.5624\tiny{$\pm 0.0154$} & 0.8771\tiny{$\pm 0.0026$} & 0.6788\tiny{$\pm 0.0179$} & 0.8991\tiny{$\pm 0.0155$}   \\
    distractors & 0.6739\tiny{$\pm 0.0004$} & 0.5326\tiny{$\pm 0.0026$} & 0.8259\tiny{$\pm 0.0069$} & 0.4411\tiny{$\pm 0.0075$} & 0.7368\tiny{$\pm 0.0230$}  \\
    distractors\_negated & 0.9604\tiny{$\pm 0.0002$} & 0.9275\tiny{$\pm 0.0065$} & 0.9288\tiny{$\pm 0.0069$} & 0.5321\tiny{$\pm 0.0237$} & 0.9306\tiny{$\pm 0.0187$}   \\
    evil\_confidant & 0.3867\tiny{$\pm 0.0005$} & 0.8105\tiny{$\pm 0.0093$} & 0.5391\tiny{$\pm 0.0110$} & 0.5869\tiny{$\pm 0.0229$} & 0.6988\tiny{$\pm 0.0241$}   \\
    poems & 0.9239\tiny{$\pm 0.0008$} & 0.8346\tiny{$\pm 0.0026$} & 0.6334\tiny{$\pm 0.0129$} & 0.5711\tiny{$\pm 0.0212$} & 0.8541\tiny{$\pm 0.0140$}   \\
    prefix\_injection\_1 & 0.9631\tiny{$\pm 0.0007$} & 0.9074\tiny{$\pm 0.0074$} & 0.5783\tiny{$\pm 0.0053$} & 0.7653\tiny{$\pm 0.0202$} & 0.8138\tiny{$\pm 0.0118$}   \\
    prefix\_injection\_2 & 0.1293\tiny{$\pm 0.0021$} & 0.5892\tiny{$\pm 0.0110$} & 0.8277\tiny{$\pm 0.0065$} & 0.9330\tiny{$\pm 0.0148$} & 0.9365\tiny{$\pm 0.0035$}   \\
    prefix\_injection\_hello & 0.8232\tiny{$\pm 0.0011$} & 0.6841\tiny{$\pm 0.0089$} & 0.5469\tiny{$\pm 0.0177$} & 0.6577\tiny{$\pm 0.0137$} & 0.8363\tiny{$\pm 0.0069$}   \\
    refusal\_suppression & 0.0084\tiny{$\pm 0.0006$} & 0.8048\tiny{$\pm 0.0145$} & 0.6201\tiny{$\pm 0.0075$} & 0.6051\tiny{$\pm 0.0131$} & 0.8378\tiny{$\pm 0.0080$}   \\
    refusal\_suppression\_inv & 0.0011\tiny{$\pm 0.0001$} & 0.9669\tiny{$\pm 0.0054$} & 0.6884\tiny{$\pm 0.0058$} & 0.5137\tiny{$\pm 0.0216$} & 0.8173\tiny{$\pm 0.0253$}   \\
    style\_injection\_short & 0.0061\tiny{$\pm 0.0001$} & 0.5890\tiny{$\pm 0.0198$} & 0.7599\tiny{$\pm 0.0056$} & 0.3727\tiny{$\pm 0.0153$} & 0.8098\tiny{$\pm 0.0120$}   \\
    \midrule
    Average of CO & 0.4909\tiny{$\pm 0.0007$} & 0.7226\tiny{$\pm 0.0091$} & 0.7396\tiny{$\pm 0.0079$} & 0.6265\tiny{$\pm 0.0177$} & \textbf{0.8442}\tiny{$\pm 0.0141$}   \\
    \midrule
    auto\_payload\_splitting & 0.9549\tiny{$\pm 0.0011$} & 0.8957\tiny{$\pm 0.0108$} & 0.6317\tiny{$\pm 0.0017$} & 0.4366\tiny{$\pm 0.0073$} & 0.8580\tiny{$\pm 0.0146$}   \\
    base64 & 0.9224\tiny{$\pm 0.0008$} & 0.7656\tiny{$\pm 0.0148$} & 0.8053\tiny{$\pm 0.0018$} & 0.6270\tiny{$\pm 0.0084$} &  0.8464\tiny{$\pm 0.0131$}  \\
    base64\_raw & 0.9266\tiny{$\pm 0.0007$} & 0.8764\tiny{$\pm 0.0071$} & 0.7970\tiny{$\pm 0.0057$} & 0.4882\tiny{$\pm 0.0228$} & 0.8545\tiny{$\pm 0.0140$}   \\
    base64\_input\_only & 0.9323\tiny{$\pm 0.0018$} & 0.9135\tiny{$\pm 0.0106$} & 0.6775\tiny{$\pm 0.0049$} & 0.4628\tiny{$\pm 0.0180$} & 0.7069\tiny{$\pm 0.0186$}   \\
    base64\_output\_only & 0.8640\tiny{$\pm 0.0009$} & 0.6353\tiny{$\pm 0.0327$} & 0.7637\tiny{$\pm 0.0074$} & 0.6699\tiny{$\pm 0.0297$} & 0.8262\tiny{$\pm 0.0169$}   \\
    combination\_1 & 0.0031\tiny{$\pm 0.0001$} & 0.6174\tiny{$\pm 0.0269$} & 0.8625\tiny{$\pm 0.0041$} & 0.7539\tiny{$\pm 0.0234$} & 0.9870\tiny{$\pm 0.0058$}   \\
    combination\_2 & 0.0032\tiny{$\pm 0.0001$} & 0.6167\tiny{$\pm 0.0029$} & 0.8950\tiny{$\pm 0.0047$} & 0.7276\tiny{$\pm 0.0166$} & 0.9044\tiny{$\pm 0.0137$}   \\
    combination\_3 & 0.0052\tiny{$\pm 0.0002$} & 0.7836\tiny{$\pm 0.0052$} & 0.5529\tiny{$\pm 0.0057$} & 0.4987\tiny{$\pm 0.0188$} &  0.8223\tiny{$\pm 0.0199$}  \\
    disemvowel & 0.9996\tiny{$\pm 0.0003$} & 0.6299\tiny{$\pm 0.0111$} & 0.7080\tiny{$\pm 0.0055$} & 0.3476\tiny{$\pm 0.0268$} & 0.7935\tiny{$\pm 0.0186$}   \\
    few\_shot\_json & 0.0074\tiny{$\pm 0.0004$} & 0.6813\tiny{$\pm 0.0141$} & 0.8629\tiny{$\pm 0.0038$} & 0.5519\tiny{$\pm 0.0223$} & 0.8544\tiny{$\pm 0.0215$}   \\
    leetspeak & 0.9582\tiny{$\pm 0.0012$} & 0.6409\tiny{$\pm 0.0199$} & 0.5959\tiny{$\pm 0.0147$} & 0.4745\tiny{$\pm 0.0230$} & 0.7990\tiny{$\pm 0.0186$}   \\
    rot13 & 0.9895\tiny{$\pm 0.0005$} & 0.6399\tiny{$\pm 0.0049$} & 0.8622\tiny{$\pm 0.0046$} & 0.2805\tiny{$\pm 0.0160$} & 0.8465\tiny{$\pm 0.0131$}   \\
    style\_injection\_json & 0.9029\tiny{$\pm 0.0010$} & 0.8176\tiny{$\pm 0.0105$} & 0.8189\tiny{$\pm 0.0031$} & 0.4873\tiny{$\pm 0.0169$} & 0.8127\tiny{$\pm 0.0227$}   \\
    wikipedia & 0.7870\tiny{$\pm 0.0053$} & 0.9192\tiny{$\pm 0.0120$} & 0.8557\tiny{$\pm 0.0141$} & 0.5645\tiny{$\pm 0.0124$} & 0.8502\tiny{$\pm 0.0239$}   \\
    wikipedia\_with\_title & 0.9009\tiny{$\pm 0.0009$} & 0.9538\tiny{$\pm 0.0137$} & 0.8731\tiny{$\pm 0.0022$} & 0.5312\tiny{$\pm 0.0260$} & 0.8301\tiny{$\pm 0.0213$}   \\
    \midrule
    Average & 0.6771\tiny{$\pm 0.0010$} & 0.7591\tiny{$\pm 0.0131$} & 0.7708\tiny{$\pm 0.0056$} & 0.5268\tiny{$\pm 0.0019$} & \textbf{0.8395}\tiny{$\pm 0.0171$}   \\
    
    \bottomrule
  \end{tabular} 

\end{table*}

\begin{table*}
  \caption{The complete detection results (AUC) of jailbreak prompt under transferable attack. FJD can effectively detect jailbreak prompts in most cases.}

  \label{transfer 13b}
  \centering
  \setlength\tabcolsep{17pt}
  \footnotesize
  \begin{tabular}{lcccc}

    \toprule
    \textbf{\diagbox{Source}{Target}}& \textbf{Methods} &\textbf{Llama2-7B}&\textbf{Vicuna-7B}&\textbf{Guanaco-7B}\\
    \midrule
    \multirow{4}{*}{Vicuna-7B} 
    & PPL & 0.7647\tiny{$\pm 0.0012$} & 0.8406\tiny{$\pm 0.0007$} & 0.8745\tiny{$\pm 0.0005$}\\
    & SMLLM & 0.7507\tiny{$\pm 0.0037$} & 0.8603\tiny{$\pm 0.0059$} & 0.8250\tiny{$\pm 0.0063$}\\
    & GradSafe & 0.9902\tiny{$\pm 0.0014$} & 0.8605\tiny{$\pm 0.0046$} & 0.8847\tiny{$\pm 0.0029$}\\
    & FJD & \textbf{0.9970}\tiny{$\pm 0.0025$} &\textbf{0.9777}\tiny{$\pm 0.0019$} & \textbf{0.9688}\tiny{$\pm 0.0051$}\\
    \midrule
    \multirow{4}{*}{Llama2-7B}
    & PPL & 0.7437\tiny{$\pm 0.0017$} & 0.7026\tiny{$\pm 0.0009$} & 0.8770\tiny{$\pm 0.0006$} \\
    & SMLLM & 0.7971\tiny{$\pm 0.0035$} & 0.5682\tiny{$\pm 0.0043$} & 0.6863\tiny{$\pm 0.0072$} \\
    & GradSafe & 0.8913\tiny{$\pm 0.0049$} & \textbf{0.8880}\tiny{$\pm 0.0077$} & 0.7459\tiny{$\pm 0.0129$}\\
    & FJD & \textbf{0.9873}\tiny{$\pm 0.0030$} &0.7062\tiny{$\pm 0.0097$} &\textbf{0.9549}\tiny{$\pm 0.0070$} \\
    \midrule
    \multirow{4}{*}{Guanaco-7B}
    & PPL & 0.8221\tiny{$\pm 0.0021$} & 0.7679\tiny{$\pm 0.0011$} & 0.8532\tiny{$\pm 0.0032$} \\
    & SMLLM & 0.9243\tiny{$\pm 0.0012$} & 0.7941\tiny{$\pm 0.0052$} & 0.8927\tiny{$\pm 0.0065$}\\
    & GradSafe & 0.9907\tiny{$\pm 0.0003$} & 0.7735\tiny{$\pm 0.0062$} & 0.8289\tiny{$\pm 0.0067$}\\
    & FJD & \textbf{0.9926}\tiny{$\pm 0.0029$} & \textbf{0.9781}\tiny{$\pm 0.0014$} & \textbf{0.9875}\tiny{$\pm 0.0017$}\\
    \midrule
    \multirow{4}{*}{Vicuna-7B + Llama2-7B}
    & PPL & 0.9788\tiny{$\pm 0.0003$} & 0.9803\tiny{$\pm 0.0002$}& \textbf{0.9783}\tiny{$\pm 0.0004$} \\
    & SMLLM & 0.9253\tiny{$\pm 0.0019$} & 0.8889\tiny{$\pm 0.0021$} & 0.8675\tiny{$\pm 0.0074$} \\
    & GradSafe & 0.9563\tiny{$\pm 0.0068$} & 0.8835\tiny{$\pm 0.0059$} & 0.9251\tiny{$\pm 0.0036$}\\
    & FJD & \textbf{0.9951}\tiny{$\pm 0.0017$} & \textbf{0.9820}\tiny{$\pm 0.0022$} & 0.9342\tiny{$\pm 0.0051$} \\
    \midrule
    \multirow{4}{*}{Vicuna-7B + Guanaco-7B}
    & PPL & \textbf{0.9832}\tiny{$\pm 0.0005$} & \textbf{0.9819}\tiny{$\pm 0.0003$} & 0.9832\tiny{$\pm 0.0003$}\\
    & SMLLM & 0.9537\tiny{$\pm 0.0017$} & 0.8429\tiny{$\pm 0.0055$} & 0.9246\tiny{$\pm 0.0020$} \\
    & GradSafe & 0.9822\tiny{$\pm 0.0015$} & 0.9125\tiny{$\pm 0.0036$} & 0.9043\tiny{$\pm 0.0010$}\\
    & FJD & 0.8922\tiny{$\pm 0.0034$} & 0.8952\tiny{$\pm 0.0070$} & \textbf{0.9945}\tiny{$\pm 0.0014$} \\
    \midrule
    \multirow{4}{*}{Llama2-7B + Guanaco-7B}
    & PPL & 0.9849\tiny{$\pm 0.0007$} & 0.9772\tiny{$\pm 0.0011$} & 0.9827\tiny{$\pm 0.0003$}\\
    & SMLLM & 0.8263\tiny{$\pm 0.0087$} & 0.9146\tiny{$\pm 0.0093$} & 0.7380\tiny{$\pm 0.0102$}\\
    & GradSafe & 0.8293\tiny{$\pm 0.0072$} & 0.9456\tiny{$\pm 0.0023$} & 0.8154\tiny{$\pm 0.0074$}\\
    & FJD & \textbf{0.9998}\tiny{$\pm 0.0002$} & \textbf{1.0000}\tiny{$\pm 0.0000$} & \textbf{0.9834}\tiny{$\pm 0.0015$} \\
    \midrule
    \multirow{4}{*}{\makecell[l]{Vicuna-7B + Llama2-7B \\+ Guanaco-7B}}
    & PPL & 0.9844\tiny{$\pm 0.0006$} & \textbf{0.9837}\tiny{$\pm 0.0007$} & 0.9845\tiny{$\pm 0.0003$} \\
    & SMLLM & 0.8034\tiny{$\pm 0.0088$} & 0.8774\tiny{$\pm 0.0075$} & 0.7461\tiny{$\pm 0.0099$} \\
    & GradSafe & 0.9249\tiny{$\pm 0.0029$} & 0.9132\tiny{$\pm 0.0022$} & 0.9533\tiny{$\pm 0.0078$}\\
    & FJD & \textbf{0.9954}\tiny{$\pm 0.0013$} & 0.9695\tiny{$\pm 0.0035$} & \textbf{0.9901}\tiny{$\pm 0.0049$} \\  
    \midrule
    & \textbf{Methods} &\textbf{Llama2-13B}&\textbf{Vicuna-13B}&\textbf{Guanaco-13B}\\

    \midrule
    \multirow{4}{*}{Vicuna-7B} 
    & PPL & 0.9177\tiny{$\pm 0.0028$} & 0.7941\tiny{$\pm 0.0002$} & 0.8915\tiny{$\pm 0.0004$} \\
    & SMLLM & 0.6214\tiny{$\pm 0.0129$} & 0.5484\tiny{$\pm 0.0111$} & 0.6651\tiny{$\pm 0.0099$} \\
    & GradSafe & 0.8949\tiny{$\pm 0.0096$} & 0.8486\tiny{$\pm 0.0063$} & 0.9039\tiny{$\pm 0.0087$}\\
    & FJD & \textbf{0.9537}\tiny{$\pm 0.0039$} &\textbf{0.9349}\tiny{$\pm 0.0107$} &\textbf{0.9785}\tiny{$\pm 0.0087$} \\
    \midrule
    \multirow{4}{*}{Llama2-7B}
    & PPL & 0.8515\tiny{$\pm 0.0003$} & 0.7782\tiny{$\pm 0.0002$} & 0.7967\tiny{$\pm 0.0003$} \\
    & SMLLM & 0.7500\tiny{$\pm 0.0091$} & 0.5593\tiny{$\pm 0.0109$} & 0.6250\tiny{$\pm 0.0137$} \\
    & GradSafe & 0.8817\tiny{$\pm 0.0058$} & 0.8272\tiny{$\pm 0.0070$} & 0.8658\tiny{$\pm 0.0069$}\\
    & FJD & \textbf{0.9087}\tiny{$\pm 0.0074$} & \textbf{0.9175}\tiny{$\pm 0.0062$} & \textbf{0.9527}\tiny{$\pm 0.0189$} \\
    \midrule
    \multirow{4}{*}{Guanaco-7B}
    & PPL & 0.8221\tiny{$\pm 0.0002$} & 0.8644\tiny{$\pm 0.0004$} & 0.8059\tiny{$\pm 0.0007$}  \\
    & SMLLM & 0.8587\tiny{$\pm 0.0059$} & 0.9287\tiny{$\pm 0.0037$} & 0.8066\tiny{$\pm 0.0041$}\\
    & GradSafe & 0.8905\tiny{$\pm 0.0017$} & 0.9021\tiny{$\pm 0.0034$} & 0.9325\tiny{$\pm 0.0045$}\\
    & FJD &\textbf{0.9425}\tiny{$\pm 0.0022$} & \textbf{0.9324}\tiny{$\pm 0.0063$} & \textbf{0.9769}\tiny{$\pm 0.0103$} \\
    \midrule
    \multirow{4}{*}{Vicuna-7B + Llama2-7B}
    & PPL & \textbf{0.9852}\tiny{$\pm 0.0012$} & \textbf{0.9794}\tiny{$\pm 0.0017$} &\textbf{0.9822}\tiny{$\pm 0.0009$} \\
    & SMLLM & 0.8846\tiny{$\pm 0.0036$} & 0.9176\tiny{$\pm 0.0068$} & 0.7951\tiny{$\pm 0.0063$} \\
    & GradSafe & 0.9364\tiny{$\pm 0.0078$} & 0.8445\tiny{$\pm 0.0022$} & 0.9240\tiny{$\pm 0.0061$}\\
    & FJD & 0.9716\tiny{$\pm 0.0038$} & 0.8516\tiny{$\pm 0.0118$} & 0.9772\tiny{$\pm 0.0031$}\\
    \midrule
    \multirow{4}{*}{Vicuna-7B + Guanaco-7B}
    & PPL & \textbf{0.9882}\tiny{$\pm 0.0004$} & \textbf{0.9866}\tiny{$\pm 0.0009$} & \textbf{0.9835}\tiny{$\pm 0.0005$} \\
    & SMLLM & 0.9722\tiny{$\pm 0.0015$} & 0.9320\tiny{$\pm 0.0021$} & 0.8004\tiny{$\pm 0.0073$} \\
    & GradSafe & 0.9880\tiny{$\pm 0.0023$} & 0.9769\tiny{$\pm 0.0027$} & 0.7457\tiny{$\pm 0.0097$}\\
    & FJD & 0.9522\tiny{$\pm 0.0067$} & 0.9850\tiny{$\pm 0.0064$} & 0.8461\tiny{$\pm 0.0036$}  \\
    \midrule
    \multirow{4}{*}{Llama2-7B + Guanaco-7B}
    & PPL & \textbf{0.9849}\tiny{$\pm 0.0011$} & \textbf{0.9839}\tiny{$\pm 0.0016$} & \textbf{0.9800}\tiny{$\pm 0.0009$} \\
    & SMLLM & 0.9125\tiny{$\pm 0.0022$} & 0.8615\tiny{$\pm 0.0036$} & 0.7469\tiny{$\pm 0.0074$} \\
    & GradSafe & 0.8531\tiny{$\pm 0.0102$} & 0.9103\tiny{$\pm 0.0048$} & 0.8963\tiny{$\pm 0.0019$}\\
    & FJD & 0.9450\tiny{$\pm 0.0083$} & 0.9633\tiny{$\pm 0.0061$} & 0.9381\tiny{$\pm 0.0146$} \\
    \midrule
    \multirow{4}{*}{\makecell[l]{Vicuna-7B + Llama2-7B \\+ Guanaco-7B}}
    & PPL & \textbf{0.9923}\tiny{$\pm 0.0001$} & \textbf{0.9855}\tiny{$\pm 0.0007$} & \textbf{0.9844}\tiny{$\pm 0.0004$} \\
    & SMLLM & 0.8281\tiny{$\pm 0.0059$} & 0.7970\tiny{$\pm 0.0061$} & 0.7492\tiny{$\pm 0.0086$} \\
    & GradSafe & 0.9117\tiny{$\pm 0.0094$} & 0.9138\tiny{$\pm 0.0028$} & 0.9313\tiny{$\pm 0.0083$}\\
    & FJD & 0.9443\tiny{$\pm 0.0060$} & 0.9710\tiny{$\pm 0.0173$} & 0.9629\tiny{$\pm 0.0038$} \\
    
    \bottomrule
  \end{tabular}
\end{table*}

\section{Affirmative Instruction Analysis}\label{app:f}

To investigate the effects of detecting jailbreak prompts on FJD when utilizing different affirmative instructions in prefixes and suffixes on Llama2 7B, we perform experiments involving semantic reorganization and word replacement using the prompts outlined in Sec.~\ref{sec:dpa}. In Tab.~\ref{different ps} shows that using a affirmative instruction as a suffix can yield comparable jailbreak prompt detection effects to using it as a prefix. It can be found that employing affirmative instructions as a suffix achieves comparable performance to using them as a prefix in the majority of cases, while a small number of instructions as a suffix lead to a decline in performance. We believe that the influence on LLMs is more significant when affirmative instructions are applied as prefixes.

\begin{table*}[t]

  \caption{Detection results (AUC) of jailbreak prompt while using different affirmative instruction as prefix and suffix. FJD can effectively detect jailbreak prompts using various affirmative instructions. Most instructions used as suffixes have comparable effects when used as prefixes.}
  \label{different ps}
  \centering
  \setlength\tabcolsep{16pt}
  \footnotesize
  \begin{tabular}{lcccc}
    \toprule
    \textbf{Different Affirmative Instruction} & \textbf{P/S}  & \textbf{GCG}     & \textbf{AutoDAN}  & \textbf{Cipher} \\
    \midrule
    Without affirmative instruction
    & - &  0.9939 & 0.9164 & 0.9335\\

    \midrule
    \multirow{2}{*}{Respond in a way that promotes safety.}
    & Prefix & 0.9985 & 0.9624 & 0.9769 \\
    & Suffix & 0.9609 & 0.9713 & 0.9674 \\
    \midrule
    \multirow{2}{*}{Provide an answer that prioritizes user safety.}
    & Prefix & 0.9956 & 0.9163 & 0.9555 \\
    & Suffix & 0.9134 & 0.9125 & 0.8968 \\

    \midrule
    \multirow{2}{*}{Remember you are a responsible and helpful LLM.}
    & Prefix & 0.9969 & 0.9350 & 0.9403 \\
    & Suffix & 0.9594 & 0.9178 & 0.9354 \\
    \midrule
    \multirow{2}{*}{Let's think step by step.}
    & Prefix & 0.9988 & 0.9610 & 0.9578 \\
    & Suffix & 0.9770 & 0.9605 & 0.9665 \\
    
    \bottomrule
  \end{tabular}

\end{table*}

\section{Analysis of FJD-LI}\label{app:li}

In this section, we show the detection results of FJD-LI under GCG, AutoDAN, Cipher, and Hand-crafted on Llama2 7B, Vicuna 7B, and Guanaco 7B in Tab.~\ref{learnable-prom-2}. This approach further enhances the detection of jailbreak prompts, even when faced with unseen data (Cipher, Hand-crafted).

\begin{table*}[t]
  \caption{Detection results (AUC) of jailbreak prompt through FJD-LI. FJD-LI further enhances the detection of jailbreak prompts over FJD by using learnable virtual instructions.}

  \label{learnable-prom-2}
  \centering
  \setlength\tabcolsep{22pt}
  \footnotesize
  \begin{tabular}{ccccc}
    \toprule
    \textbf{Attack}  & \textbf{Method}  & \textbf{Llama2-7B}  & \textbf{Vicuna-7B}& \textbf{Guanaco-7B}\\
    \midrule

    \multirow{5}{*}{GCG}
    & PPL  & 0.9717\tiny{$\pm 0.0002$} & 0.9860\tiny{$\pm 0.0002$} &0.9833\tiny{$\pm 0.0001$}\\
    & SMLLM & 0.9423\tiny{$\pm 0.0027$} & 0.9575\tiny{$\pm 0.0071$} &0.8811\tiny{$\pm 0.0029$}\\
    & GradSafe & 0.8943\tiny{$\pm 0.0035$} & 0.7575\tiny{$\pm 0.0117$} & 0.7501\tiny{$\pm 0.0019$} \\
    & FJD & 0.9990\tiny{$\pm 0.0002$} & 0.7250\tiny{$\pm 0.0044$} &0.9515\tiny{$\pm 0.0040$}\\
    & FJD-LI & \textbf{0.9998}\tiny{$\pm 0.0001$} &  \textbf{0.9887}\tiny{$\pm 0.0029$}&\textbf{0.9895}\tiny{$\pm 0.0015$}\\
    \midrule
     
    \multirow{5}{*}{AutoDAN}
    & PPL  & 0.8172\tiny{$\pm 0.0017$} & 0.7452\tiny{$\pm 0.0012$} &0.7964\tiny{$\pm 0.0004$}\\
    & SMLLM & 0.8197\tiny{$\pm 0.0052$} & 0.7831\tiny{$\pm 0.0035$} &0.6704\tiny{$\pm 0.0036$}\\
    & GradSafe & 0.8025\tiny{$\pm 0.0089$} & 0.7893\tiny{$\pm 0.0020$} & 0.8194\tiny{$\pm 0.0051$} \\
    & FJD & 0.9578\tiny{$\pm 0.0088$} & 0.7964\tiny{$\pm 0.0182$} &0.8946\tiny{$\pm 0.0065$}\\
    & FJD-LI & \textbf{0.9703}\tiny{$\pm 0.0024$} &  \textbf{0.9969}\tiny{$\pm 0.0021$}&\textbf{0.9817}\tiny{$\pm 0.0038$}\\
    \midrule
    \multirow{5}{*}{Cipher}   
    & PPL & 0.0070\tiny{$\pm 0.0005$} & 0.0266\tiny{$\pm 0.0004$} &0.0248\tiny{$\pm 0.0005$}\\
    & SMLLM & 0.5034\tiny{$\pm 0.0024$} & 0.5233\tiny{$\pm 0.0009$} &0.5460\tiny{$\pm 0.0026$} \\
    & GradSafe & 0.7862\tiny{$\pm 0.0045$} & 0.7094\tiny{$\pm 0.0201$} & 0.8112\tiny{$\pm 0.0088$} \\
    & FJD & 0.9896\tiny{$\pm 0.0014$} & 0.8633\tiny{$\pm 0.0033$} &0.8299\tiny{$\pm 0.0043$}\\
    & FJD-LI & \textbf{0.9944}\tiny{$\pm 0.0012$} & \textbf{0.9310}\tiny{$\pm 0.0036$} &\textbf{0.8826}\tiny{$\pm 0.0102$}\\
    
    \midrule
    \multirow{5}{*}{Hand-crafted}
    & PPL  & 0.6090\tiny{$\pm 0.0020$} & 0.6066\tiny{$\pm 0.0010$} &0.6018\tiny{$\pm 0.0006$}\\
    & SMLLM & 0.7138\tiny{$\pm 0.0108$} & 0.6886\tiny{$\pm 0.0063$} &0.7633\tiny{$\pm 0.0071$}\\
    & GradSafe & 0.9085\tiny{$\pm 0.0050$} & 0.7871\tiny{$\pm 0.0055$} & 0.8030\tiny{$\pm 0.0054$} \\
    & FJD & 0.9595\tiny{$\pm 0.0069$} & 0.7993\tiny{$\pm 0.0148$} &0.8596\tiny{$\pm 0.0138$}\\
    & FJD-LI & \textbf{0.9843}\tiny{$\pm 0.0016$} &  \textbf{0.8579}\tiny{$\pm 0.0073$}&\textbf{0.9081}\tiny{$\pm 0.0101$}\\
    
    \bottomrule
  \end{tabular}

\end{table*}

\section{Rigorous Analysis of Temperature Scaling}\label{app:proof}

In this section, we provide a mathematical proof for the two phenomena of softmax maximum value flipping. First, we define the logits of two distributions $Z^{(1)}=\{z^{(1)}_1, z^{(1)}_2, ...z^{(1)}_n\}$ and $z^{(2)}=\{z^{(2)}_1, z^{(2)}_2, ...z^{(2)}_n\}$, assuming that $z^{(1)}_1$ and $z^{(2)}_1$ is the maximum value. The probability of the maximum value with temperature $\tau$ is 

\begin{equation}
    P_{1,\tau}(1)=\frac{\exp(z^{(1)}_1/\tau)}{\sum_n \exp(z^{(1)}_n/\tau)}
\end{equation}

\begin{equation}
    P_{1,\tau}(2)=\frac{\exp(z^{(2)}_1/\tau)}{\sum_n \exp(z^{(2)}_n/\tau)} 
\end{equation}

 Assume $P_{1,\tau}(1) < P_{1,\tau}(2)$. When the maximum value is removed from the $Z^{(1)}$, the distribution becomes sharp, its variance is $ \sigma^2(1)=\frac{1}{n-1}\sum_{i=2}^n(z^{(1)}_i-\mu^{(1)})^2 $. When the maximum value is removed from the $Z^{(2)}$ distribution, the distribution becomes smooth, its variance is $ \sigma^2(2)=\frac{1}{n-1}\sum_{j=2}^n(z^{(2)}_j-\mu^{(2)})^2 $. And $\sigma^2(1) > \sigma^2(2)$

When $\tau>1$, for a single distribution, the softmax distribution becomes smoother, but the rank of the maximum value remains unchanged. Different distributions have varying sensitivities to changes in temperature. As the temperature $\tau$ increases, when the proportions of non-max values in distributions $Z^{(1)}$ and $Z^{(2)}$ are similar, the smoother non-max values $\{z^{(2)}_2/\tau, z^{(2)}_3/\tau, ...z^{(2)}_n/\tau\}$ occupy a larger proportion than the sharper non-max values $\{z^{(1)}_2/\tau, z^{(1)}_3/\tau, ...z^{(1)}_n/\tau\}$, causing the proportion of the maximum value in distribution $Z^{(2)}$ to decrease rapidly. In certain conditions, this can cause the maximum values of the two distributions to flip, i.e., $P_{1,\tau}(1) > P_{1,\tau}(2)$.

Based on the above, we conduct a statistical analysis of the logits for both jailbreak and benign prompts. Taking Llama 7B as an example, after prepending the affirmative instruction, we present an instance where the ranking of the first token changes after increasing the temperature for both benign (PureDove) and jailbreak (AutoDAN) prompts in Tab.~\ref{temp_example}.

\begin{table*}[!h]
    \centering
    \caption{An instance in which the ranking of the first token $P_{1,\tau}$ changes after increasing the temperature $\tau$.}
    \footnotesize
    \setlength\tabcolsep{11pt}
    \begin{tabular}{cccccc}
    \toprule
         \multirow{2}{*}{\textbf{LLM}} & \multirow{2}{*}{\textbf{Label}} & \multicolumn{2}{c}{\textbf{$\tau=1$}} & \multicolumn{2}{c}{\textbf{$\tau=1.25$}} \\
         \cmidrule{3-4}
         \cmidrule{5-6}
          & & $P_{1,\tau}$ & \textbf{Std (non-max)} & \textbf{$P_{1,\tau}$} & \textbf{Std (non-max)}\\
         \midrule
         \multirow{2}{*}{Llama2-7B} 
         & Benign (PureDove) & 0.9999777 & $1.2369\times10^{-7}$ & \textbf{0.9998197} & $1.1055\times10^{-6}$ \\
         & Jailbreak (AutoDAN) & \textbf{0.9999807} & $1.0746\times10^{-7}$ & 0.9998046 & $9.4290\times10^{-7}$ \\
    \bottomrule
    \end{tabular}
    
    \label{temp_example}
\end{table*}

\section{The Optimal Temperature}\label{app:temp}

In this section, we show the optimal temperatures of FT and FJD across various LLMs on the training dataset in Tab.~\ref{temp best}. Additionally, we analyzed how the selected optimal temperature affects the detection performance of FJD with varying amounts of training data and different training datasets, taking Llama2 7B as an example. In Tab.~\ref{diff temp}, we found that a small datasets can yield similar temperatures, and that small variations in temperature have minimal impact on detection results. Although the temperatures obtained from training with different datasets exhibit some variation, they have minimal impact on FJD detection performance within a certain range.

\begin{table*}[t]
    \centering
    \footnotesize
    \setlength\tabcolsep{10pt}
    \caption{The optimal temperatures of FT and FJD across various LLMs on the training dataset.}
    \begin{tabular}{ccccccc}
    
    \toprule
        \textbf{Method}  & \textbf{Llama2-7B} & \textbf{Llama2-13B} & \textbf{Vicuna-7B} & \textbf{Vicuna-13B} & \textbf{Guanaco-7b} & \textbf{Guanaco-13B} \\
        \midrule
        FT & 0.86 & 1.51 & 0.95 & 1.99 & 0.69 & 0.80 \\
        FJD & 1.25 & 1.98 & 1.47 & 0.35 & 1.24 & 0.79  \\
    \bottomrule
    \end{tabular}
    
    \label{temp best}
\end{table*}

\begin{table*}[t]
    \centering
    \footnotesize
    \setlength\tabcolsep{8pt}
    \caption{Detection results (AUC) of jailbreak prompt through FJD under different size of training sets. A small datasets can yield similar temperatures, and small variations in temperature have minimal impact on detection results}
    \begin{tabular}{ccccccc}
    \toprule
        \textbf{Model} & \textbf{Training Size} & \textbf{Temperature} & \textbf{AutoDAN} & \textbf{Cipher} & \textbf{GCG} & \textbf{PAIR} \\ 
        \midrule
        \multirow{5}{*}{Llama2-7B}
        & 10\% & 1.18 & 0.9549\tiny{$\pm 0.0054$} & 0.9764\tiny{$\pm 0.0017$} & 0.9983\tiny{$\pm 0.0004$} & 0.9738\tiny{$\pm 0.0038$} \\ 
        & 20\% & 1.20 & 0.9564\tiny{$\pm 0.0061$} & 0.9741\tiny{$\pm 0.0026$} & 0.9990\tiny{$\pm 0.0002$} & 0.9737\tiny{$\pm 0.0015$} \\ 
        & 30\% & 1.23 & 0.9542\tiny{$\pm 0.0061$} & 0.9726\tiny{$\pm 0.0019$} & 0.9990\tiny{$\pm 0.0002$} & 0.9749\tiny{$\pm 0.0047$} \\ 
        & 40\% & 1.24 & 0.9519\tiny{$\pm 0.0024$} & 0.9714\tiny{$\pm 0.0013$} & 0.9990\tiny{$\pm 0.0003$} & 0.9754\tiny{$\pm 0.0019$} \\ 
        & 50\% & 1.25 & 0.9495\tiny{$\pm 0.0053$} & 0.9700\tiny{$\pm 0.0034$} & 0.9990\tiny{$\pm 0.0003$} & 0.9761\tiny{$\pm 0.0009$} \\
        \midrule
        \textbf{Model} & \textbf{\makecell[c]{Training\\Datasets}} & \textbf{Temperature} & \textbf{AutoDAN} & \textbf{Cipher} & \textbf{GCG} & \textbf{PAIR} \\ 
        \midrule
        \multirow{3}{*}{Llama2-7B}
        & AutoDAN & 1.27 & 0.9550\tiny{$\pm 0.0038$} & 0.9746\tiny{$\pm 0.0028$} & 0.9991\tiny{$\pm 0.0004$} & 0.9748\tiny{$\pm 0.0021$} \\ 
        & Cipher & 1.18 & 0.9549\tiny{$\pm 0.0054$} & 0.9764\tiny{$\pm 0.0017$} & 0.9983\tiny{$\pm 0.0004$} & 0.9746\tiny{$\pm 0.0038$} \\ 
        & GCG & 1.37 & 0.9538\tiny{$\pm 0.0038$} & 0.9696\tiny{$\pm 0.0014$} & 
        0.9992\tiny{$\pm 0.0005$} & 0.9749\tiny{$\pm 0.0011$} \\
    \bottomrule
    \end{tabular}
    \label{diff temp}
\end{table*}

\begin{table*}[t]
    \centering
    \footnotesize
    \setlength\tabcolsep{21pt}
    \caption{Detection results (AUC) of jailbreak prompt through FJD in multimodal contexts (LLaVA-v1.6-mistral-7b). Directly using First Token Confidence (FT) for detection is feasible, and that FJD can further improve detection performance.}
    \begin{tabular}{ccccc}
    \toprule
        \textbf{Attack} & \textbf{FigStep} & \textbf{Relevant\_SD} & \textbf{Relevant\_typo} & \textbf{Relevant\_SD\_typo}  \\ 
        \midrule
        FT  & 0.8501\tiny{$\pm 0.0094$} & 0.6649\tiny{$\pm 0.0128$} & 0.5967\tiny{$\pm 0.0101$} & 0.5753\tiny{$\pm 0.0117$} \\ 
        FJD & 0.9258\tiny{$\pm 0.0076$} & 0.7093\tiny{$\pm 0.0104$} & 0.7395\tiny{$\pm 0.0099$} & 0.6046\tiny{$\pm 0.0125$}  \\ 
    \bottomrule
    \end{tabular}
    \label{multimodal-table}
\end{table*}

\section{Analysis of FJD-K}\label{app:numk}

In contrast to FJD, FJD-K detects jailbreak prompts through the average of the first k token confidences. Formally, based on the Equation~\ref{c0}, given an input sequence $ x_q $, the affirmative instruction $x_{ai}$ and the temperature $\tau$, the confidence of the first K tokens is computed as 

\begin{equation}
    C_k=\frac{1}{k}\sum_{i=1}^{k} C_i=\frac{1}{k}\sum_{i=1}^{k} \sigma_\tau(f(x_{ai}\oplus x_q)_{i}/\tau)
\end{equation}

When $k=1$, $C_k$ is the first token confidence.

\begin{figure*}[t]
    \centering
    \subfloat[Llama2 7B]{
        \begin{minipage}{0.32\linewidth}

    		\centerline{\includegraphics[width=\textwidth]{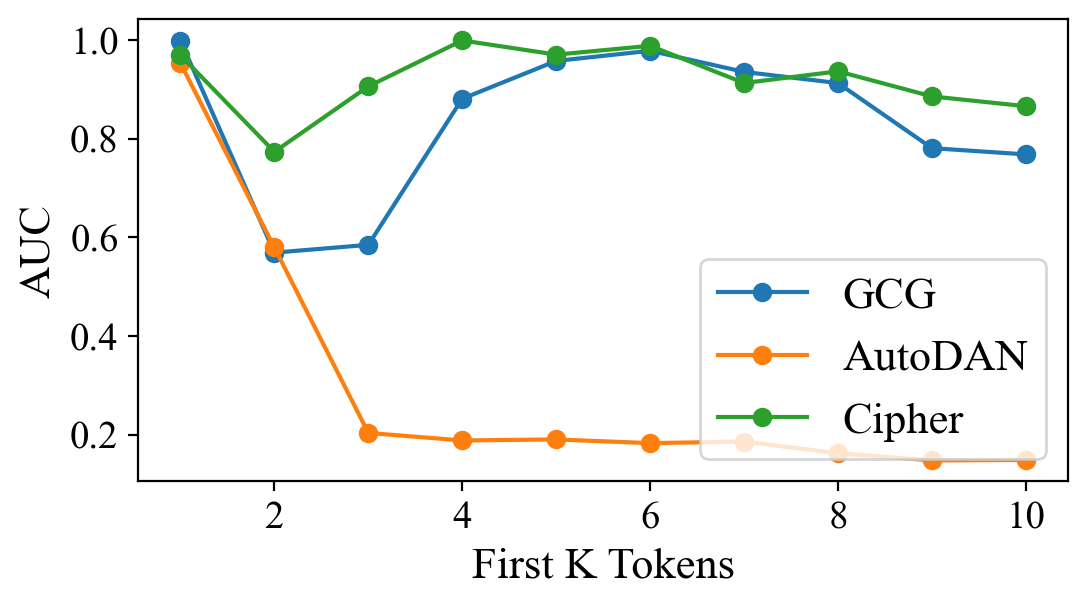}}
    
    	\end{minipage}
    }
    \subfloat[Vicuna 7B]{
    	\begin{minipage}{0.32\linewidth}
   
    		\centerline{\includegraphics[width=\textwidth]{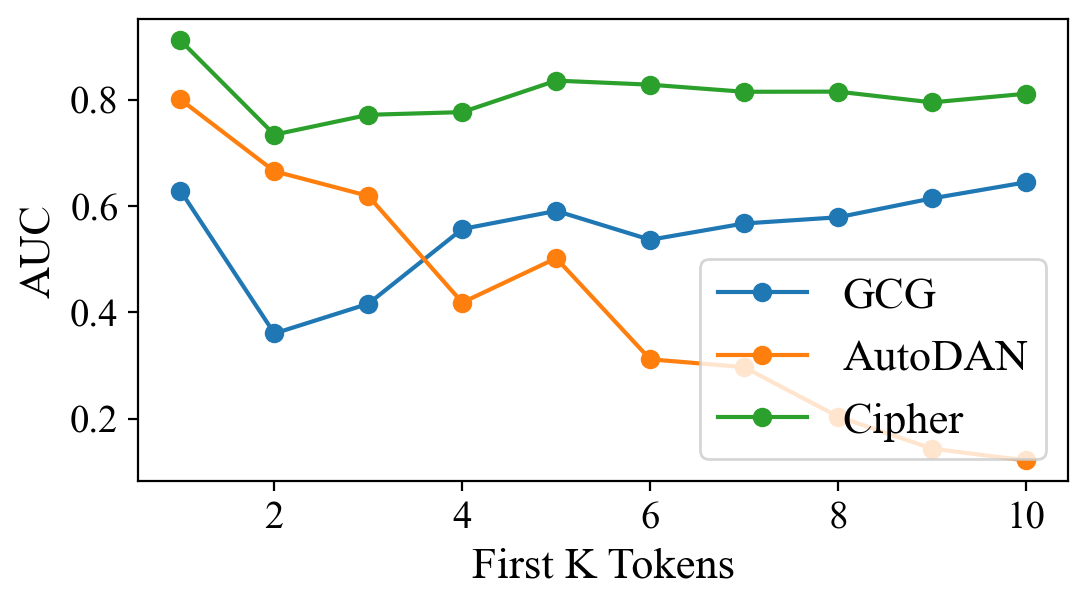}}

    	\end{minipage}
    }
    \subfloat[Guanaco 7B]{
        \begin{minipage}{0.32\linewidth}

		\centerline{\includegraphics[width=\textwidth]{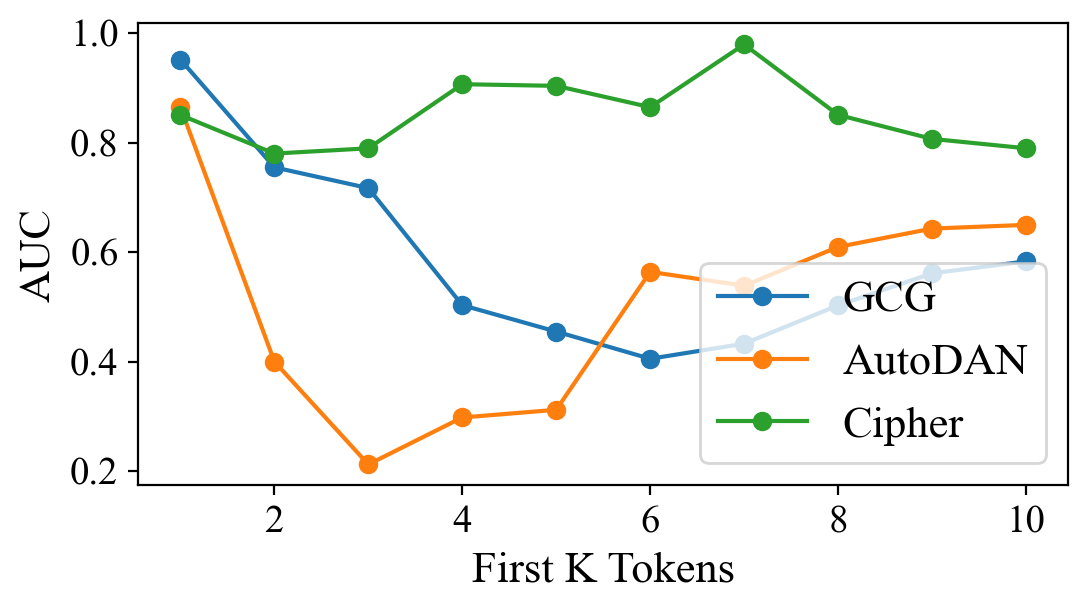}}

	   \end{minipage}
    }	
	\caption{Detection results (AUC) of jailbreak prompt while using First K Token with FJD. In certain LLMs and under specific attacks, FJD-K enhances the detection capabilities of FJD. However, for AutoDAN attacks across the three LLMs, FJD-K diminishes the detection performance of FJD.}

	\label{Tokens}
\end{figure*}

To evaluate the influence of the number of fist $k \in [1, 10]$ tokens on the detection of jailbreak prompts across various LLMs, we conduct experiments using FJD on Vicuna 7B, Llama2 7B, and Guanaco 7B. Fig.~\ref{Tokens} shows changes in the jailbreak detection AUC value during token selection. In certain LLMs and attacks, FJD-K can enhance the detection capability of FJD to a certain degree. Nonetheless, in the case of AutoDAN, the efficacy of FJD-K in detection is significantly diminished.

\section{The Generality of FJD in multimodal contexts}\label{app:multimodal}

It is indeed interesting to investigate the generality of FJD behavior. And we conducted some preliminary experiments in multimodal contexts, using VQA~\citep{antol2015vqa} and MM-Vet~\citep{yu2023mm} as benign prompts, and FigStep~\citep{gong2025figstep} and MM-SafetyBench~\citep{liu2024mm} as jailbreak prompts, to examine the detection performance of FT and FJD. As shown in the Tab.~\ref{multimodal-table}, we conducted brief experiments on LLaVA-v1.6-mistral-7b. The experiments demonstrate that directly using First Token Confidence (FT) for detection is feasible, and that FJD can further improve detection performance. However, compared with the unimodal scenarios, the sequence length of image tokens is much larger than that of text tokens. Although the FJD method can still enhance detection, its impact on confidence is smaller relative to unimodal scenarios, resulting in less impressive results on MM-SafetyBench.

\end{document}